\documentclass[fleqn,usenatbib]{mnras}

\usepackage[T1]{fontenc}

\DeclareRobustCommand{\VAN}[3]{#2}
\let\VANthebibliography\thebibliography
\def\thebibliography{\DeclareRobustCommand{\VAN}[3]{##3}\VANthebibliography}

\usepackage{graphicx}	
\usepackage{amsmath}	
\usepackage{amssymb}	
\usepackage{amsfonts}
\usepackage{sansmath}
\usepackage{newtxtext,newtxmath}

\newcommand{\vs}{$v \sin i$}
\newcommand{\kms}{km\,s$^{-1}$}
\newcommand{\ms}{m\,s$^{-1}$}

\newcommand{\kinj}{K$_{\rm{inj}}$}
\newcommand{\porb}{P$_{\rm{orb}}$}
\newcommand{\php}{$\phi_{\rm{p}}$}

\newcommand{\kest}{k$_{\rm{est}}$}

\newcommand{\vobs}{\mathbfit{v}$_{\rm{obs}}$}
\newcommand{\vpar}{\mathbfit{v}$_{\parallel}$}
\newcommand{\vper}{\mathbfit{v}$_{\perp}$}

\defcitealias{cameron2021}{CC21}

\title[Eight-year study of solar activity]{Investigating stellar activity through eight years of Sun-as-a-star observations}

\author[B. Klein et al.]{
Baptiste Klein,$^{1}$\thanks{E-mail: baptiste.klein@physics.ox.ac.uk}
Suzanne Aigrain,$^{1}$
Michael Cretignier,$^{1}$
Khaled Al Moulla,$^{2}$
Xavier Dumusque,$^{2}$
\newauthor
Oscar Barrag\'an,$^{1}$
Haochuan Yu,$^{1}$
Annelies Mortier,$^{3}$
Federica Rescigno,$^{4}$
Andrew Collier Cameron,$^{5}$
\newauthor
Mercedes L\'opez-Morales,$^{6}$
Nad\`ege Meunier,$^{7}$
Alessandro Sozzetti,$^{8}$
Niamh K. O'Sullivan$^{1}$
\\
$^{1}$Department of Physics, University of Oxford, Oxford OX1 3RH, UK\\
$^{2}$Observatoire Astronomique de l’Université de Genève, Chemin Pegasi 51, 1290 Versoix, Switzerland\\
$^{3}$School of Physics \& Astronomy, University of Birmingham, Edgbaston, Birmingham B15 2TT, UK\\
$^{4}$Department of Astrophysics, University of Exeter, Stocker Rd, Exeter, EX4 4QL, UK\\
$^{5}$Centre for Exoplanet Science / SUPA, School of Physics \& Astronomy,
University of St Andrews, North Haugh ST ANDREWS, Fife, KY16 9SS, UK\\
$^{6}$Center for Astrophysics | Harvard \& Smithsonian, 60 Garden Street, Cambridge, MA 02138, USA\\
$^{7}$Univ. Grenoble Alpes, CNRS, IPAG, 38000 Grenoble, France\\
$^{8}$INAF – Osservatorio Astrofisico di Torino, via Osservatorio 20, 10025 Pino Torinese, Italy
}

\date{Accepted 2024 May 20. Received 2024 May 20; in original form 2024 February 28}

\pubyear{2024}

\begin{document}
\label{firstpage}
\pagerange{\pageref{firstpage}--\pageref{lastpage}}
\maketitle

\begin{abstract}
Stellar magnetic activity induces both distortions and Doppler-shifts in the absorption line profiles of Sun-like stars. Those effects produce apparent radial velocity (RV) signals which greatly hamper the search for potentially habitable, Earth-like planets. In this work, we investigate these distortions in the Sun using cross-correlation functions (CCFs), derived from intensive monitoring with the high-precision spectrograph HARPS-N. We show that the RV signal arising from line-shape variations on time-scales associated with the Sun’s rotation and activity cycle can be robustly extracted from the data, reducing the RV dispersion by half. Once these have been corrected, activity-induced Doppler-shifts remain, that are modulated at the solar rotation period, and that are most effectively modelled in the time domain, using Gaussian Processes (GPs). Planet signatures are still best retrieved with multi-dimensonal GPs, when activity is jointly modelled from the raw RVs and indicators of the line width or of the Ca II H \& K emission. After GP modelling, the residual RVs exhibit a dispersion of 0.6-0.8 m\,s$^{-1}$, likely to be dominated by signals induced by super-granulation. Finally, we find that the statistical properties of the RVs evolve significantly over time, and that this evolution is primarily driven by sunspots, which control the smoothness of the signal. Such evolution, which reduces the sensitivity to long-period planet signatures, is no longer seen in the activity-induced Doppler-shifts, which is promising for long term RV monitoring surveys such as the Terra Hunting Experiment or the PLATO follow-up campaign. 
\end{abstract}


\begin{keywords}
Sun: activity -- techniques: radial velocities -- planets and satellites: detection -- line: profiles -- methods: statistical 
\end{keywords}


\section{Introduction}

Doppler spectroscopy is one of the bedrocks of past, present and future exoplanetary science. As of 2024, it remains the preferred technique for confirming transiting planet candidates and the most prolific method for detecting non-transiting planets. It provides one of the most fundamental parameters of an exoplanet, its mass, crucial to constrain its inner composition \citep{mordasini2012} and to characterise its atmosphere \citep{batalha2019}. Since the detection of 51~Pegasi~b by \citet{mayor1995}, the radial velocity (RV) accuracy of optical high-resolution \'echelle spectrographs has dramatically increased, to the point where planets with RV semi-amplitudes below 1~\ms\ can be reliably detected \citep[e.g.][]{faria2022,john2023}. New-generation spectrographs like ESPRESSO \citep{pepe2021}, EXPRES \citep{jurgenson2016}, HARPS-3 \citep{thompson2016}, NEID \citep{schwab2016} and KPF \citep{gibson2016} have been designed to provide RVs with sub 0.3-\ms\ precision, suggesting that a detection of an Earth-like planet around a Sun analog may be possible in the coming decade.

The intrinsic variability of stars is currently the main limitation to the detection of low-mass and long-period planets with Doppler spectroscopy \citep[see][]{fischer2016,crass2021,meunier2021}. Various processes related to photospheric flows \citep[e.g. granulation, super-granulation, meridional circulation;][]{dumusque2011,meunier2015,cegla2018,meunier2019,meunier2020} and magnetic activity \citep[e.g., active regions, flares, magnetic cycles;][]{saar1997,desort2007,meunier2010,lovis2011,gomes2012} distort stellar line profiles, giving rise to RV signals that hamper the search for planet signatures. The accurate modelling of these signals has become an extremely active area of research, with promising state-of-the-art methods currently under investigation \citep[see][for a review]{zhao2022}. The mathematically tractable and flexible framework of Gaussian processes \citep{aigrain2022} is now widely used to model the signals induced by stellar activity on the RVs and other activity indicators \citep[e.g.][]{haywood2014,rajpaul2015,delisle2022,barragan2022}. On the other hand, more and more innovative methods aim to correct line distortions directly from the cross-correlation functions \citep[e.g.][]{cameron2021,klein2022,deBeurs2022} or from the spectra \citep[e.g.][]{jones2017,dumusque2018,rajpaul2020,lienhard2022,cretignier2022}. Yet, none of the methods is currently able to filter stellar activity RV signals significantly below the symbolic barrier of $\sim$1~\ms\ (after averaging over signals induced by P-mode oscillations and granulation), and remain poorly sensitive to low-amplitude ($\lesssim$0.5\,\ms) long-period ($\gtrsim$100\,d) planet signals.


The Sun is our best laboratory to better understand stellar activity signals in high-resolution spectra. It is the only star whose surface can be resolved at high resolution and for which planet-induced RV variations can be integrally removed. Since 2015, the Sun has been intensively monitored with the High Accuracy Radial-velocity Planet Searcher in the Northern hemisphere \citep[HARPS-N;][]{cosentino2012}, making it possible to investigate the full contributions of active regions on high-resolution spectra \citep{milbourne2019,thompson2020,haywood2022,lienhard2023}. These high-cadence high-precision solar spectra are ideally suited to assess the ability of a given method to filter stellar variability \citep[][Dumusque et al. in prep.]{dumusque2021}. Additionally, continuous monitoring of the solar surface with, for example, the Helioseismic and Magnetic Imager (HMI) instrument onboard the Solar Dynamics Observatory \citep[SDO;][]{pesnell2012} makes it possible to better understand the physical processes driving the observed disc-integrated RVs \citep[e.g.][Rescigno et al., submitted]{milbourne2019,haywood2022}.

In this study, we investigate how the HARPS-N solar activity RV signals can be modelled using information in both wavelength and time domains. In Sec.~\ref{sec:sec2}, we present our framework to separate a time series of cross-correlation functions (CCFs) into two components whose temporal variations are caused either by pure Doppler shifts or by shape-distortions. After describing the HARPS-N solar data set in Sec.~\ref{sec:observations_red}, we investigate, in Sec.~\ref{sec:results}, the effects of filtering the shape-driven distortions on the solar RVs, and assess the planet detectability in the activity-filtered RVs through a large number of planet injection-recovery tests. In Sec.~\ref{sec:gp_sec}, we couple this framework with Gaussian processes and, notably, study the effect of combining newly-extracted shape-driven activity indicators with the HARPS-N solar RVs in a multi-dimensional Gaussian process framework. Finally, in Sec.~\ref{sec:section:discussion}, we discuss the implications of our results in terms of activity modelling and planet search in the light of forthcoming long-term missions.

\section{Method}\label{sec:sec2}

Our approach builds on the Doppler-constrained principal component analysis (PCA) framework of \citet{jones2017} and on the \texttt{SCALPELS} formalism introduced by \citet{cameron2021}, borrowing elements from both to separate, as simply as possible, the components of the RV variations that can be explained by a pure Doppler shift from the rest. Using the same designations as \citet{cameron2021}, we refer to these two components as the shift- and shape-driven velocities, respectively.

\subsection{Framework}\label{ssec:framework}

We start from CCFs produced by, for example, the HARPS-N data reduction software (see Sec.~\ref{ssec:observations}). We consider a time series of $n$ CCFs spanning $m$ velocity bins $\bmath{v}$~=~($v_{1}$,...,$v_{m}$) and the associated RV time-series \mathbfit{v}$_{\rm{obs}}$ (typically taken from the reduction pipeline). Our framework requires the computation of a non-noisy and activity-low reference CCF, \mathbfit{C}$_{\rm{R}}$, obtained by computing the inverse-variance-weighted mean of all the CCFs in the stellar rest frame\footnote{One alternative could be to compute \mathbfit{C}$_{\rm{R}}$ by averaging just a subset of CCFs, around the activity minimum of the star. However, we caution that robustly identifying activity minimums can be challenging in some cases, and that stars still exhibit signs of activity during the cycle minimum, as shown in Sec.~\ref{ssec:1D_gp_new}. In practice, we find that the computation of \mathbfit{C}$_{\rm{R}}$ marginally impacts the results.}. We use the simple analytic framework of \citet{jones2017} to decompose the time series of reference-subtracted CCFs, \mathbfss{C}, into a pure Doppler component, \mathbfss{C}$_{\rm{D}}$, and residuals, \mathbfss{C}$_{\rm{S}}$, whose variations are driven by distortions in the profile shape. Assuming that the Doppler shifts are small compared to the line width, we approximate the Doppler component by the first-order term of the Taylor expansion of \mathbfss{C} with respect to the CCF velocity bins, $\bmath{v}$ \citep[see][]{bouchy2001}, matching to the spectrum's resolution. Therefore, noting \mathbfit{D}~$ =  \left[ \boldsymbol{C_{\rm{R}}}' \left(v_{1}\right) ,...,\boldsymbol{C_{\rm{R}}}'\left(v_{m}\right)  \right]^\intercal$, the first derivative of the reference CCF with respect to the velocity bins, we have

\begin{align}
    \textsf{\textbf{C}}_{\rm{D}}  =  \textsf{\textbf{C}} \cdot \frac{\boldsymbol{D} \boldsymbol{D}^\intercal  }{\boldsymbol{D}^\intercal \boldsymbol{D}} \quad &  & 
    \textsf{\textbf{C}}_{\rm{S}}  =  \textsf{\textbf{C}} - \textsf{\textbf{C}}_{\rm{D}}.
    \label{eq:dcpca}
\end{align}

\noindent
The time series of residuals, \mathbfss{C}$_{\rm{S}}$, contains all the shape distortions of the input CCFs, but are, in principle, insensitive to Doppler shifts. Following the method introduced by \citet{cameron2021}, we use singular-value decomposition (SVD) to build a set of $k$ independent vectors, $\{ \boldsymbol{U_{1}},...,\boldsymbol{U_{k}} \}$, from \mathbfss{C}$_{\rm{S}}$. Each of these vectors is a time series of $n$ points associated with a score that scales with its contribution to the variance of \mathbfss{C}$_{\rm{S}}$. In what follows, the basis vectors are sorted by decreasing score value, so that $\boldsymbol{U_{1}}$ and $\boldsymbol{U_{k}}$ contribute the most and the least to the variance of \mathbfss{C}$_{\rm{S}}$, respectively.

Finally, we project our reference-subtracted CCFs \mathbfss{C} onto the basis \mathbfss{U}\,=\,$\left( \boldsymbol{U_{1}},...,\boldsymbol{U_{k}} \right)$, where each column $i$ is equal to vector $\boldsymbol{U_{i}}$. Using the same notations as \citet{cameron2021}, we call \mathbfss{C}$_{\parallel}$ the projection of \mathbfss{C} on \mathbfss{U}, such that

\begin{equation}
\textsf{\textbf{C}}_{\parallel} = \textsf{\textbf{U}} \cdot \textsf{\textbf{U}}^\intercal \cdot   \textsf{\textbf{C}}.
\label{eq:projection}
\end{equation}

\noindent
We extract a RV time-series, \mathbfit{v}$_{\parallel}$ (called shape-driven RVs), by fitting a Gaussian function to each CCF of \mathbfss{C}$_{\parallel}$~+~\mathbfss{C}$_{\rm{R}}$. Finally, we build a Doppler-driven (hereafter called shift-driven) RV time-series, \mathbfit{v}$_{\perp}$, by subtracting \mathbfit{v}$_{\parallel}$ from the initial RV time-series \mathbfit{v}$_{\rm{obs}}$.

The number $k$ of principal components used to build \mathbfss{U} must be chosen with caution. Basis vectors marginally contributing to the variance of the input CCF time-series are primarily composed of white noise. Thus, including them in the projection (Eq.~\ref{eq:projection}) will increase the dispersion of the input RVs and interfere with potential planetary signals. In practice, we adapted the method described in the Section~3 of \citet{klein2024} to choose the number of components used in the projection. We independently generate about 10 matrices, matching \mathbfss{C}$_{\rm{S}}$ in shape, but containing only white noise, drawn from the formal uncertainties on the CCF. We then apply SVD to each of these matrices and use the highest eigenvalue S$_{\mathrm{max}}$ of all matrices as a proxy for noise-dominated components. When we apply SVD to \mathbfss{C}$_{\rm{S}}$, components with eigenvalues greater than S$_{\mathrm{max}}$ are assumed to enclose a significant fraction of correlated noise (e.g. activity-induced shape distortions) and are used in the basis definition. Conversely, components with eigenvalues smaller than S$_{\mathrm{max}}$ are assumed to be dominated by white noise and discarded. Note that, in practice, the selected principal components used in the basis are consistent with the "elbow" heuristic method and account for an explained variance score typically greater than 95\%. Assuming that \mathbfss{U} is noise free, all RV variations induced by distortions of the input CCFs are enclosed in \vpar. Therefore, in principle, \vper\ will be composed of planet signatures, Doppler shifts of unknown origin (e.g. residual stellar activity or instrument systematics) and white noise. As a word of caution, we stress that our theoretical framework is limited to spots and faculae, which do affect the shape of the CCFs. Other processes like super-granulation or meridional circulation may induce Doppler shifts without distorting the CCFs \citep[see][and references therein]{meunier2021}.

\subsection{Application to synthetic data}\label{ssec:synthetic}

\begin{table}
    \centering
    \caption{Description of the parameters used to generate time series of CCFs with \texttt{SOAP2}. In each case, we give the number of spots (N$_{\rm{spot}}$) and faculae (N$_{\rm{fac}}$), as well as their corresponding stellar rotational phase $\phi_{\mathrm{r}}$ and latitudes (Lat). The size of the active regions are set to 0.18\% (faculae) and 0.1\% (spot) with respect to the area of the visible hemisphere, in order to produce a RV signal of about 1\,\ms. The last column gives the number N$_{\rm{pl}}$ of pure Doppler signals injected in the CCFs.}
    \label{tab:info_simu}
    \begin{tabular}{cccccccc}
       \hline
       Case  & N$_{\rm{spot}}$ & $\phi_{\mathrm{r}}$ & Lat &  N$_{\rm{fac}}$ & $\phi_{\mathrm{r}}$ & Lat & N$_{\rm{pl}}$  \\
       -- & -- & -- & [deg]  & -- & -- & [deg] & -- \\
       \hline
       (1) & 1 & 0.5 & 0 & 0 & -- & -- & 0  \\
       (2) & 0 & -- & -- & 1 & 0.5 & 0 & 0 \\
       (3) & 2 & 0.0, 0.5 & 30 & 2 & 0.25, 0.75 & 30 & 0 \\
       (4) & 0 & -- & -- & 0 & -- & -- & 1 \\
       \hline
    \end{tabular}
\end{table}

\begin{figure}
    \centering
    \includegraphics[width=\linewidth]{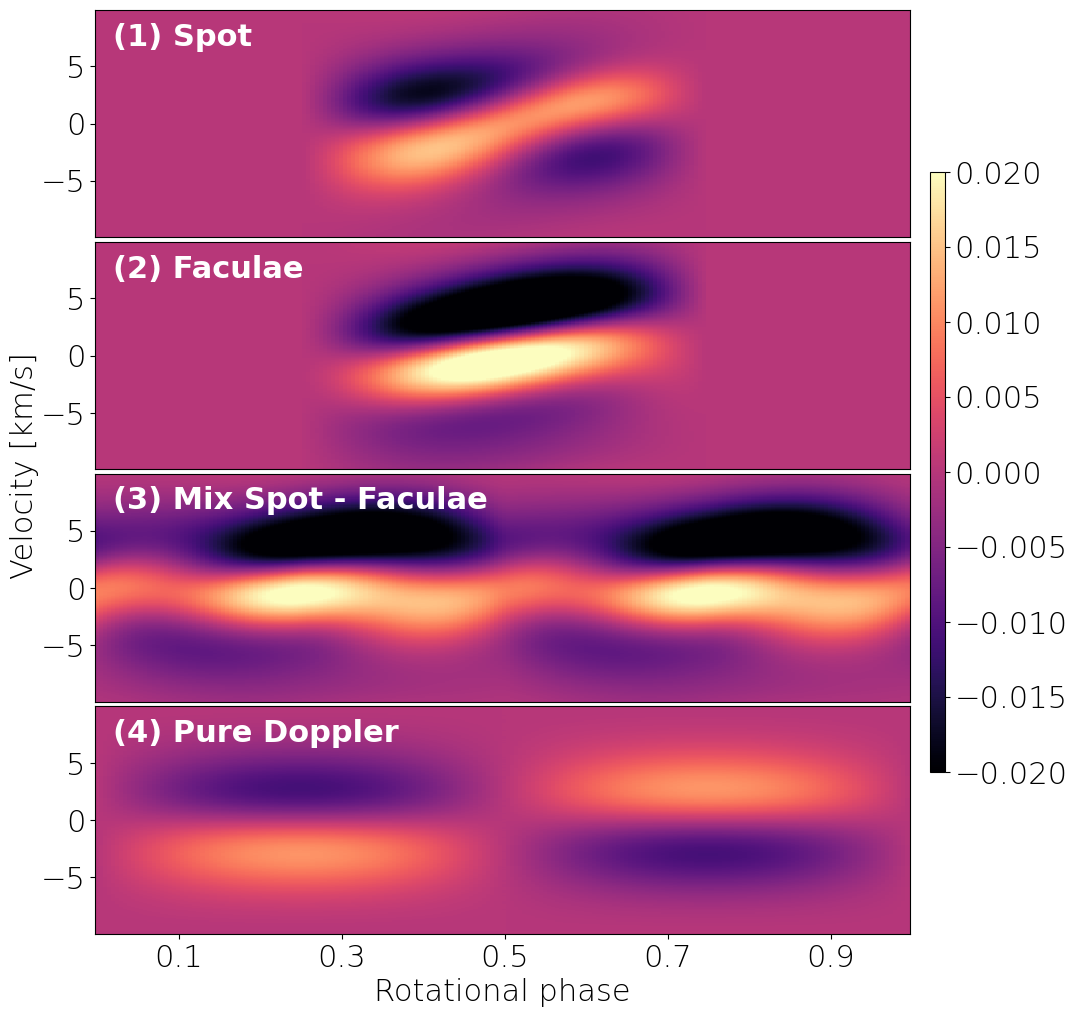}
    \caption{Dynamical spectrum of the different time series of CCFs generated with \texttt{SOAP2} using the parameters listed in Tab.~\ref{tab:info_simu}. Each panel depicts the relative flux of the synthetic CCFs (after subtraction of a reference line profile, computed in absence of activity), in percent, as a function of the rotational phase of the star (X-axis) and line velocity (Y-axis).}
    \label{fig:dynamic_spectrum_soap}
\end{figure}


Before applying the framework introduced in Sec.~\ref{ssec:framework} to real solar data, we conduct some simple tests on synthetic CCFs computed using the \texttt{SOAP2} online tool \citep{dumusque2014,boisse2012}. This code allows the user to simulate the effects of spots (i.e., pure brightness inhomogeneities) and faculae (via the inhibition of the convective blueshift) on the absorption line profile. We consider three different configurations of active regions, listed in Tab.~\ref{tab:info_simu}. Cases (1) and (2) correspond to a single spot and a single facula, at the equator of the star. For case~(3), we consider a mix of four active regions, two spots and two faculae, of equal latitude (30$^{\circ}$) and evenly spread in stellar longitude. Following \citet{berdyugina2005} and \citet{meunier2010}, the temperature of the spots is set to $\sim$650~K below the Sun's effective temperature. The size of the spots (resp. faculae) is set to 1\% (resp. 0.18\%)  of the visible stellar hemisphere, so that the semi-amplitude of the corresponding RV signal is about 1~\ms, which is the typical order of magnitude of activity-induced RV variations for the Sun \citep[e.g.][]{meunier2010}.

Finally, in case (4), we consider the case of a pure Doppler signal in absence of stellar activity, in order to double check  that our method does not affect planetary signatures. The CCFs are generated by interpolating a non-spotted line profile computed with \texttt{SOAP2}, and shifting the entire profile according to the RV signature induced by a 1-\ms\ sine-wave with a period equal to the stellar rotation period to the data (note that no stellar activity is considered in this case). In all cases, we generate a time series of 200 CCFs evenly spanning one stellar rotation cycle, and add a white noise of 10$^{-5}$ with respect to the normalised continuum, which is of the same order of magnitude as the photon noise in HARPS-N solar CCFs. Note that a adding white noise significantly larger than the planet-induced Doppler shift is crucial in case~(4). Since the Taylor expansion of Eq.~\ref{eq:dcpca} is only an approximation of the first derivative of the CCFs with respect to the wavelength, a small fraction of the planet signature will still be present in the residuals of Taylor expansion (i.e. the matrix \mathbfss{C}$_{\rm{S}}$ in Sec.~\ref{ssec:framework}). If the amplitude of this contribution is significantly weaker than the noise, it will naturally be discarded during the dimension reduction process (i.e. the SVD). For example, a 1-\ms\ Doppler signal induces residuals of about 10$^{-8}$ in \mathbfss{C}$_{\rm{S}}$, about 3-to-4 orders of magnitude lower than the typical white noise in HARPS-N spectra.  A way to account for these variations, described in \citet{cameron2021}, \citet{wilson2022} and \citet{john2022}, is to perform the dimensional reduction and search for planet-induced Doppler signatures simultaneously, which we apply in Sec.~\ref{ssec:results_planet}. Note however that the planet signatures considered in this study are small enough not to affect significantly the shape-driven components of the Taylor expansion.

\begin{figure}
    \centering
    \includegraphics[width=\linewidth]{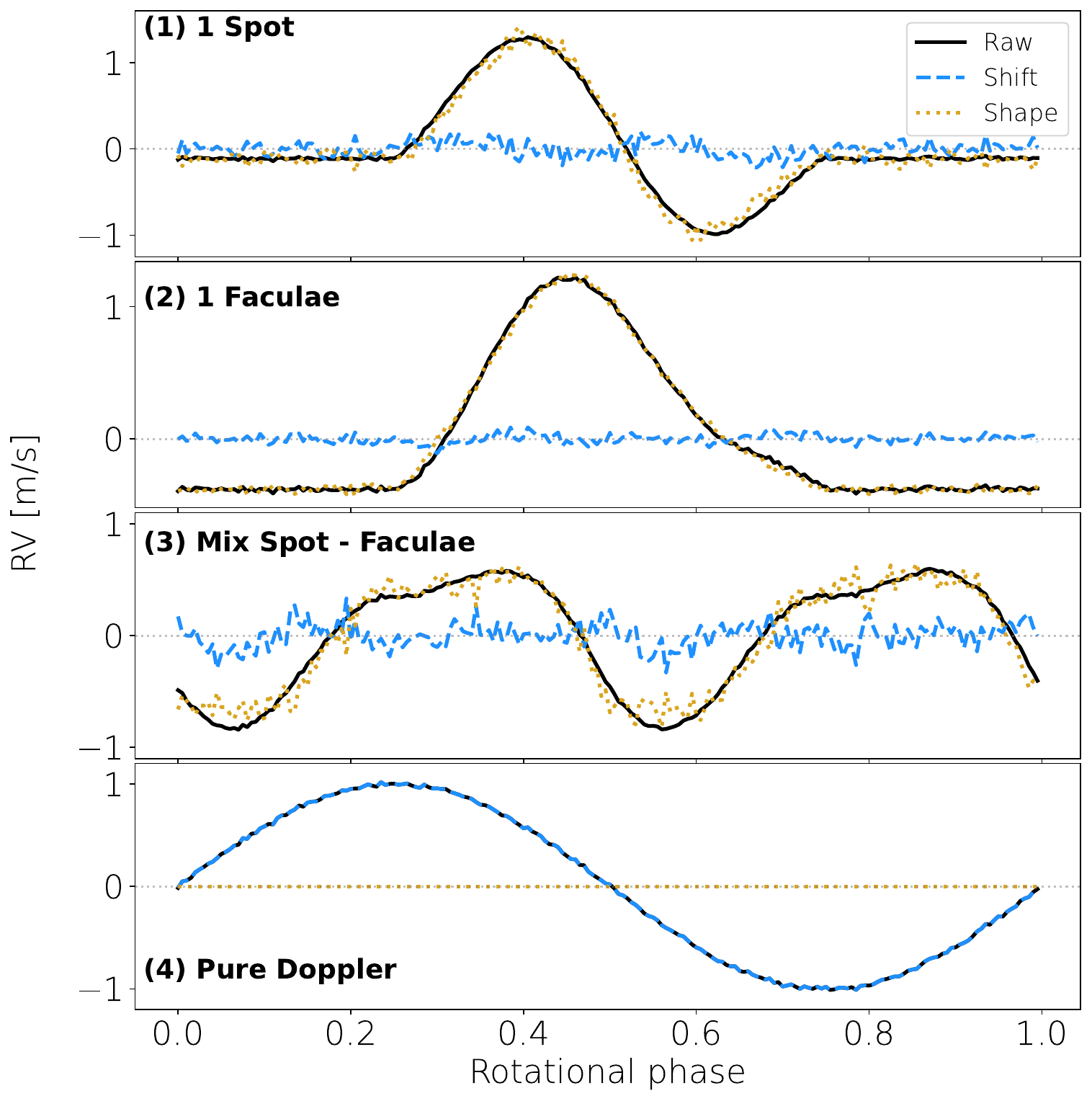}
    \caption{Time series of raw RVs (black solid lines), shift-driven RVs (\vper; blue dashed lines) and shape-driven RVs (\vpar; yellow dotted lines), as a function of the rotational phase of the star, for the simulations described in Tab.~\ref{tab:info_simu}.}
    \label{fig:rv_fit_soap}
\end{figure}

We then apply the framework of Sec.~\ref{ssec:framework} to the simulated data. We use the unspotted line profile generated by \texttt{SOAP2} as reference (\mathbfit{C}$_{\rm{R}}$ in Sec.~\ref{ssec:framework}). A total of 3, 4 and 4 principal components is used to contruct the basis \mathbfss{U} in case (1), (2) and (3), respectively. As for case (4), eigenvectors are found to be dominated by white noise and are therefore discarded, which is expected as no shape distortion was introduced in the data set. In this last case, adding components in the matrix \mathbfss{U} introduces white noise in both \vpar\ and \vper, but the amplitude of the planet signature is only affected starting from $\boldsymbol{U_{20}}$ ($\sim$3\% of the explained variance in a data set dominated by white noise). The time series of noise-free reference-subtracted CCFs are shown in Fig.~\ref{fig:dynamic_spectrum_soap} for the cases listed in Tab.~\ref{tab:info_simu}. The shape- and shift-driven RVs (i.e. resp. \vpar\ and \vper\ in Sec.~\ref{ssec:framework}) extracted in each case are shown in Fig.~\ref{fig:rv_fit_soap}. Unsurprisingly, the Doppler component is well recovered in case (4). We also find that RV signals induced by active regions are well filtered, with a suppression of 86, 94 and 78\% of the stellar activity signals in cases (1), (2) and (3), respectively. However, even in the ideal case considered in this section, \vper\ still exhibits activity-induced fluctuations of up to 0.1\,\ms\ RMS in case (3), significantly larger than the level of white noise ($\sim$0.01\,\ms\ RMS). This is due to the fact that Doppler shifts and shape distortions are not purely orthogonal in the Taylor expansion and, therefore, cannot be entirely separated in Eq.~\ref{eq:dcpca}. This can be understood by the fact that the first derivative of the CCF has no reason to be orthogonal to higher-order derivatives (especially with odd-order derivatives). By analysing the periodicities in \vpar\ and \vper\ for cases (1) and (2), we find that, in the case of a single spot, the shift-driven RVs, \vper, still exhibits modulations at harmonics of the star's rotation period, whereas, in the case of a single facula, \vper\ does not exhibit any explicit sign of periodicity. Finally, we note that including more SVD components in the basis projection affects only marginally the recovered RV signals and that none of the additional principal components exhibit significant periodic modulation.

\section{Data}\label{sec:observations_red}

\begin{figure*}
    \centering
    \includegraphics[width=\linewidth]{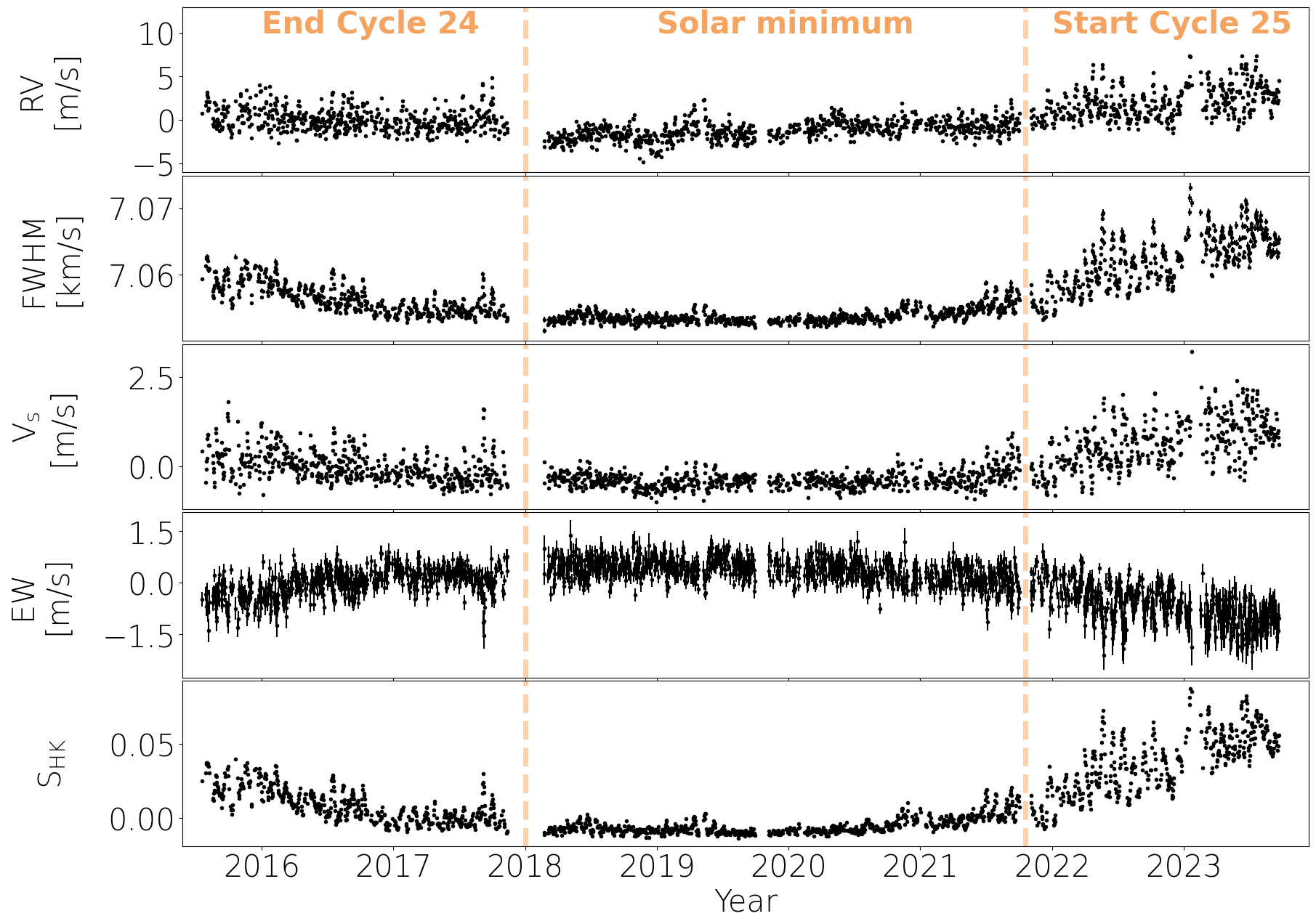}
    \caption{YVA solar dataset. From top to bottom: time series of HARPS-N solar RVs, FWHM, velocity span and equivalent width of the HARPS-N solar CCFs, and S index. The vertical dashed lines delimitate the three seasons defined in Tab.~\ref{tab:rms_rvs}.}
    \label{fig:samplings}
\end{figure*}

\subsection{Solar observations and data selection}\label{ssec:observations}

Since 2015, disc-integrated solar spectra were collected at a 5-min cadence with the stabilised cross-dispersed \'echelle spectrograph HARPS-N \citep[][]{cosentino2012,dumusque2015,phillips2016,cameron2019,dumusque2021}, mounted at the 3.58-m Telescopio Nazionale Galileo (TNG) at Roque de Los Muchachos observatory (La Palma, Spain). These observations leverage the high-resolving power ($\mathcal{R}$~=~115\,000) and the large spectral coverage in the $V$ band (from 383 to 690~nm) of the instrument to achieve RV uncertainties lower than 0.5~\ms. Our input data set was collected between July 2015 and September 2023  and processed with the ESPRESSO data-reduction software \citep[DRS v.2.3.5;][]{pepe2021}, adapted to the HARPS-N solar telescope in \citet{dumusque2021} and Dumusque et al., submitted. 

Since the Sun's surface is resolved, any partial obstruction of the solar disk by, for example, clouds in Earth's atmosphere, will likely induce substantial distortions in the CCFs, and, thus, large spurious RV deviations. In order to flag cloud-contaminated spectra, \citet{cameron2019} perform a daily linear fit to the apparent solar magnitude as a function of airmass (corresponding to the expected extinction law in optical observation conditions). The magnitude is given, as in \citet{cameron2019}, by $m\,{=}\,{-}5\log_{10}{\mathrm{S/N}_{60}}$, where $\mathrm{S/N}_{60}$ is the signal-to-noise ratio (S/N) in the 60\textsuperscript{th} spectral order (corresponding to echelle order 98 which has a central wavelength of 625\,nm). The deviation of each data point to the best-fitting extinction law is estimated through a Bayesian mixture model \citep{hogg2010}, assigning a probability $p$ that the spectrum is not contaminated by clouds. Following the recommendations of \citet{dumusque2021} and \citet{cameron2021}, we only consider spectra with $p$\,$>$\, 0.95 in the following analysis and airmass observations lower than 2.25. After rejecting such spectra, our initial data set of 147\,741 solar spectra is reduced down to 104\,573 (70.8\%). The median S/N in the 60\textsuperscript{th} spectral order is 355 and the median RV uncertainty per spectrum is 0.28~\ms.The fraction of retained spectra are in agreement with \citet{almoulla2023}, where in their Fig. A.2 it is shown that for most days there are only about 0-2 outliers per day among the dozens of daily HARPS-N solar spectra. By keeping only the spectra least affected by clouds, we are able to compute daily stacked spectra (see the following section) that are minimally affected by differential extinction.


\subsection{Spectra post-processing}\label{ssec:post_processing}

The analysis of solar activity with HARPS-N is made complex since Sun-as-a-star observations are different from real stellar observations. For instance, since the Earth is turning around the Sun and since the Earth is not perfectly aligned with the solar equator, a substantial change in the Sun's projected rotational velocity (\vs) is observed, which introduces 1-year and half-a-year signals in the times series of the full-width at half maximum (FWHM) of the CCFs \citep{cameron2019}. Furthermore, the HARPS-N cryostat was changed on the $4^{\rm th}$ October 2021, which also introduced an offset in several CCF moments since the focus of the instrument was realigned, slightly improving the resolution. Those instrumental interventions are expected to have RV contributions smaller than 0.5\,\ms\ since no clear offset is observed by eye in either the solar RVs or standard stars RVs, but likely affected the spectra in some way.

In order to remove those signals and better isolate stellar activity contributions, we post-processed the S1D spectra time-series with YARARA \citep{Cretignier2021}. Note that we did not use any time-domain empirical corrections developed for the atmospheric extinction or \vs\ \citep{cameron2019,dumusque2021}, since those corrections were developed for the time-domain and their ability to perform in the wavelength domain is unclear. However, as shown below, YARARA itself is able to correct for them. YARARA is a post-processing methodology that aims to split the different contaminations coming from the instruments, such as ghosts, stitchings, interference patterns, ThAr bleeding, point spread function (PSF) defocus, from the tellurics and from the stellar activity directly in the spectra.

The main steps of the pipeline are to (i)~daily stack the S1D spectra in the heliocentric rest-frame, (ii)~normalise the continuum of the spectra with RASSINE \citep{Cretignier2020b} and (iii)~correct the continuum-normalised spectra with YARARA \citep{Cretignier2021,cretignier2023}. The dataset after daily stacking the spectra and removing anomalous residuals among them\footnote{Anomalous residuals spectra can be due to spectra with unperfect corrections or lower SNR observations.} consists of 1880 daily-stacked spectra. Note that we only use the first version of YARARA (hereafter "YV1"), which performs systematics correction in the wavelength domain as described in \citet{Cretignier2021}. The pipeline was slightly upgrade on HARPS-N to better correct the cryostat change (see Appendix \ref{appendix:yarara}).

The final RVs were obtained from CCFs with a line list tailored for the Sun as described in \citet{cretignier2020}, formally described as "Custom" in \citet{Bourrier2021} and "Kit-Cat" line list in \citet{Cretignier2022t}. The CCF's FWHM, bisector velocity span, \citep[V$_{\mathrm{s}}$;][]{hatzes1996,queloz2001}, and equivalent width (EW; product of FWHM and contrast), as well as the Mount Wilson S index \citep[S$_{\mathrm{HK}}$;][]{noyes1984} are also computed from the processed spectra. The CCF EW (i.e. the area of the CCF) is a known tracer of stellar temperature and metallicity \citep{malavolta2017}. As the Sun's metallicity is expected to remain roughly constant during our observations, the CCF EW is a good tracer of global temperature changes in the solar photosphere \citep[e.g. magnetic intensification and/or spot-induced flux variations][]{cameron2019}. On the other hand, V$_{\mathrm{s}}$, defined as the RV difference beween the center of a Gaussian fit to the whole CCF and the center of a parabola fit on the core of the CCF (i.e. within $\pm$2.3\,\kms), is highly sensitive to velocity suppressions on the solar disk (mostly due to faculae).

Because YARARA is working with spectra interpolated on a 0.01-\AA\ wavelength grid, the CCFs are slightly oversampled to velocity bins of 530\,\ms, compared to the 820\,\ms\ of the usual DRS \citep{dumusque2021}. This is not critical, but it means that two consecutive velocity bins are not fully independent and share some covariance. Flux uncertainties on the CCF are artificially boosted by the square-root of the oversampling factor to cancel the oversampling effect. The natural output of the YARARA post-processing (spectra, CCFs or time-series) is usually already corrected from the stellar activity component. Since the purpose of this work is to study the solar activity signals, we introduced back the correction related to activity by YARARA to form an "YVA" dataset in order to follow the terminology introduced in \citet{Dalal2024}. The time-series obtained with the YVA dataset are shown in Fig.\ref{fig:samplings}.

\section{Results}\label{sec:results}

\subsection{Planet-free case}\label{ssec:planet-free-sun}

\begin{table}
    \centering
    \caption{RMS of the time series of \vobs, \vper\ (shift-driven RVs) and \vpar\ (shape-driven RVs), over different periods of time. In each case, the first four principal components are used to build the time series of shape-driven CCFs, \mathbfss{C}$_{\parallel}$, from which \vpar\ is extracted.}
    \label{tab:rms_rvs}
    \begin{tabular}{ccccc}
    \hline
    Name & Period & \vobs & \vper & \vpar  \\
    \hline
     Season 1 & 2015 - 2018   & 1.34\,\ms & 1.08\,\ms & 0.97\,\ms \\
     Season 2 & 2018 - 2021.8   & 1.07\,\ms & 1.00\,\ms & 0.50\,\ms \\
     Season 3 & 2021.8 - 2023.7   & 1.98\,\ms & 1.11\,\ms & 1.69\,\ms \\
     Altogether & 2015 - 2023.7 & 2.05\,\ms & 1.06\,\ms & 1.76\,\ms \\ 
    \hline
    \end{tabular}
\end{table}

\begin{figure}
    \centering
    \includegraphics[width=\linewidth]{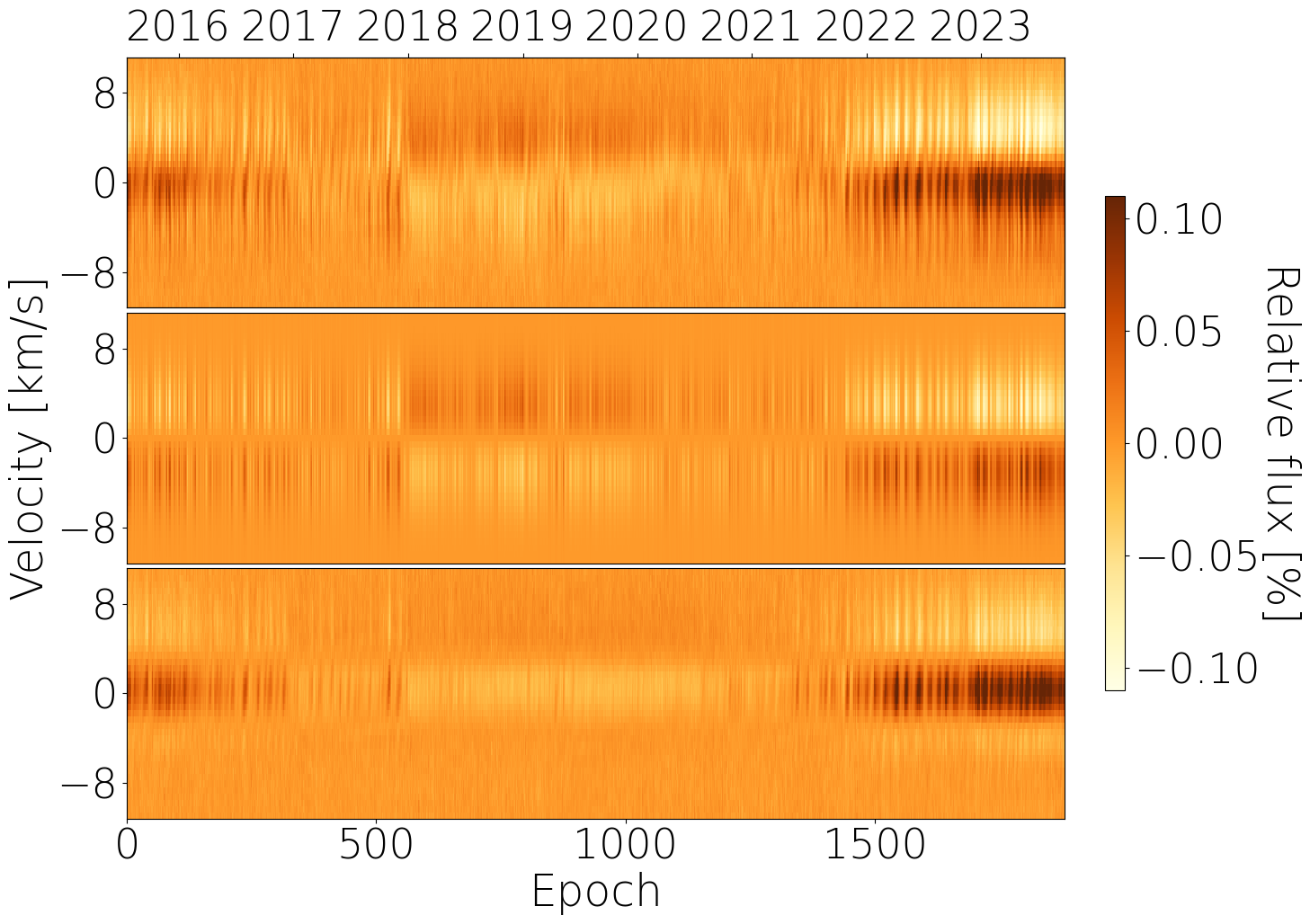}
    \caption{Time series of reference-subtracted Solar CCFs (top panel), Doppler components computed using Eq.~\ref{eq:dcpca} (\mathbfss{C}$_{\rm{D}}$, middle panel), and shape-driven residuals (\mathbfss{C}$_{\rm{S}}$, bottom panel).}
    \label{fig:ccf_timeseries}
\end{figure}

We apply the framework defined in Sec.~\ref{ssec:framework} to the time series of daily-binned solar CCFs. Using the mean line profile as reference profile \mathbfit{C}$_{\rm{R}}$, we compute the time series of reference-subtracted CCFs (see the first panel of Fig.~\ref{fig:ccf_timeseries}). From Eq.~\ref{eq:dcpca}, we derive a time series of shift-driven (i.e. Doppler component in Sec.~\ref{ssec:framework}) and shape-driven CCFs, shown in the middle and bottom panels of Fig.~\ref{fig:ccf_timeseries}. As expected, the Doppler component only contains information on the first derivative of the CCFs, as evidenced by the fact that, at each epoch, the CCF variations are symmetrical with respect the center of the line (i.e. zero-velocity point). On the other hand, shape-driven residuals exhibit a more complex structure, non symmetrical with respect to the line center, suggesting that it probes higher-order derivatives of the CCF variations. We also double checked that the width and asymmetry of the shift-driven profiles remain constant over time.

\begin{figure*}
    \centering
    \includegraphics[width=\linewidth]{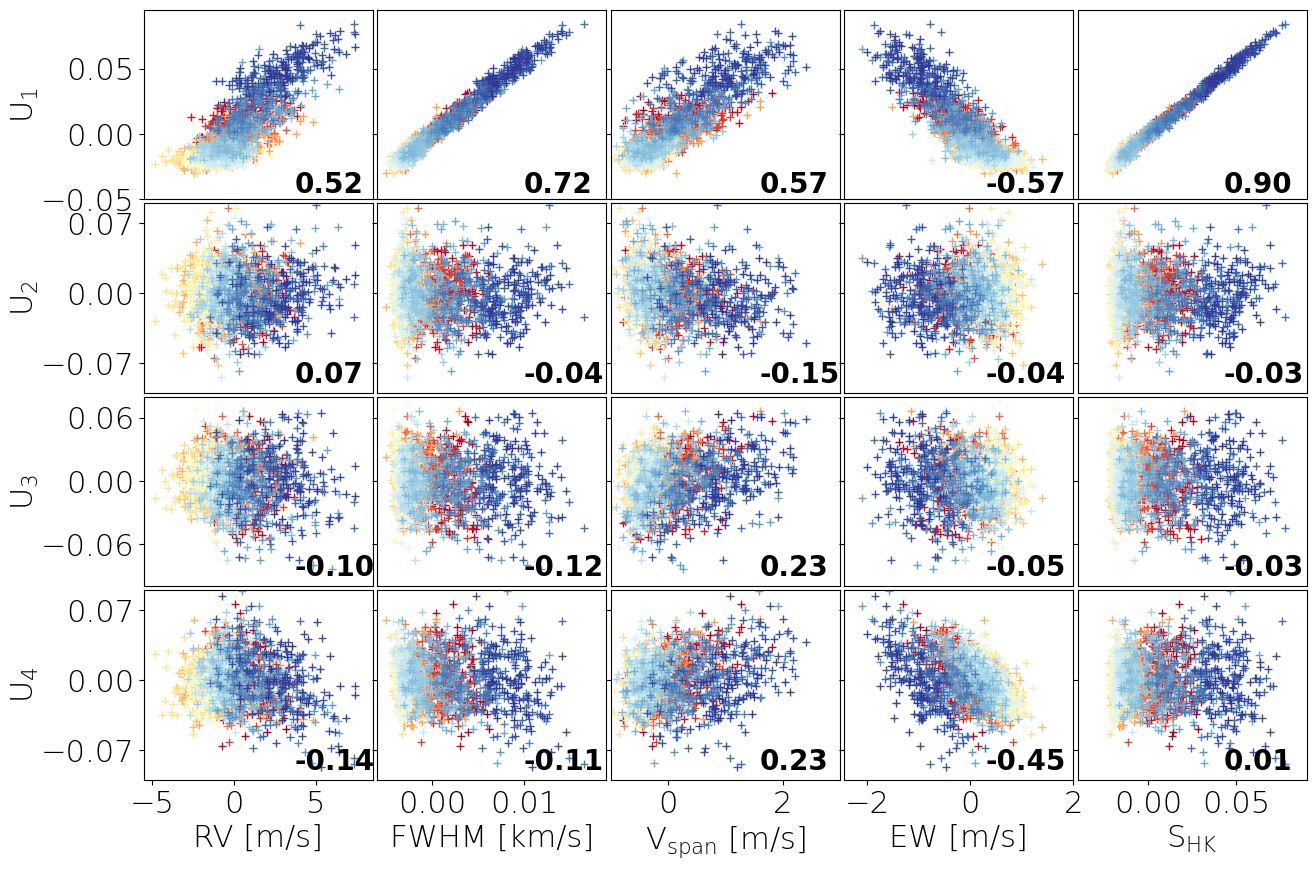}
    \caption{First four SVD components ($\boldsymbol{U_{1}}$ to  $\boldsymbol{U_{4}}$) plot against the median-subtracted HARPS-N solar RVs, the FWHM, bisector velocity span and EW of the CCFs, and the S$_{\mathrm{HK}}$ time series. The individual points are colour-coded by date of observation, bluer (redder) points corresponding to more recent (older) observations. In the bottom right of each panel, we indicate the Pearson correlation coefficient of the two time series (weigthed by the uncertainties on the X axis). Note that all correlation coefficients are associated with $p$-values lower than 0.05, and are thus considered to be significant.}
    \label{fig:correl}
\end{figure*}

We then apply SVD to the shape-driven CCFs and extract a basis of orthogonal vectors tracing the main non-Doppler-induced line profile variations. Applying the procedure described in Sec~\ref{sec:sec2}, we use the first four principal components in our reconstruction. As shown in Fig.~\ref{fig:correl}, the first eigenvector $\boldsymbol{U_{1}}$ correlates strongly with \vobs\ and usual activity indicators. In particular, we obtain a weighted Pearson correlation coefficient of 0.90 between $\boldsymbol{U_{1}}$ and S$_{\mathrm{HK}}$, which suggests that the variations of these two quantities are driven by similar processes (see Sec.~\ref{sec:gp_sec} for a more detailed comparison). Higher-order SVD components tend to exhibit weaker correlations with CCF indicators, with the exception of $\boldsymbol{U_{4}}$, which exhibits a relatively strong anti-correlation with the EW of the CCF, indicating that this indicator could be a good proxy of the photospheric temperature \citep[see][]{cameron2019}.

\begin{figure}
    \centering
    \includegraphics[width=\linewidth]{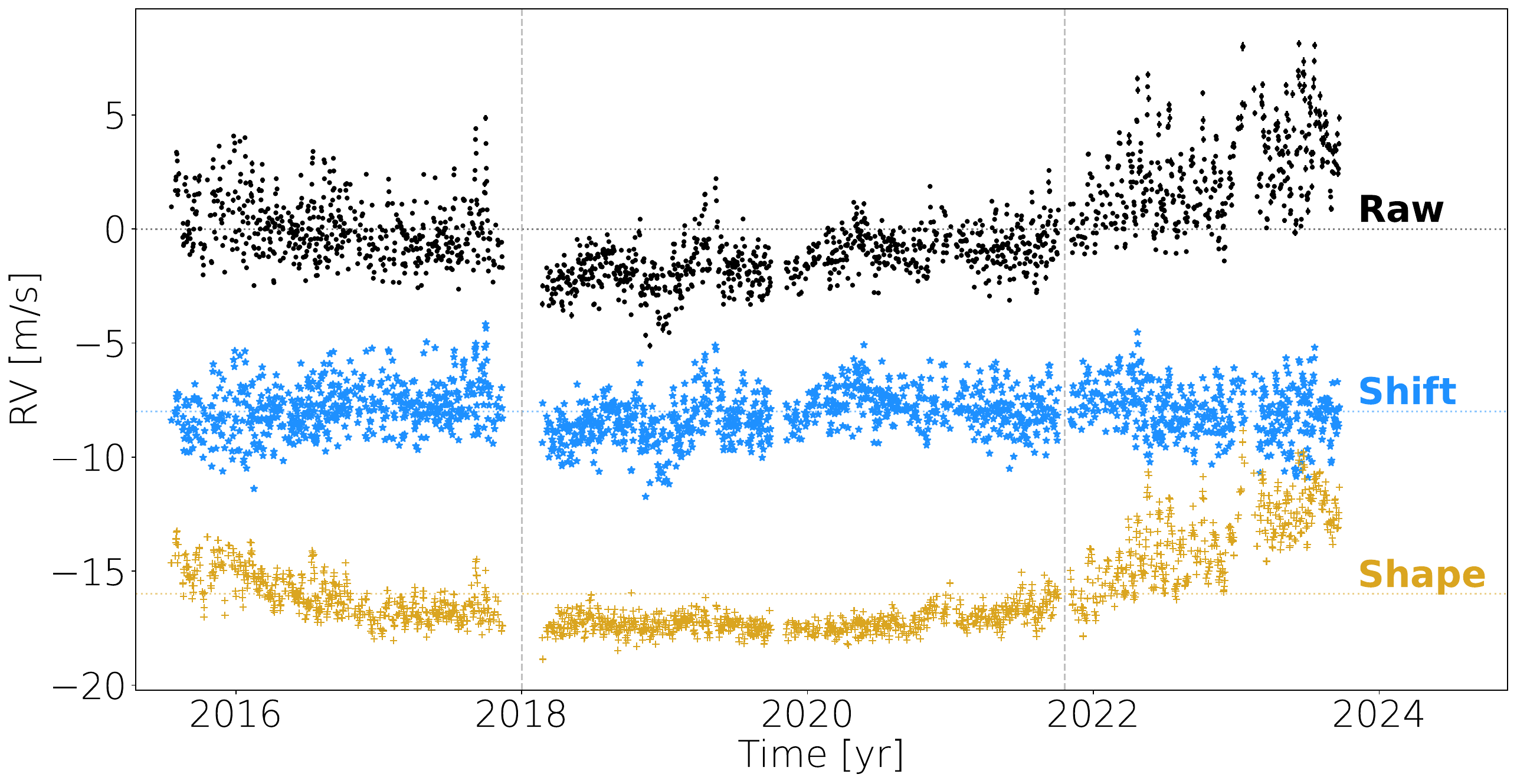}
    \caption{Time series of HARPS-N solar RVs (top curve), shift-driven RVs (\vper; curve in the middle) and shape-driven RVs (\vpar; bottom curve). The typical formal error bar on each RV point is about 0.16\,\ms. The vertical dashed lines divide the data into the three periods shown in Fig.~\ref{fig:samplings}.}
    \label{fig:rv_fit_sun}
\end{figure}

We project the input CCFs onto the basis defined by components $\boldsymbol{U_{1}}$ to $\boldsymbol{U_{4}}$, and derived shape- and shift-driven RVs (resp. \vpar\ and \vper) using the method described in Sec.~\ref{ssec:framework}. Both time-series are shown in Fig.~\ref{fig:rv_fit_sun} and their root-mean-square (RMS) deviations are given in Tab.~\ref{tab:rms_rvs}. Starting from a dispersion of 2.1~\ms\ in the input data, the shape- and shift-driven RVs exhibit RMS dispersions of 1.8 and 1.1~\ms, respectively\footnote{Note that the quadratic sum of the RMSs of the shift- and shape-driven RVs is not strickly equal to the RMS of \vobs, which reflects the fact that the decomposition defined in Sec.~\ref{ssec:framework} is not entirely orthogonal.}. In order to assess the effectiveness of the method for different levels of solar activity, we divide our input data set in the three seasons shown in Fig.~\ref{fig:samplings}. From 2015 to 2018 (Season~1), the Sun reaches the end of its activity cycle but still exhibits clear signs of activity. The Sun enters then a long period of activity minimum (Season~2, 2018-2021.8) which lasts until the end of 2021. With the start of Cycle 25 (Season~3, 2021.8-2024), the Sun exhibits activity signals with significantly larger amplitude than in Season~1. Note that this division is motivated by two factors. Firstly, the cycle-induced evolution of the solar activity, evidenced, for example, by the amplitude of the fluctuations of the disk-integrated magnetic flux density \citep[lower than 0.5\,G in Season~2, according to SDO/HMI data extracted with \texttt{SOLASTER}; see][]{ervin2022} allows to to separate three different activity regimes. Secondly, important instrument maintenance operations, namely the change of Fabry-P\'erot interferometer, between Seasons~1 and 2, and the refurbishment of the CCD camera, between Seasons~2 and 3, justify the definition of the three seasons adopted in this study. From Tab.~\ref{tab:rms_rvs}, we note that our activity-filtering framework performs best when the Sun is more active, with a 44\% reduction in RV RMS in Season~3, compared to only 7\% in Season~2 (solar minimum). We also note that, in Season~1, our results are similar to those obtained with \texttt{SCALPELS}, on a similar data set \citep{cameron2021}.

\begin{figure}
    \centering
    \includegraphics[width=\linewidth]{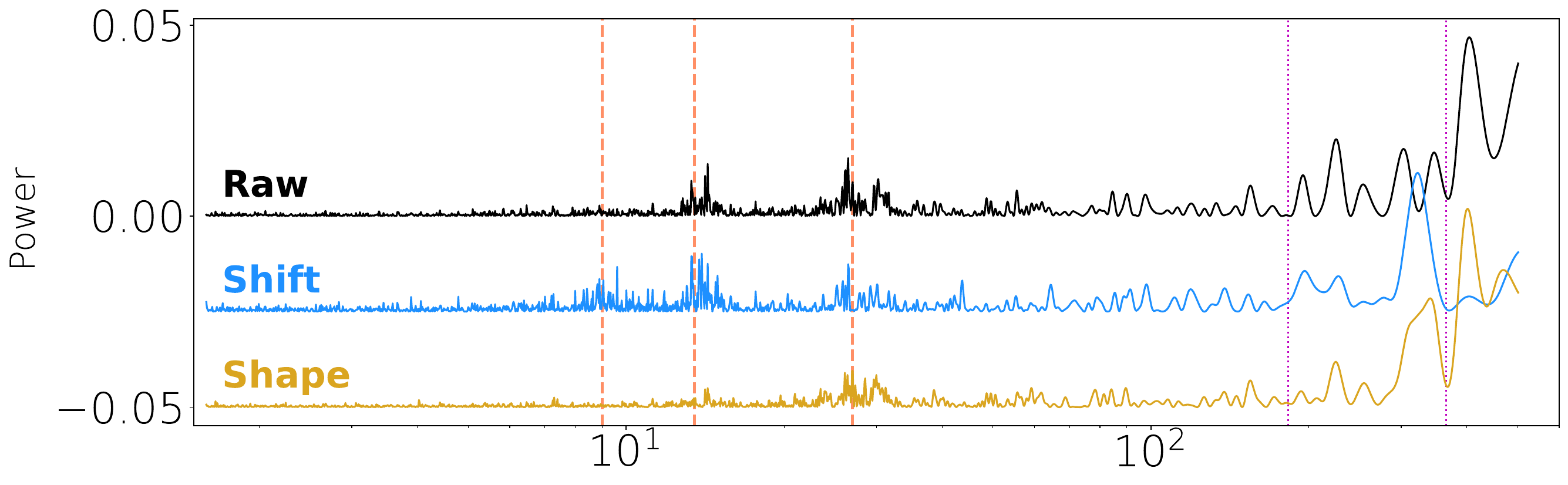}
    \includegraphics[width=\linewidth]{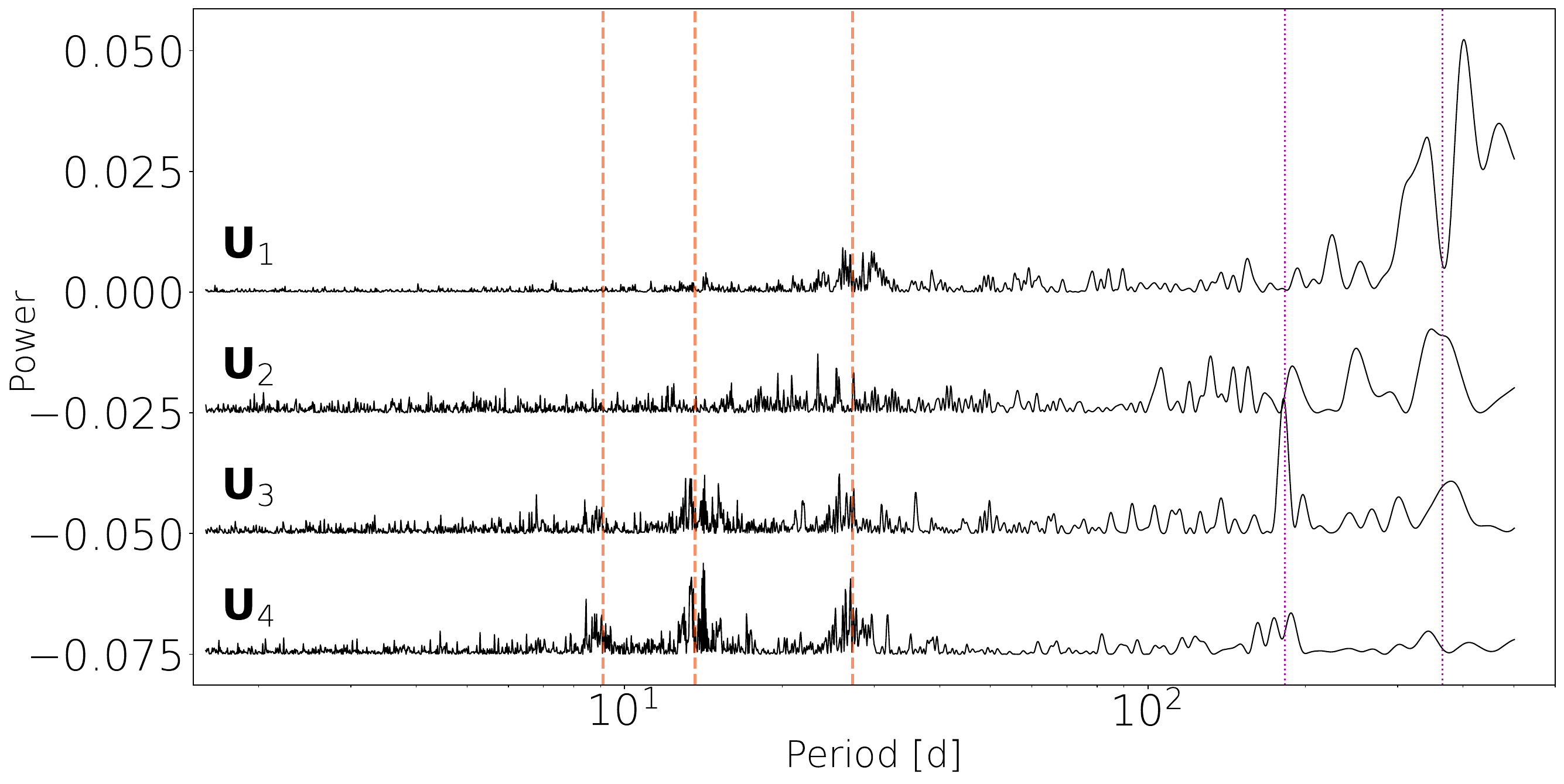}
    \caption{\textit{Top panel:} GLS periodogram of the HARPS-N solar RVs (top line), the shift-driven RVs (\vper, middle line) and shape-driven RVs (\vpar, bottom line). \textit{Bottom panels:} GLS periodogram of the first four principal components, $\boldsymbol{U_{1}}$ to $\boldsymbol{U_{4}}$, of the shape-driven CCFs. From right to left, the vertical orange dashed and magenta dotted lines indicate the Solar rotation period (and first two harmonics) and the Earth orbital period (and first harmonic). The periodograms were computed using the \texttt{astropy} python module \citep{astropy2013,astropy2018,astropy2022}.}
    \label{fig:periodogram_solar}
\end{figure}

As shown in Fig.~\ref{fig:rv_fit_sun}, the shape-driven RVs seem to enclose most of the long-term RV variations, primarily driven by the magnetic cycle. This is further evidenced in the generalised Lomb-Scargle \citep[GLS;][]{zechmeister2009} periodograms of the RV time series (see bottom curve on the top panel of Fig.~\ref{fig:periodogram_solar}), which are dominated by peaks at periods larger than $\sim$100~d. We note also a significant power near the Earth orbital period, also present in the periodogram of component $\boldsymbol{U_{2}}$, attributable to residuals of the effects of Earth's orbital eccentricity and solar obliquity on the FWHM of the CCF, as described in \citet{cameron2019}. These effects will not be observed in other stars. We also note that the power at the Sun's rotation period and harmonics remains strong in the shift-driven RVs, despite a fraction of it being transferred to the shape-driven RVs. Surprisingly, components $\boldsymbol{U_{3}}$ and $\boldsymbol{U_{4}}$ exhibit a significant modulation of the Sun's rotation period and first harmonics and, thus, could potentially act as good proxies of quasi-periodic stellar activity signals (see Sec~\ref{sec:gp_sec}), although no strong correlation with CCF activity indicators is observed.

\subsection{Effects on planet recovery}\label{ssec:results_planet}

In order to assess the ability of our activity-filtering procedure to preserve planet RV signatures, we create 1\,000 data sets from the solar HARPS-N observations, each containing a single planetary signal directly injected in the CCF time-series with the same temporal sampling. As we assume the planet orbit to be circular, the corresponding RV signature, $v_{\rm{p}}$, as a function of the time, $t$, is given by

\begin{eqnarray}
    v_{\rm{p}}(t) = K_{\rm{inj}} \sin{ 2 \pi \left[  \frac{T_{0} - t}{P_{\rm{orb}}}  + \phi_{\rm{p}}  \right] }
    \label{eq:rvp}
\end{eqnarray}

\noindent
where the reference time, $T_{0}$, is set to BJD~=~2\,457\,223.49054 (i.e., the time of our first observation), and where \kinj, \porb\ and \php\ are respectively the semi-amplitude, orbital period and orbital phase of the injected signature. These three parameters are randomly drawn using log-uniform laws (for \kinj\ and \porb) and uniform laws (for \php), as described in Tab.~\ref{tab:planet_param}.

To inject the planet signature, we interpolate each CCF in the solar rest frame using a centered square-exponential Gaussian Process \citep[GP;][]{rasmussen2006,aigrain2022} with covariance function $k$ between each pair of velocity bins $v_{i}$ and $v_{j}$ given by

\begin{equation}
k(v_{i},v_{j}) = A \exp{ \left[ - \frac{(v_{i}-v_{j})^{2}}{2 \lambda_{\rm{e}}^{2}}   \right]}   + \eta_{i}^{2} \delta_{i,j} 
\label{eq:SE_kernel}
\end{equation}

\noindent
where $\eta_{i}$, is the uncertainty on the CCF flux associated with pixel $i$, and $\delta$ stands for the Kronecker delta. The two free hyperparameters of Eq.~\ref{eq:SE_kernel}, namely the GP amplitude $A$ and correlation length $\lambda_{\rm{e}}$, are estimated by maximising the likelihood of the data, computed using the \texttt{GEORGE} python module \citep{ambikasaran2015}. As a sanity check, we measured the RV and FWHM of each planet-injected CCF by fitting a Gaussian profile to it. The resulting RV/FWHM time-series differ by no more than 0.1\,$\sigma$ from their DRS-provided counterparts, confirming that this interpolation procedure only marginally affects the shape and position of the line profile. As a word of caution, we note that injecting the planet signature at the CCF level rather than at the spectrum level assumes that the Doppler shift is integrally preserved in the cross-correlation process. This assumption is motivated by recent studies demonstrating that reliable planet parameters were extracted from the CCF \citep[e.g.][]{john2023,debeurs2024}.

Our procedure to retrieve the injected planet signatures is similar of that used in \citet{cameron2021} and \citet{wilson2022}. For each data set, we jointly apply the dimensional reduction of Sec.~\ref{ssec:framework} and fit for a planet Doppler motion with Eq.~\ref{eq:rvp}. As detailed in the Appendix~A of \citet{wilson2022}, projecting the observed RVs \vobs\ onto the matrix $\textsf{\textbf{P}} = \boldsymbol{I} - \textsf{\textbf{U}} \cdot \textsf{\textbf{U}}^\intercal$, where $\boldsymbol{I}$ is the identity matrix and $\textsf{\textbf{U}}$ the basis defined in Sec.~\ref{ssec:framework} (using four principal components), is equivalent to subtracting \vpar\ from \vobs. The likelihood $\mathcal{L}$ of the data given the model parameters $\boldsymbol{\theta}$ is then given by

\begin{align}
\ln \mathcal{L} = \quad & -\frac{1}{2} \left[ \textsf{\textbf{P}} \cdot (\boldsymbol{v_{\mathrm{obs}}} - \boldsymbol{v_{\mathrm{p}}}(\boldsymbol{\theta})) \right]^\intercal \cdot  \boldsymbol{\Sigma}^{-1}  \left[ \cdot  \textsf{\textbf{P}} \cdot (\boldsymbol{v_{\mathrm{obs}}} - \boldsymbol{v_{\mathrm{p}}}(\boldsymbol{\theta})) \right]  \notag \\
 & -\frac{1}{2} \ln \lvert \boldsymbol{\Sigma} \rvert  -\frac{N}{2} \ln 2 \pi
\label{eq:likelihood}
\end{align}

\noindent
where $\boldsymbol{v_{\mathrm{p}}}$ is the planet RV model given by Eq.~\ref{eq:rvp} and $N$ is the number of data points. The covariance matrix $\boldsymbol{\Sigma}$ is assumed 
diagonal, with $\Sigma_{kk} = \sigma_{k}^{2} + \sigma_{j}^{2}$, where $\sigma_{k}$ is the formal RV uncertainty on the $k$-th data point and $\sigma_{j}$ is a free parameter of the model to absorb RV variations that are not captured by Eq.~\ref{eq:rvp} or the formal RV uncertainties (e.g. activity and granulation signals, instrumental stability). We assume that the planet orbital period and phase are known, leaving the planet RV semi-amplitude and the jitter term $\sigma_{j}$ as the only free parameters of the model. Their posterior probablity given the data is estimated, in the Bayesian framework, using the affine-invariant Markov Chain Monte Carlo (MCMC) process \texttt{emcee} \citep[5\,000 iterations of 100 chains;][]{foreman2013}. The best-fitting parameters and 1$\sigma$-uncertainties are estimated from the median and from the 16$^{\mathrm{th}}$ and 84$^{\mathrm{th}}$ percentiles of the chain after removing a burn-in period significantly longer than the auto-correlation time of the chain (about 100 iterations). As a reference, we also fit Eq.~\ref{eq:rvp} to the input RV time-series, \vobs, before filtering the shape-induced contributions.



\begin{figure}
    \centering
    \includegraphics[width=\linewidth]{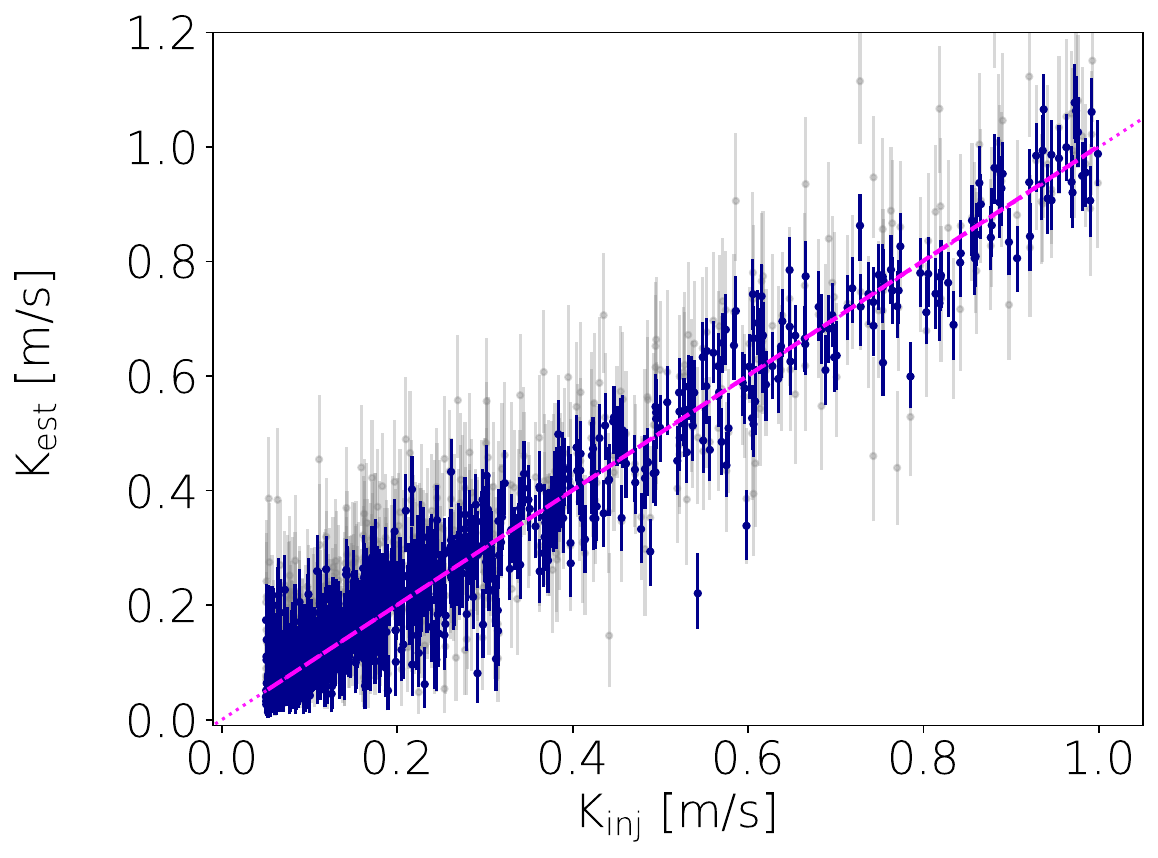}\\
    \includegraphics[width=\linewidth]{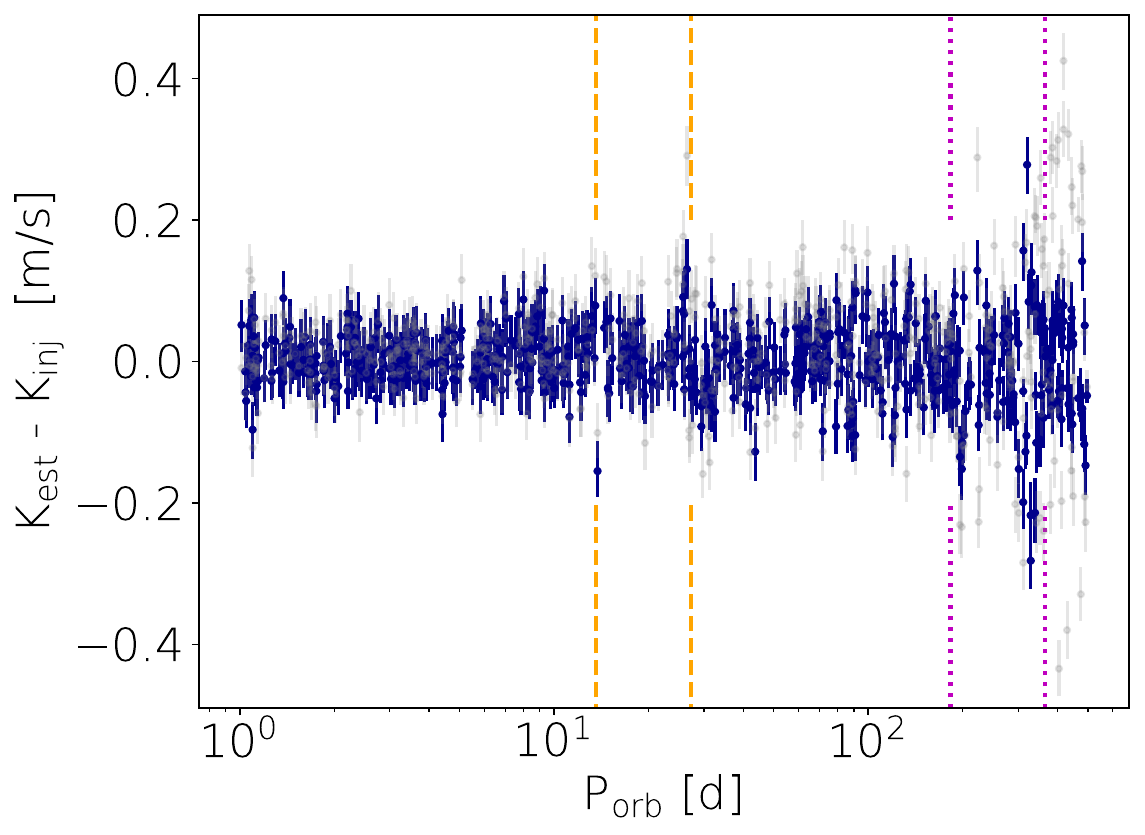}
    \caption{\textit{Top:} Planet RV semi-amplitude, \kest, recovered from the activity-filtered RVs (\vper, dark blue dots) and from the raw RVs (\vobs, light gray dots), as a function of the RV semi-amplitude \kinj\ of the planets injected in Sec~\ref{ssec:results_planet}. The magenta dotted and dashed lines indicate the identity and the best-fitting straight line, respectively. \textit{Bottom:} Difference between \kest\ and \kinj\ as a function of the orbital period \porb\ of the injected planet. The vertical orange dashed and magenta dotted lines indicate the Solar rotation period (and first harmonic) and the Earth orbital period (and first harmonic).}
    \label{fig:kp_distrib}
\end{figure}

The distribution of recovered RV semi-amplitudes, \kest, is shown as a function of \kinj\ in the top panel of Fig.~\ref{fig:kp_distrib} for \kinj$<$1\,\ms. In most cases, the recovered values of the planet RV semi-amplitude match their injected counterpart, with a slope of 1.03\,$\pm$\,0.1. As shown in the bottom panel of Fig.~\ref{fig:kp_distrib}, devivations from the identity are observed when the planet orbital period lies close to the Sun's rotation period or to the Earth orbital period (and first harmonic), which roughly corresponds to the location of the peaks in the periodogram of \vper\ (see Fig.~\ref{fig:periodogram_solar}). No particular trend is observed between \kest\ and the injected planet RV semi-amplitude or orbital phase.

Let us assume, conservatively, that a planet is detected if (i)~\kest\ differs from \kinj\ by less than 1$\sigma$, and (ii)~\kest\ differs from 0 by at least 3$\sigma$. Using this criterion, about 45\} of planets with RV semi-amplitudes lower than 1.0\,\ms\ are detected from the activity-filtered RVs. This detection rate slowly decreases until \kinj\,$\approx$\,0.2\,\ms, where 29\% of the injected planets are still recovered, and drops near zero for lower values of \kinj. The planet detection rate decreases roughly linearly with the planet orbital period, and is about 30\% and (resp. 20\%) for planet with orbital periods greater than 100\,d (resp. 300\,d). The fraction of detected planets from \vobs\ and \vper\ is shown in the (\kinj, \porb) space in Fig.~\ref{fig:sensitivity_maps}. We find that, on average, our sensitivity to planet signatures increases by about 20\% from \vobs\ to \vper, and that this increase is the most spectacular for Earth-mass planets with orbital periods larger than 100\,d, where the detection rate skyrockets by $\sim$50\%. However, Earth-mass planets with orbital periods larger than 300\,d remain mostly undetected in both time series, which is in line with the predictions of \citet{meunier2023}\footnote{Note also that the evolution of the planet detection rate from \vper\ with the planet mass is found roughly similar to that shown, for example, in the figure~8 of \citet{meunier2023}}.


\begin{figure}
    \centering
    \includegraphics[width=\linewidth]{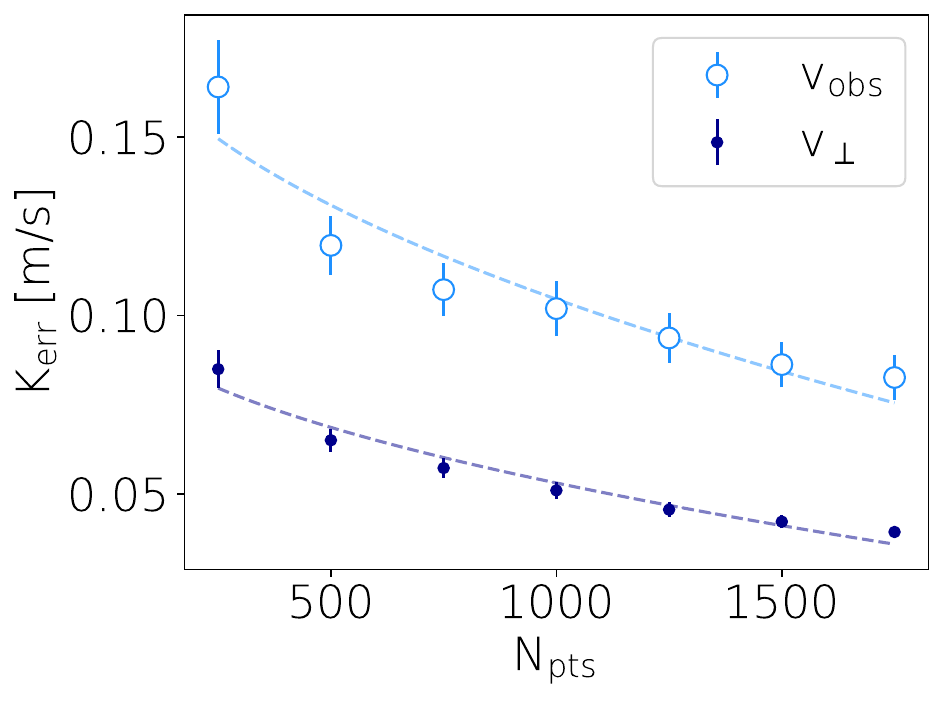}
    \caption{Mean error $K_{\mathrm{err}}$ on the planet RV semi-amplitude estimated from the activity-filtered RVs (\vper, dark blue dots) and from the raw RVs (\vobs, light blue open circles) as a function of the number of data points N$_{\mathrm{pts}}$. For each case, the dashed line indicates the the best-fitting square-root law (reduced $\chi^{2}$ of 1.0 and 1.1 for \vobs\ and \vper, respectively).}
    \label{fig:err_kp}
\end{figure}


\begin{table}
    \centering
    \caption{Probability laws used to generate the semi-amplitude, orbital period and orbital phase of the planet RV signatures injected in the HARPS-N Solar CCFs (see Sec.~\ref{ssec:results_planet}). $\mathcal{U}$ and $\log \mathcal{U}$ stand respectively for the Uniform and log-Uniform probability laws. The last two columns give the minimum and maximum values used for each parameter.}
    \label{tab:planet_param}
    \begin{tabular}{ccccc}
    \hline
    Parameter  & Notation & Probability law & Min & Max  \\
    \hline
    Semi-amplitude & \kinj & $\log \mathcal{U}$ & 0.05\,\ms & 5.0\,\ms\ \\
    Orbital period & \porb & $\log \mathcal{U}$ & 1.0\,d & 500\,d \\
    Orbital phase  & \php & $\mathcal{U}$ & 0.0 & 1.0 \\    
    \hline
    \end{tabular}
\end{table}

In order to assess how the sampling strategy affects our activity-filtering framework, we perform the planet injection-recovery for different numbers of randomly-selected data points, between 250 and 1750. For each sampling, we repeat the injection-recovery of the same 1\,000 planet signals (using the same sampling for all planets). Fig.~\ref{fig:err_kp} shows the mean error K$_{\mathrm{err}}$ on the RV semi-amplitude recovered with and without activity filtering. On average, K$_{\mathrm{err}}$ is reduced by a factor $\sim$2 between \vobs\ and \vper. In both cases, the evolution of the detection rate with the number of points is well described by a square-root law (with reduced $\chi^{2}$ of 1.0 and 1.1 for \vobs\ and \vper; see Fig.~\ref{fig:err_kp}). This suggests that the precision on \kest\ is mostly driven by the number of data points used in the fit, despite the presence of super-granulation, activity residuals and instrumental systematics.



\begin{table*}
    \centering
    \caption{Best-fitting evolution time scale ($\lambda_{\mathrm{e}}$), inverse harmonic complexity ($\lambda_{\mathrm{p}}$) and period (P$_{\mathrm{GP}}$)} of the one-dimension GP fit to the different time series analysed in this study.
    \label{tab:parameters_1d}    
    \begin{tabular}{cccccccccc}
    \hline
        \vspace{0.1cm}
         &  \multicolumn{3}{c|}{\textbf{End Cycle 24: 2015-2018}} & \multicolumn{3}{|c|}{\textbf{Solar minimum: 2018-2022}} & \multicolumn{3}{|c|}{\textbf{Start Cycle 25: 2022-2024}}   \\
   &  $\lambda_{\mathrm{e}}$ [d] &   $\lambda_{\mathrm{p}}$ & P$_{\mathrm{GP}}$ [d]     &  $\lambda_{\mathrm{e}}$ [d] &   $\lambda_{\mathrm{p}}$ & P$_{\mathrm{GP}}$ [d] &  $\lambda_{\mathrm{e}}$ [d] &   $\lambda_{\mathrm{p}}$ & P$_{\mathrm{GP}}$ [d]  \\ 
    \hline
\vobs & 18.0 $\pm$ 1.6 & 0.22 $\pm$ 0.02 & 26.37$_{-0.44}^{+0.45}$ & 21.8 $\pm$ 1.5 & 0.32 $\pm$ 0.03 & 27.25 $\pm$ 0.35 & 21.4$_{-1.4}^{+1.3}$ & 0.32 $\pm$ 0.02 & 27.39 $\pm$ 0.36  \\
FWHM & 22.2 $\pm$ 1.1 & 0.51 $\pm$ 0.03 & 27.70$_{-0.42}^{+0.43}$ & 27.7$_{-1.7}^{+1.8}$ & 0.54 $\pm$ 0.04 & 27.58$_{-0.35}^{+0.37}$ & 22.9$_{-1.3}^{+1.2}$ & 0.56 $\pm$ 0.04 & 27.59$_{-0.44}^{+0.43}$  \\
V$_{\mathrm{s}}$ & 19.8 $\pm$ 1.1 & 0.28 $\pm$ 0.02 & 26.43$_{-0.33}^{+0.37}$ & 22.7$_{-1.5}^{+1.6}$ & 0.33 $\pm$ 0.02 & 27.16 $\pm$ 0.35 & 23.0$_{-1.3}^{+1.2}$ & 0.32 $\pm$ 0.02 & 27.45$_{-0.27}^{+0.28}$  \\
S$_{\mathrm{HK}}$ & 21.6 $\pm$ 1.2 & 0.59 $\pm$ 0.04 & 28.01$_{-0.51}^{+0.49}$ & 23.9$_{-1.4}^{+1.3}$ & 0.55 $\pm$ 0.04 & 27.81$_{-0.45}^{+0.44}$ & 22.7$_{-1.4}^{+1.3}$ & 0.57$_{-0.03}^{+0.04}$ & 28.75$_{-0.45}^{+0.46}$  \\
\vpar & 22.1 $\pm$ 1.5 & 0.52 $\pm$ 0.04 & 28.75 $\pm$ 0.62 & 25.5$_{-2.3}^{+2.5}$ & 0.70$_{-0.08}^{+0.10}$ & 25.67$_{-0.64}^{+0.66}$ & 22.3$_{-1.4}^{+1.3}$ & 0.60$_{-0.04}^{+0.05}$ & 27.63$_{-0.48}^{+0.50}$  \\
U$_{\mathrm{1}}$ & 22.2 $\pm$ 1.4 & 0.55 $\pm$ 0.04 & 28.65$_{-0.53}^{+0.52}$ & 26.8$_{-2.2}^{+2.4}$ & 0.67$_{-0.06}^{+0.07}$ & 26.23$_{-0.57}^{+0.56}$ & 22.2 $\pm$ 1.5 & 0.59 $\pm$ 0.04 & 28.47$_{-0.59}^{+0.51}$  \\
U$_{\mathrm{3}}$ & 22.7$_{-2.9}^{+4.1}$ & 0.21 $\pm$ 0.04 & 26.11$_{-0.40}^{+0.50}$ & 20.7 $\pm$ 3.8 & 0.49$_{-0.13}^{+0.28}$ & 26.58$_{-1.21}^{+1.88}$ & 18.2 $\pm$ 2.3 & 0.21$_{-0.02}^{+0.03}$ & 27.55$_{-0.49}^{+0.51}$  \\
U$_{\mathrm{4}}$ & 23.0$_{-2.4}^{+2.8}$ & 0.20 $\pm$ 0.02 & 26.75$_{-0.29}^{+0.31}$ & 23.4$_{-4.3}^{+5.6}$ & 0.32$_{-0.06}^{+0.09}$ & 26.93$_{-0.83}^{+1.15}$ & 24.1$_{-2.2}^{+2.6}$ & 0.21 $\pm$ 0.02 & 26.99$_{-0.23}^{+0.24}$  \\
\vper & 18.8$_{-2.4}^{+3.7}$ & 0.18$_{-0.02}^{+0.03}$ & 26.01$_{-0.54}^{+0.58}$ & 20.7$_{-1.8}^{+1.9}$ & 0.35$_{-0.04}^{+0.05}$ & 27.32$_{-0.59}^{+0.62}$ & 21.9$_{-2.3}^{+2.2}$ & 0.23 $\pm$ 0.02 & 27.36$_{-0.43}^{+0.38}$  \\
    \hline
    \end{tabular}
\end{table*}

\begin{figure*}
      \centering
    \includegraphics[width=\linewidth]{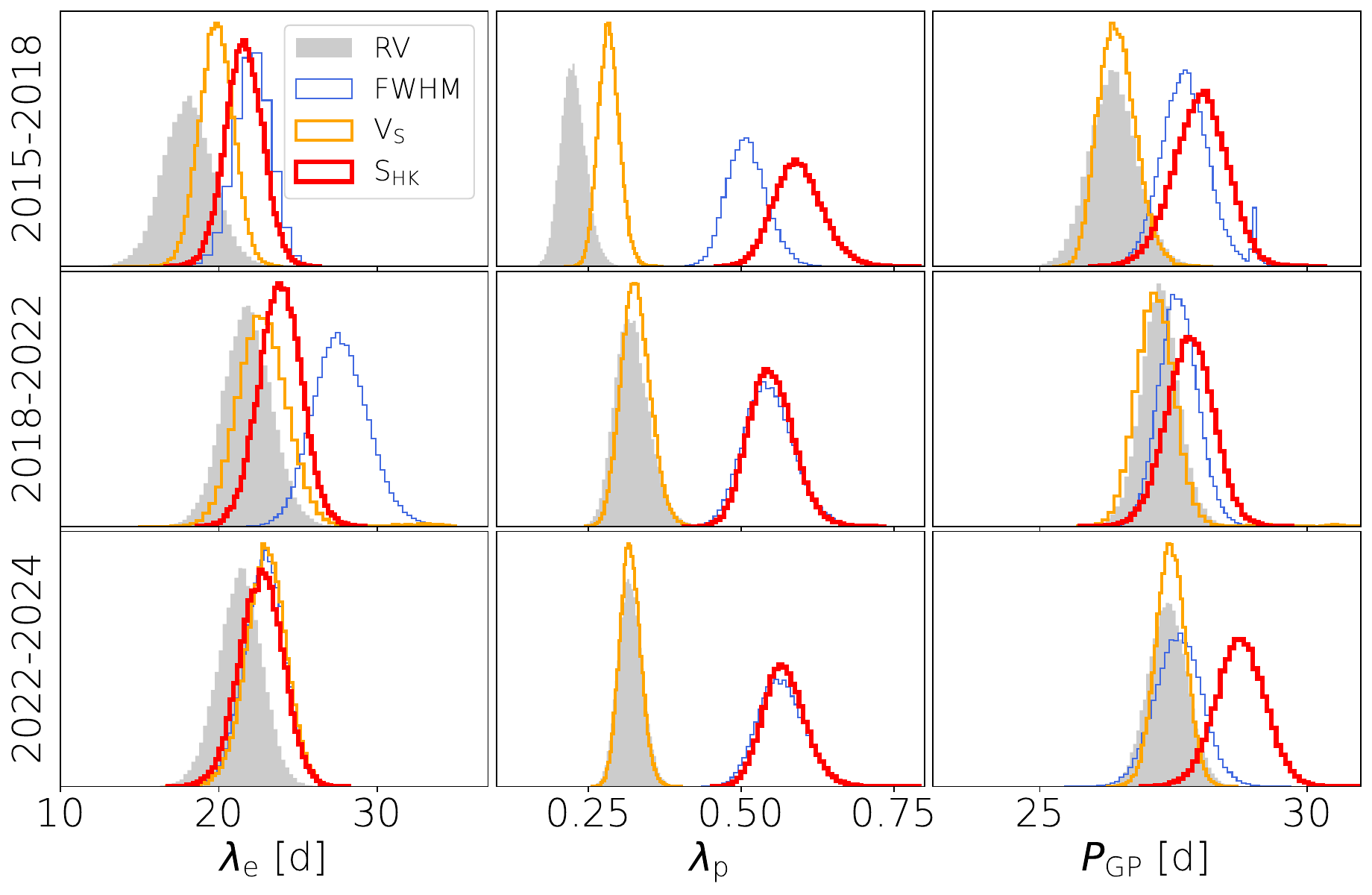}  
    \caption{One-dimensional posterior densities of the hyper-parameters of the GP fit to the HARPS-N solar RVs (filled grey histograms), and to the time series of FWHM (thin blue lines), bisector velocity span (gold lines) and S$_{\mathrm{HK}}$ (thick red lines), for the three seasons listed in Tab.~\ref{tab:rms_rvs}. Columns 1, 2 and 3 show the posterior densities of the GP evolution time scale ($\lambda_{\mathrm{e}}$), inverse harmonic complexity ($\lambda_{\mathrm{p}}$) and period (P$_{\mathrm{GP}}$).}    
    \label{fig:posterior_usual}
\end{figure*}

\section{Gaussian process regression}\label{sec:gp_sec}

Our activity-filtering framework provides a new time series of RVs, \vper, less dispersed than the input RVs, \vobs. In the last section, we demonstrated that planet signatures in \vobs\ would be largely preserved in \vper. However, while the activity-filtering algorithm turns out to be efficient at filtering long-term activity effects, it leaves a significant rotationally-modulated component in \vper, which also needs to be filtered out in order to improve the detectability of small planet signatures. Gaussian Process regression has become widely used to model quasi-periodic signals \citep{haywood2014,rajpaul2015,barragan2022,nicholson2022,aigrain2022}. In this section, we investigate how GPs can be used to model the stellar activity contributions in \vobs\ and \vper. In Sec.~\ref{ssec:1D_gp_usual}, we first investigate how the statistical properties of \vobs\ and usual stellar activity indicators compare to each other and evolve over time. In sec.~\ref{ssec:1D_gp_new}, we assess the ability of GPs to model the quasi-periodic activity signals in \vper, and in $\boldsymbol{U_{1}}$ to $\boldsymbol{U_{4}}$. In Sec.~\ref{ssec:GP_planet_injection}, we finally investigate how a joint multi-dimensional GP-based activity filtering and planet search algorithm will perform for different indicators and input RV signals. All GP modellings in this section were performed using the open-source software \texttt{pyaneti} \citep{barragan2019,barragan2022A} with non-informative priors for all parameters.

\subsection{Modelling of usual activity proxies}\label{ssec:1D_gp_usual}

We start by independently modelling the HARPS-N RV time series, \vobs, as well as the time series of usual activity proxies (FWHM, V$_{\mathrm{s}}$, S$_{\mathrm{HK}}$), using a one-dimensional (1D) GP with quasi-periodic covariance kernel $k$ between each pair of epochs $t_{i}$ and $t_{j}$ given by

\begin{equation}
    k(t_{i},t_{j}) = A^2 \exp \left[ - \frac{\sin^{2} \left( \pi \left(t_{j} - t_{i} \right)/P_{\rm{GP}} \right)}{2 \lambda_{\rm{p}}}  - \frac{\left(t_{j} - t_{i} \right)^{2}}{2 \lambda_{\rm{e}}}  \right],
    \label{eq:cov_QP}
\end{equation}

\noindent
where $P_{\rm{GP}}$ is the GP period, which corresponds to the stellar rotation period \citep{angus2018,nicholson2022}, $\lambda_{\rm{p}}$ is the inverse harmonic complexity and $\lambda_{\rm{e}}$ is the GP evolution time scale (already defined in Eq.~\ref{eq:SE_kernel}). Each time series is modelled by six parameters: three hyper-parameters in the covariance matrix (Eq.~\ref{eq:cov_QP}), the GP amplitude $A$, a constant offset, and one additional uncorrelated jitter term $\sigma_{\mathrm{j}}$, added in quadrature to the diagonal of the covariance matrix, to absorb any variations not accounted for by the GP. The parameter space is explored using the Bayesian Markov chain Monte Carlo (MCMC) sampler defined in \citet{barragan2019}, with uninformative priors for the parameters. To ensure that the MCMC process has converged, \texttt{pyaneti} iteratively runs 250 independent chains of 5\,000 steps until the chains converge, which is assessed using the statistics of \citet{gelman1992}. As detailed in the Section~2.5 of \citet{barragan2019}, the code compares the variances between chains and within chains and uses the scaling factor $\hat{R}$ introduced in \citet{gelman2004} to assess convergence (typically with $\hat{R}$\,$<$\,1.02). We then run a last MCMC process with 5\,000 steps of 250 independent chains, corresponding to 125\,000 independent samples for each parameter, from which we estimate the best-fitting value with 1$\sigma$ uncertainties.

Since the HARPS-N solar data set covers more than half of the Sun's activity cycle, it is likely that the statistical properties (i.e. evolution time scale, harmonic complexity, period) of the activity RV signal vary over the course of the observations. We therefore independently analyse each of the three seasons defined in Tab.~\ref{tab:rms_rvs} (i.e. 2015-2018, 2018-2021, 2021-2023). The best-fitting GP hyper-parameters for each season are given in Tab.~\ref{tab:parameters_1d}, and their associated 1D posterior distributions are shown in Fig.~\ref{fig:posterior_usual}.

Significant variations are observed from one season to the next. The RV GP period increases from $\sim$26.4\,d, in the 2015-2018 season, to $\sim$27.4\,d after 2022 (i.e. $\sim$2$\sigma$ increase). This is in line with expectations as active regions are found at higher latitudes in the start of the activity cycle and migrate near the equator at the end of the cycle \citep[see][for a review of solar activity cycle]{hathaway2010}. The difference in rotation periods then simply reflects the latitudinal differential rotation of the Sun \citep[e.g.][]{howard1970,snodgrass1990,beck2000}. For example, the average latitudes of sunspots computed from the Solar Optical Observing Network of the US Air Force (USAF), with the help of the US National Oceanic and Atmospheric Administration (NOAA), is about 12$^{\circ}$ in 2015-2018 and 20$^{\circ}$ in 2022-2024\footnote{\url{http://solarcyclescience.com/index.html}}, corresponding to a difference of about 1d in rotation period \citep[using the latitudinal differential rotation profile of][]{snodgrass1990}, consistent with our estimates in Tab.~\ref{tab:parameters_1d}.




\begin{figure*}
    \centering
    \includegraphics[width=\linewidth]{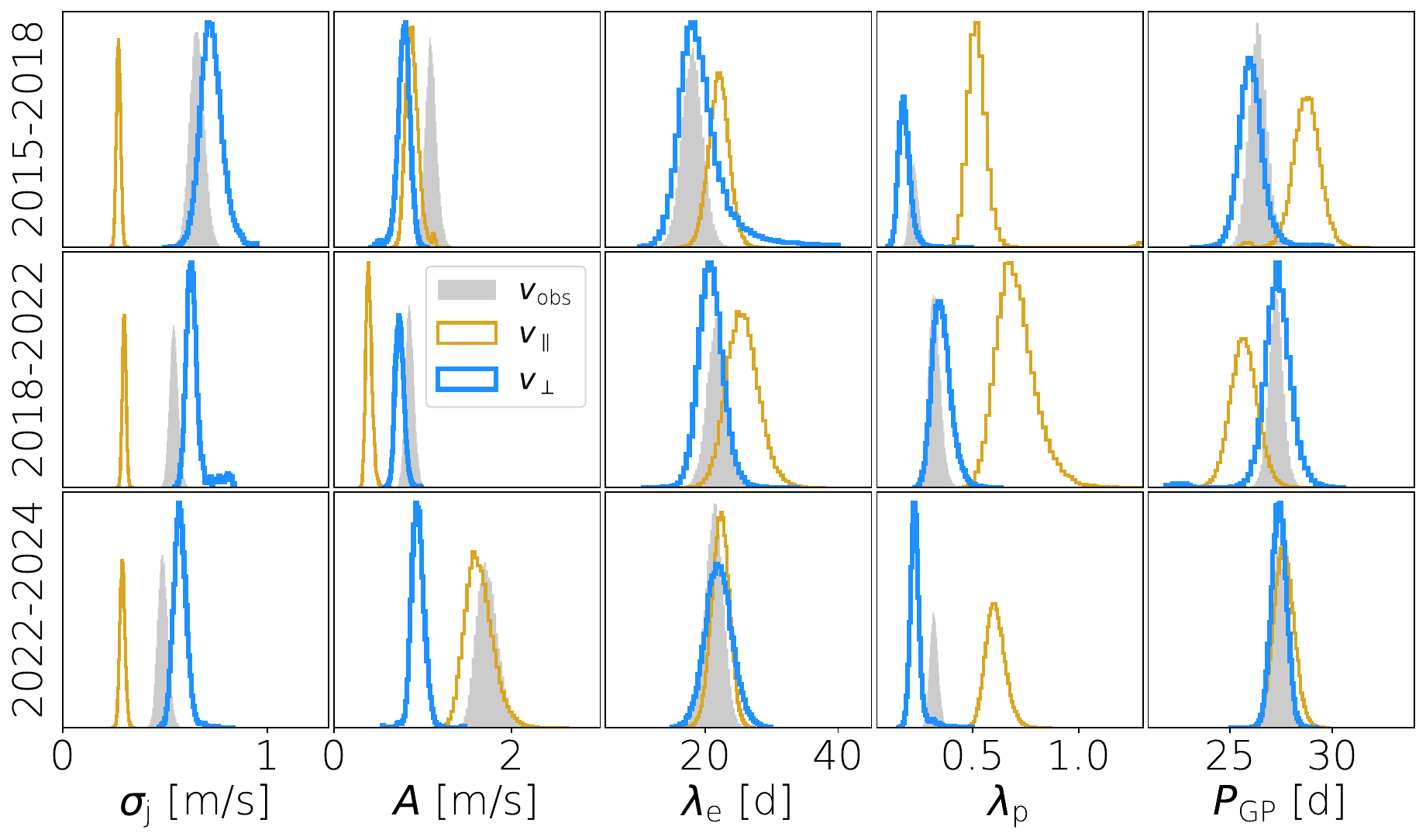}
    \caption{Uni-dimensional posterior densities of the jitter term, GP amplitude and GP hyper-parameters of the fit to the HARPS-N solar RVs (filled gray distributions), the shape-driven RVs \vpar\ (thin gold lines), and the shift-driven RVs \vper\ (thick blue lines).}
    \label{fig:posterior_new}
\end{figure*}

We also note that the RV GP evolution time scales ($\lambda_{\rm{e}}$) are almost 2$\sigma$ larger (i.e. $\sim$3\,d) past 2018. Disk-resolved solar observations found larger active regions in the 2022-2024 period than in 2015-2018\footnotemark[\value{footnote}], which is consistent with our observations as the active region evolution time scale roughly scales with the surface area of the features \citep[e.g.][]{leighton1964,wang1989,foukal1998}. The higher harmonic complexity (i.e. lower $\lambda_{\rm{p}}$) in the 2015-2018 season can be explained by the fact that equatorial active regions perturbs the wings of the profile more than higher-latitude regions and thus have an higher impact on the flux derivative which induce a sharper RV signature. As expected, $\lambda_{\rm{p}}$ is about twice as large for indicators of the photospheric flux (FWHM, S$_{\mathrm{HK}}$) than for indicators of the first temporal derivative of the flux, like RV V$_{\mathrm{s}}$ \citep[e.g.][]{aigrain2012,klein2019,nicholson2022}. The evolution of the GP covariance kernel over time, and its effect on the modelling of RV time series, is discussed in Sec~\ref{ssec:discussion_covariance}.

\subsection{Modelling activity-filtered RVs with Gaussian Processes}\label{ssec:1D_gp_new}

Using the 1D GP framework described in Sec.~\ref{ssec:1D_gp_usual}, we model the time series of \vper, \vpar, as well as $\boldsymbol{U_{1}}$ to $\boldsymbol{U_{4}}$. Except $\boldsymbol{U_{2}}$, for which the MCMC process does not converge, all time series exhibit a quasi-periodic modulation well captured by the GP. For these time series, the best-fitting GP hyper-parameters are given in Tab.~\ref{tab:parameters_1d} and the posterior distributions of the fit parameters are shown for \vpar\ and \vper\ in Fig.~\ref{fig:posterior_new}.

As already intuited from Fig.~\ref{fig:rv_fit_sun}, we find that the shape-driven RVs, \vpar, are systematically smoother (i.e. with larger values of $\lambda_{\mathrm{p}}$) than \vobs, which suggests that the activity signal within \vpar\ is mostly driven by variations of the photospheric flux (typically induced by faculae and plages). This is further evidenced by the fact that \vpar\ tends, in general, to evolve on longer time scales than \vobs. On the other hand, the shift-driven RVs \vper, exhibit a higher harmonic complexity and evolve on time scales similar to \vobs, which suggests that the stellar activity signal in \vper\ is sensitive to the variations of the first derivative of the photospheric flux (e.g. induced by spots). As expected, the best-fitting GP amplitudes are consistent with the RMS of the different time series given in Tab.~\ref{tab:rms_rvs}.

In all three seasons, a value of 0.25-0.3\,\ms\ is obtained for jitter term in \vpar, much lower than its counterpart in \vobs. This is expected as, by definition, \vpar\ contains only a small fraction of the white noise of HARPS RV and, therefore, the estimate of $\sigma_{\mathrm{j}}$ reflects more what the GP cannot model than the real dispersion of white noise in \vpar. On the other hand, jitter terms of 0.5 to 0.6\,\ms\ are found for \vobs\ and \vper \footnote{The typical error bar on the jitter term is about 0.04\,\ms.}. In each case, the jitter term is considerably larger than the photon noise of $\sim$\,0.1\,\ms, on the daily-binned RVs, which suggests that a significant fraction of the RV variation can be explained neither by the GP nor by formal uncertainties. In our daily-binned data set, both oscillation- and granulation-driven RV variations are expected to be averaged out due to their short evolution time scales \citep[e.g.][]{dumusque2011,chaplin2019}. In contrast, super-granulation, which evolves on time scale of days, will be only partially reduced by our binning process. Recently, \citet{almoulla2023} and \citet{lakeland2024} measured RV RMS contributions of $\sim$0.7\,\ms\ and $\sim$0.9\,\ms\ for super-granulation in the HARPS-N solar data, consistent with the simulations of \citet{meunier2015}. Therefore, the dispersion budget in the RV residuals is likely a mix of super-granulation and long-term stability ($\sim$0.4\,\ms\ for HARPS-N).

The posterior densities of the hyper-parameters of the GP fit to $\boldsymbol{U_{1}}$, $\boldsymbol{U_{3}}$ and $\boldsymbol{U_{4}}$ are shown in Fig.~\ref{app:fig:posterior_sha}, and their best-fitting values are given in Tab.~\ref{tab:parameters_1d}. As $\boldsymbol{U_{1}}$ is by far the dominant component of the SVD to the shape-driven CCFs, it follows a very similar behaviour to \vpar\ and to S$_{\mathrm{HK}}$ (as expected from Fig.~\ref{fig:correl}). On the other hand, $\boldsymbol{U_{3}}$ and $\boldsymbol{U_{4}}$ exhibit similar statistical properties as \vper, which suggests that they could be good proxies of the stellar activity signal in the latter time series. This is investigated in the next section.

\subsection{Multi-dimensional GP analysis}\label{ssec:GP_planet_injection}

In the practical case of the search for planet signatures in observed RVs, the shape-driven RVs, \vpar\ could potentially be used as planet-independent proxies to mitigate stellar activity signals. The multi-dimensional GP framework of \citet{rajpaul2015} is one of the most robust ways to simultaneously model RVs and activity proxies, thereby boosting the number of constraints on the stellar activity parameters. On the other hand, the shift-driven RVs, \vper, are less dispersed than \vobs, which could boost the sensitivity to low-ampitudes planet signatures. Furthermore, we have shown in Sec.~\ref{ssec:1D_gp_new} that the residual quasi-periodic activity signals in \vper\ would be well modelled by a GP with quasi-periodic kernel. In this section, we explore how \vobs\ and \vper\ perform in the search for long-period low-amplitude planet signatures.

\subsubsection{Method}\label{sssec:gpmethod}

\renewcommand{\arraystretch}{1.2}
\begin{table*}
    \centering
    \caption{Estimates of the planet RV semi-amplitude (\kest) and RV jitter ($\sigma_{\rm{j}}$) for the different cases considered in Sec.~\ref{ssec:GP_planet_injection}. Note that the semi-amplitude of the planet RV signature injected in the data is \kinj\,=\,0.4\,\ms.}
    \label{tab:estimates_kp}
    \begin{tabular}{cccccccc}
    \hline
        \vspace{0.1cm}
      \textbf{Case} & \textbf{Time series}   &  \multicolumn{2}{c|}{\textbf{2015-2018}} & \multicolumn{2}{|c|}{\textbf{2018-2022}} & \multicolumn{2}{|c|}{\textbf{2022-2024}}   \\
  &  &  \kest [\ms] &   $\sigma_{\mathrm{j}}$ [\ms] &  \kest [\ms] &   $\sigma_{\mathrm{j}}$ [\ms] &  \kest [\ms] &   $\sigma_{\mathrm{j}}$ [\ms] \\ 
    \hline
(R$_{0}$) & \vobs & 0.40 $\pm$ 0.13 & 0.65$_{-0.02}^{+0.03}$  & 0.39 $\pm$ 0.12 & 0.54 $\pm$ 0.02  & 0.43$_{-0.19}^{+0.22}$ & 0.49$_{-0.02}^{+0.03}$  \\
(R$_{\perp}$) &  $v_{\perp}$ & 0.41 $\pm$ 0.09 & 0.71$_{-0.04}^{+0.06}$  & 0.40 $\pm$ 0.12 & 0.62 $\pm$ 0.02  & 0.40 $\pm$ 0.11 & 0.58$_{-0.02}^{+0.03}$ \\
(1) & \vobs + FWHM & 0.40 $\pm$ 0.06 & 0.89$_{-0.02}^{+0.05}$  & 0.39$_{-0.06}^{+0.07}$ & 0.81 $\pm$ 0.02  & 0.40$_{-0.06}^{+0.07}$ & 0.94$_{-0.02}^{+0.03}$  \\
(2) & \vobs + V$_{\mathrm{s}}$ & 0.40 $\pm$ 0.07 & 0.90$_{-0.02}^{+0.05}$  & 0.38$_{-0.11}^{+0.12}$ & 0.62 $\pm$ 0.02  & 0.41 $\pm$ 0.07 & 0.80$_{-0.03}^{+0.02}$  \\
(3) &  \vobs + S$_{\mathrm{HK}}$ & 0.39 $\pm$ 0.07 & 0.95 $\pm$ 0.03  & 0.38 $\pm$ 0.06 & 0.83 $\pm$ 0.02  & 0.40 $\pm$ 0.08 & 1.03 $\pm$ 0.02   \\
(4) & \vobs + $v_{\perp}$ & 0.38 $\pm$ 0.07 & 0.93 $\pm$ 0.03  & 0.38 $\pm$ 0.11 & 0.55 $\pm$ 0.02  & 0.40 $\pm$ 0.07 & 0.98 $\pm$ 0.03  \\
(5) &  v$_{\perp}$ + FWHM & 0.40 $\pm$ 0.07 & 1.01$_{-0.02}^{+0.03}$  & 0.39 $\pm$ 0.06 & 0.90 $\pm$ 0.02  & 0.40$_{-0.07}^{+0.08}$ & 1.06$_{-0.03}^{+0.02}$  \\
(6) &  v$_{\perp} + V_{\mathrm{s}}$ & 0.42 $\pm$ 0.07 & 0.97 $\pm$ 0.02  & 0.36 $\pm$ 0.09 & 0.76 $\pm$ 0.03  & 0.40$_{-0.07}^{+0.09}$ & 1.02$_{-0.04}^{+0.03}$  \\
(7) & v$_{\perp} + S_{\mathrm{HK}}$ & 0.39 $\pm$ 0.07 & 1.01$_{-0.02}^{+0.03}$  & 0.39$_{-0.07}^{+0.06}$ & 0.92 $\pm$ 0.02  & 0.40$_{-0.07}^{+0.08}$ & 1.08$_{-0.02}^{+0.03}$  \\
(8) & v$_{\perp} + v_{\parallel}$ & 0.41 $\pm$ 0.07 & 1.01 $\pm$ 0.02  & 0.40 $\pm$ 0.14 & 0.67 $\pm$ 0.04  & 0.40 $\pm$ 0.07 & 1.07 $\pm$ 0.02  \\
 \hline
\end{tabular}
\end{table*}

\begin{figure}
    \centering
    \includegraphics[width=\linewidth]{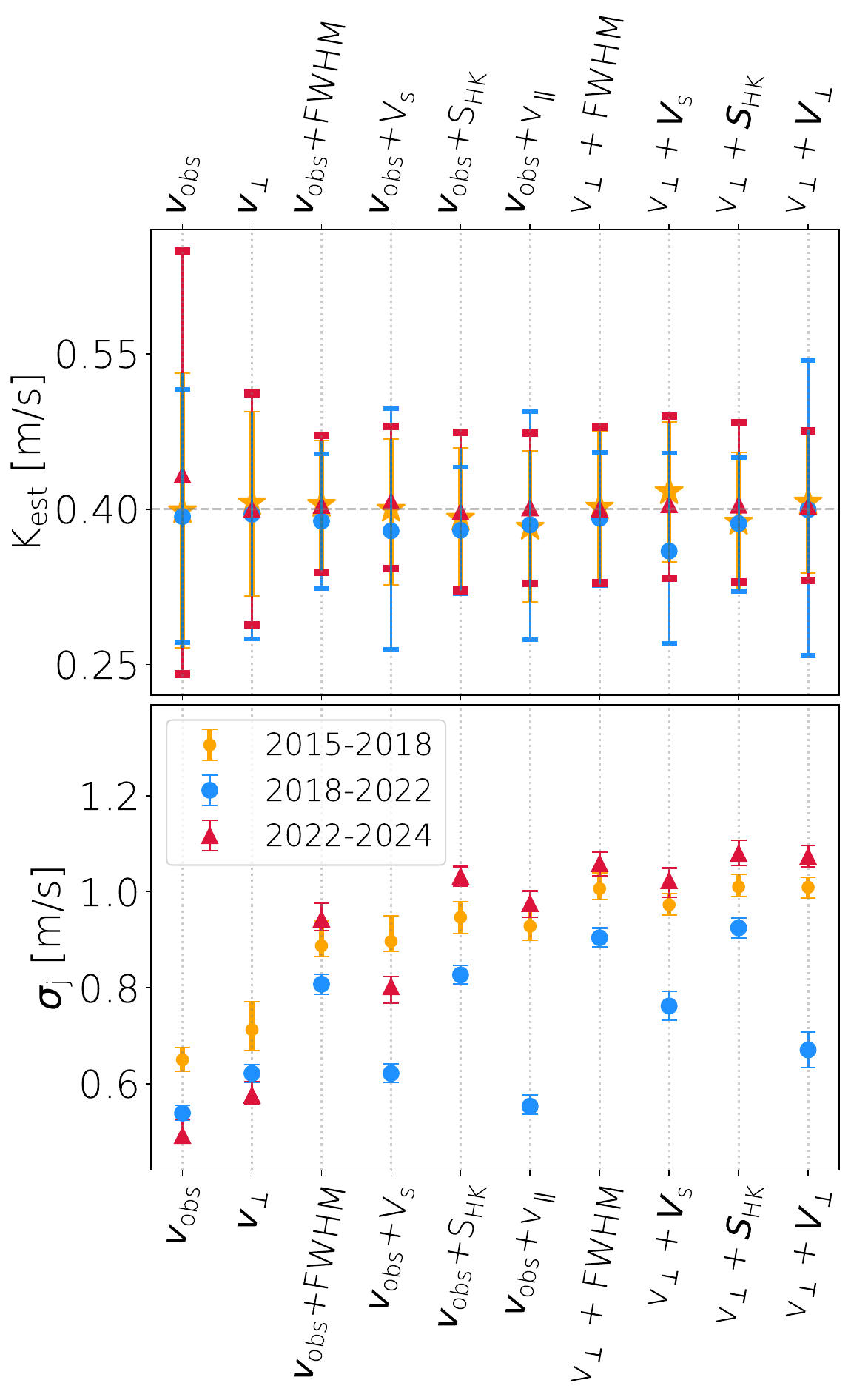}
    \caption{Best estimates of the planet RV semi-amplitude (\kest, top panel) and jitter term ($\sigma_{\mathrm{j}}$, bottom panel) for the different cases listed in Tab.~\ref{tab:estimates_kp}. The orange dots, light blue filled circles and red triangles indicate the results obtained for the 2015-2018, 2018-2022 and 2022-2024 season, respectively. The dashed horizontal line on the top panel indicates the semi-amplitude of the planet RV signature injected in the data.} 
    \label{fig:final_kp}
\end{figure}

In our model, we assume that the activity signal in the RVs and indicators follow a FF'-like relation \citep{aigrain2012,rajpaul2015}. Each time series $\boldsymbol{\alpha}$ is expressed as a linear combination of a latent variable $G$ (typically the square of the photospheric flux) and its first temporal derivative $\dot{G}$, such that, at time $t$,

\begin{equation}
\alpha (t) = A_{\alpha} G(t) + B_{\alpha} \dot{G} (t) + C_{\alpha},
\label{eq:multi_gp}
\end{equation}

\noindent
where the amplitudes $A_{\alpha}$, $B_{\alpha}$ and $C_{\alpha}$ are free parameters of the model. The latent variable $G$ is modelled as a GP with the quasi-periodic covariance kernel defined in Eq.~\ref{eq:cov_QP}.

We inject the RV signature of a single planet on a circular orbit to the HARPS-N solar CCFs. We assume an orbital period of 100\,d and a RV semi-amplitude of 0.4\,\ms, which corresponds to a planet mass of 2.9\,M$_{\earth}$. In order to limit bias, three different orbital phases ($\phi_{\mathrm{p}}$ in Eq.~\ref{eq:rvp}) of 0.0, 0.3 and 0.7 are considered. We apply the framework of Sec.~\ref{ssec:framework} to generate time series of shift- and shape-driven RVs, from the input RVs. We then use the multi-dimensional GP framework to model the activity signal within (1)~\vobs\ and FWHM, (2)~\vobs\ and V$_{\mathrm{s}}$, (3)~\vobs\ and S$_{\mathrm{HK}}$, (4)~\vobs\ and \vpar. These indicators exhibit relatively similar evolution time scales and period to \vobs, and are therefore promising candidates for multi-dimensional GP modelling. Differences in $\lambda_{\mathrm{p}}$ are naturally accounted for in the framework of \citet{barragan2022A}.


Since we have demonstrated that planet signatures are mostly preserved in \vper, the planet search can as well be performed directly from this time series\footnote{Note that, as discussed in Sec.~\ref{sec:sec2}, small-amplitude planet signatures are expected to be preserved in the shift-driven RVs, \vper. Coupling a multi-dimensional GP framework with the likelihood model of Eq.~\ref{eq:likelihood} will be implemented in a forthcoming work.}. We therefore consider four additional cases, namely (5)~\vper\ and FWHM, (6)~\vper\ and V$_{\mathrm{s}}$, (7)~\vper\ and S$_{\mathrm{HK}}$, and (8)~\vper\ and \vpar. As a reference, we also model the stellar activity RV signals in \vobs\ (case R$_{0}$) and in \vper\ (case R$_{\perp}$) using uni-dimensional GPs with QP and SE kernels, respectively. In all cases, the GP is jointly modelled with a planet RV signature of fixed orbital period and phase, to simulate the RV monitoring of a transiting planet (e.g. unveiled by the PLATO space mission). Note also that, in all cases, the same planetary signal has been injected to the data, but the GP model differs from one case to the next. As shown in Sec.~\ref{ssec:1D_gp_usual}, the statistical properties of the time series vary significantly from one season to the next. We therefore choose to perform the planet injection-recovery tests independently on each of the three seasons defined in Tab.~\ref{tab:rms_rvs}. For each season, our input data sets contain the same amount of data points (460), spread evenly over a $\sim$2-yr period. The parameter space is sampled using the procedure described in Sec.~\ref{ssec:1D_gp_usual}.

\subsubsection{Results}

The best estimates of the planet RV semi-amplitude and of the jitter term $\sigma_{\mathrm{j}}$ are given in Tab.~\ref{tab:estimates_kp} and shown in Fig.~\ref{fig:final_kp}. The recovered values of \kest\ are consistent with the injected one in all cases (note that, in all cases, the same planetary signal has been injected to the data), but the precision of the estimate varies significantly from one case to the next. Unsurprisingly, the 1D GP modelling of the HARPS-N RVs performs the least well, especially when the Sun is more active, in which case only a marginal $\sim$2\,$\sigma$ planet detection can be claimed. This is due to the fact that the long-term activity evolution and the planet signature cannot be easily separated with a simple 1D GP, which leads to larger error bars on \kest.

We find that multi-dimensional GPs give, in most cases, much smaller error bars on \kest. In particular, cases (1) and (3), where the FWHM of the CCF and the S index are used as activity proxies of \vobs, perform best in all three seasons. These two indicators capture well the activity signatures sensitive to the photospheric flux, and thereby complement well the RV signals, sensitive to variations of the flux and its first derivative. In these two cases, the multi-dimensional GP leverages all the constraints available, which results in more precise estimates of \kest. This also leads to higher values of $\sigma_{\mathrm{j}}$, since the GP is now less flexible (due to the increased number of constraints), and less likely to absorb non-activity-induced variations.

We also note that cases (2) and (4), where \vobs\ is combined with V$_{\mathrm{s}}$ and \vpar, perform well when the Sun is more active but yield significantly larger error bars on \kest\ during solar minimum. During this period, there are fewer and smaller active regions than during more active epochs. It is therefore more likely that the statistical properties of the stellar activity RV signal vary significantly from one solar rotation to the next, and are thereby more difficult to model with a quasi-periodic GP (see Sec.~\ref{ssec:discussion_covariance}). During solar minimum, neither V$_{\mathrm{s}}$ nor \vpar\ bring any additional constraint on the quasi-periodic activity RV signal and, therefore, the multi-dimensional GP modelling leads to results similar as the 1D GP modelling of \vobs.

More precise values of \kest\ are obtained in case R$_{\perp}$ than in case R$_{0}$, which just reflects the fact that activity has been partly filtered from \vper. However, these values are significantly less precise than those obtained in the multi-dimensional GP framework, even when the computation of \vper\ is bypassed (i.e. cases 1 to 4). Cases where multi-dimensional GPs are used with \vper\ (i.e. cases 5 to 8) tend to perform similar as their \vobs\ counterpart (cases 1 to 4). In addition, we note that none of the SVD components, $\boldsymbol{U_{1}}$, $\boldsymbol{U_{3}}$ and $\boldsymbol{U_{4}}$, outperforms \vpar\ in the multi-dimensional GP framework. In particular, despite their periodic modulation, $\boldsymbol{U_{3}}$ and $\boldsymbol{U_{4}}$ do not improve the RV modelling compared to case R$_{0}$.


\section{Discussion and conclusions}\label{sec:section:discussion}

In this paper, we analysed the activity-induced distortions in the absorption lines of the Solar spectrum, intensively monitored with HARPS-N over the last 8 years. From the DRS-processed systematic-corrected CCFs, we constructed a dimensionally-reduced basis, which allowed us to separate the observed RVs (\vobs) into complementary time series, namely \vper\ (i.e. shift-driven RVs) and \vpar\ (shape-driven RVs). In the first one, the variations are largely dominated by Doppler shifts of the entire CCF, induced, for example, by planets or granulation at the solar surface. In the second one, the variations are only driven by CCF distortions and, thereby, probe the effects of active regions whilst being independent of planet signatures. When we apply this framework to the HARPS-N solar spectra, the RMS of the observed RVs goes from 2.05 to 1.06\,\ms, hence a decrease of about 50\%. We find that planet signatures are mostly preserved in \vper, and that the efficiency of the activity filtering is not affected by the temporal sampling. However, as shown in Fig.~\ref{fig:periodogram_solar}, \vper\ exhibits significant rotation-induced variations, suggesting that the shift-driven RVs are still affected by stellar activity.



\subsection{Evolution of the GP covariance structure}\label{ssec:discussion_covariance}

\begin{figure}
    \centering
    \includegraphics[width=\linewidth]{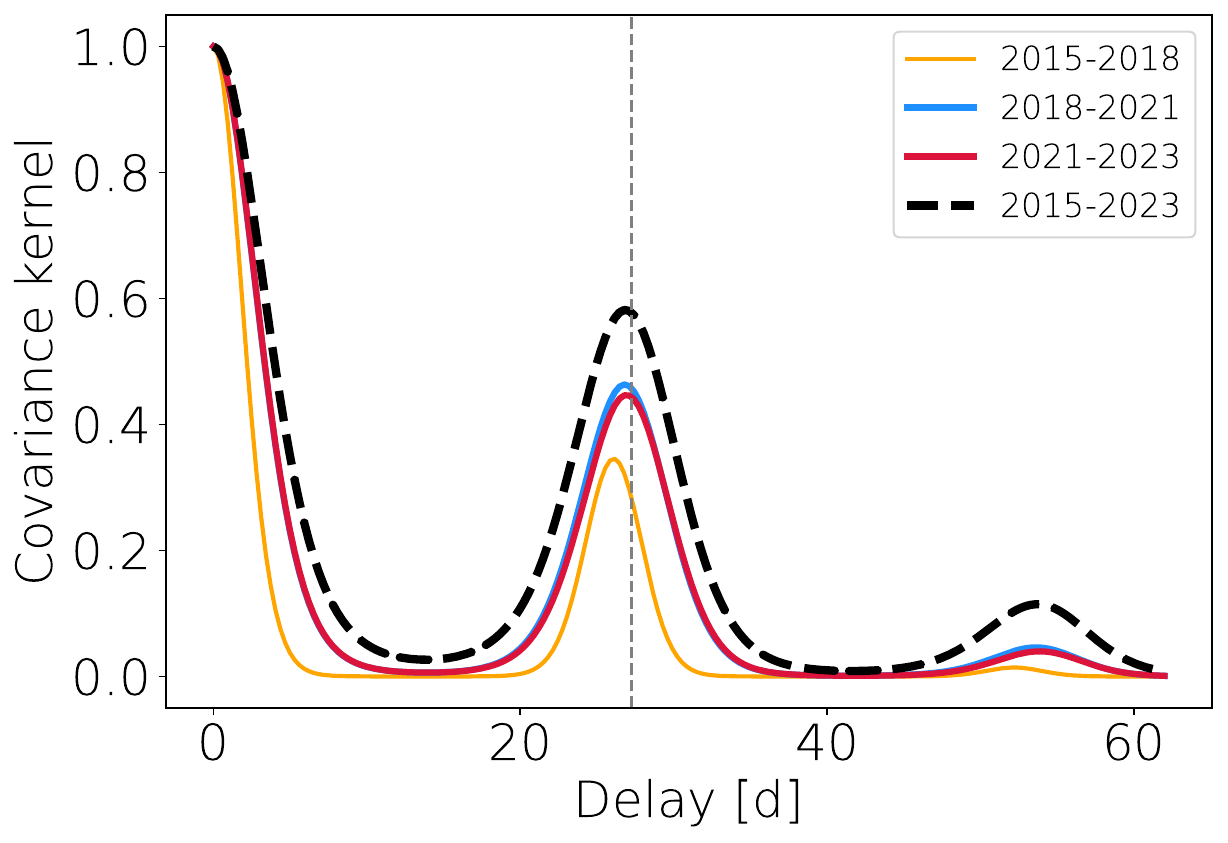}
    \caption{Evolution of the covariance kernel of the stellar activity RV signal, computed using Eq.~\ref{eq:cov_QP} with the best-fitting hyperparameters of the GP modelling. The thick dashed line is obtained by modelling all RVs together and the vertical dotted line indicate the average synodic rotation period of the Sun (27.2753\,d).}
    \label{fig:covariance}
\end{figure}

In order to better understand the nature of the rotational modulation within \vobs, \vper\ and \vpar, all time series were modelled using a GP with quasi-periodic covariance kernel. As shown in Fig.~\ref{fig:covariance}, the structure of the covariance matrix of the time series evolves significantly over time. The RV variations are smoother and more slowly evolving at the start of Cycle~25 than at the end of Cycle~24, which reflects the fact that the Sun's active regions are larger and at higher-latitude in the start of the magnetic cycle. We found that the quasi-periodic RV activity signals driven by flux variations as well as the long term cycle evolution are well captured by \vpar. On the other hand, \vper\ is still plagued by high-complexity quasi-periodic activity signals, likely reflecting variations in the first temporal derivative of the photospheric flux.

To visualise how the temporal evolution of the GP covariance kernel affects our GP fit, we model the full HARPS-N RV time series with a single 1D GP, using the quasi-periodic covariance kernel of Eq.~\ref{eq:cov_QP}. The best-fitting covariance kernel, shown in dashed black line in Fig.~\ref{fig:covariance}, differs significantly from the kernels obtained individually for the different seasons. The signal is now found smoother ($\lambda_{\rm{p}}~\approx~0.4$) and more slowly evolving ($\lambda_{\rm{e}}~\approx$~26\,d). On the other hand, the uncorrelated jitter term $\sigma_{\mathrm{j}}$ is now $\sim$1.8\,\ms, almost three times greater than on individual seasons. As explained in Sec.~\ref{ssec:GP_planet_injection}, we expect $\sigma_{\mathrm{j}}$ to increase, since we have tripled the number of constraints in the fit. On the other hand, the values of $\sigma_{\mathrm{j}}$ obtained with the multi-dimensional GP in Tab.~\ref{tab:estimates_kp} are about 0.8-1.0\,\ms, significantly smaller than 1.8\,\ms. Moreover, such a high value can be explained neither by super-granulation or long-term instrument stability. We therefore conclude that, as things stand, quasi-periodic stellar activity RV signals are better modelled by GPs on chunks during which cycle-driven variations remain small.

If we now model the full time series of shift-driven RVs, \vper, with a single 1D GP, the best-fitting hyper-parameters are fully consistent with those obtained on individual seasons, despite the variations of P$_{\rm{GP}}$ between 2015 and 2024. The best-fitting evolution time scale and GP period lie in between the estimates listed in Tab.~\ref{tab:parameters_1d}, and a model with high harmonic complexity ($\lambda_{\rm{p}} = 0.23 \pm 0.01$) is preferred. The value of $\sigma_{\mathrm{j}}$ (0.62\,$\pm$\,0.02\,\ms) lies in between the values obtained on individual seasons (between 0.57 and 0.72\,\ms), which indicates that our fit is as good on the whole data set as it is on individual seasons. This can be explained by the fact that most of the long-term activity variations (e.g. induced by the magnetic cycle), are well captured by \vpar\ and, thus, well filtered out from \vper. Therefore, we expect the GP covariance matrix to exhibit fewer variations in the case of \vper\ than in the case of \vobs.


To investigate what drives the evolution of the quasi-periodic covariance kernel in \vobs, we divide the HARPS-N RVs into 270-d chunks (i.e. 10 rotation periods), using a moving window with a step of 100 days. After discarding chunks containing large gaps in the data, we model each subset of data using a GP with the quasi-periodic kernel given in Eq.~\ref{eq:cov_QP}. The best-fitting hyper-parameters are shown in Fig.~\ref{fig:distrib_chunks}. The variation of the GP amplitude, which is well described by a parabola, is unsurprisingly a good tracer of the solar magnetic cycle. We do not observe significant variations either in the time scale of the evolution of the GP or in the period, due to the relatively large uncertainties over these two parameters (about 4\,d and 1\,d for $\lambda_{\mathrm{e}}$ and P$_{\mathrm{GP}}$, respectively). On the other hand, inverse harmonic complexity, $\lambda_{\mathrm{p}}$, varies significantly from one chunk to the next. When discarding the values obtained during solar minimum, often plagued with large uncertainties due to the weak activity signal, we find a positive correlation with the GP amplitude (Pearson correlation coefficient of 0.7).

\begin{figure}
    \centering
    \includegraphics[width=\linewidth]{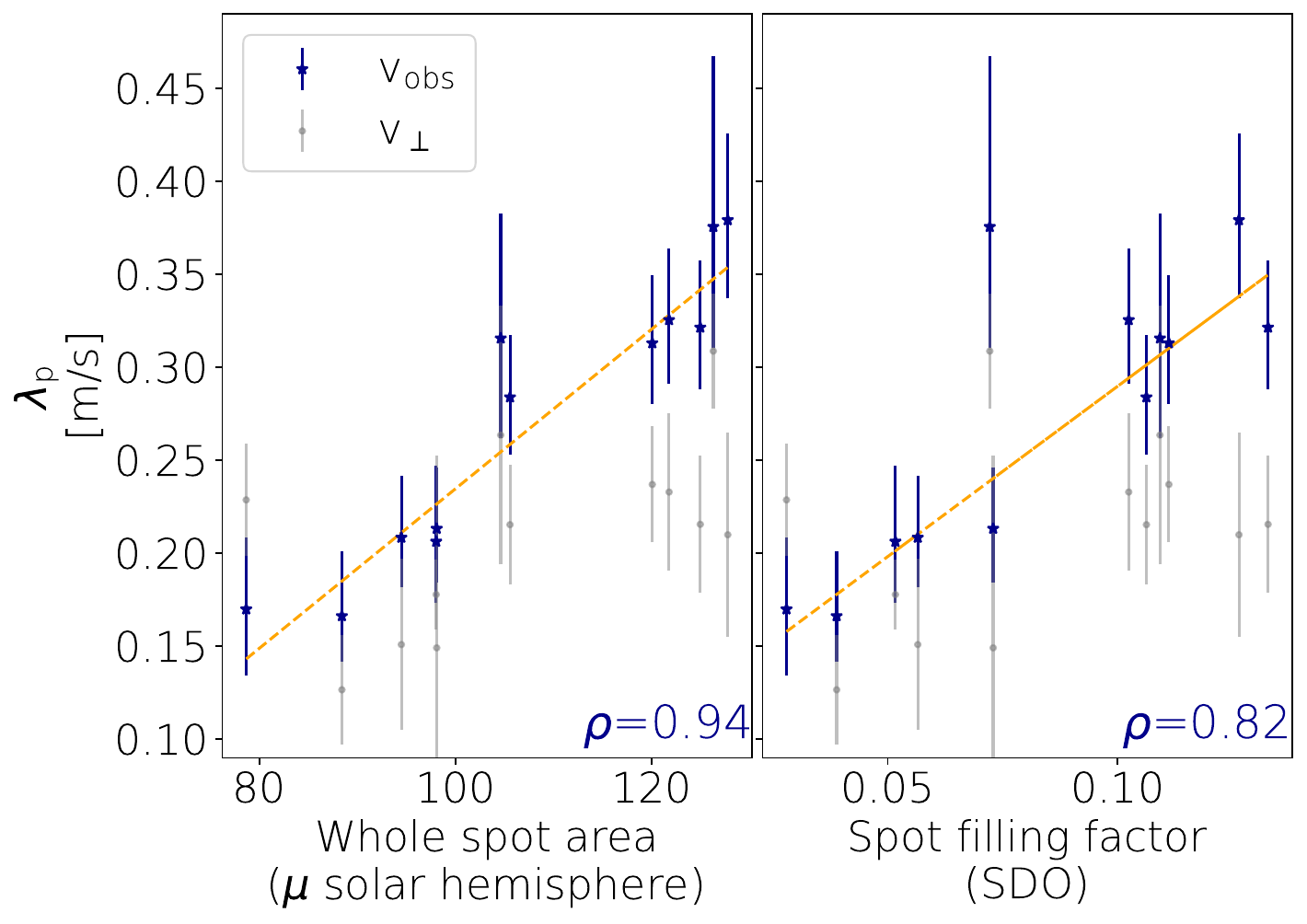}
    \caption{GP inverse harmonic complexity extracted from the HARPS-N RVs (\vobs, dark blue stars) as a function of the whole sunspot area (in millionths of solar hemisphere; left-hand panel), and of the SDO spot filling factor (right-hand panel). In each panel, we give the Pearson correlation coefficient $\rho$, and show the best-fitting straight line (orange dashed lines). For comparison, we also show the inverse harmonic complexities extracted from the shift-driven RVs (\vper) in gray dots, with Pearson correlation coefficients of 0.55 and 0.39 with the whole spot area and SDO spot filling factor, respectively.}
    \label{fig:correl_lp}
\end{figure}

In order to better understand the origin of this correlation, we compare in Fig.~\ref{fig:correl_lp} the values of $\lambda_{\mathrm{p}}$ to the whole sunspot area observed by USAF/NOAA\footnotemark[\value{footnote}], averaged over each chunk outside the solar minimum, and to the spot filling factor derived from SDO HMI observations \citep[using the method of][]{milbourne2019,haywood2022,ervin2022,lakeland2024}. Both quantities appear to strongly correlate with $\lambda_{\mathrm{p}}$ (Pearson correlation coefficients of 0.94 and 0.82; see Fig.~\ref{fig:correl_lp}). When the spot filling factor is large, sunspots are likely to be more evenly distributed in longitude at the stellar surface, which results in a smoother RV signal (i.e. greater values of $\lambda_{\mathrm{p}}$). Conversely, for small filling factors, sunspots are more likely to be longitudinally isolated and create sharper RV variations. We also find that $\lambda_{\mathrm{p}}$ exhbits a weaker correlation with the filling factor of faculae and plages (Pearson correlation coefficient of $\sim$0.6). This suggests that long-term variations in $\lambda_{\mathrm{p}}$ are due more to sunspots than to the magnetic cycle, known to be best described by the filling factor of faculae/plages over the course of our observations \citep{cretignier2024}. The inhibition of convective blueshift in faculae induces an RV contribution that evolves, to the first order, as the square of the disk-integrated photospheric flux \citep{aigrain2012}. Therefore, in absence of spots, the RV activity signals should not vary less smoothly than the FWHM or the S index. On the other hand, spot-induced activity signals, which vary as the derivative of the photospheric flux, are characterised by a significantly higher harmonic complexity. Therefore, despite the fact that faculae dominate the stellar activity RV budget, spots remain the main driver of the smoothness of the signal, probed by $\lambda_{\mathrm{p}}$. Despite the fact that this result has been intuited in the literature \citep[e.g.][]{nicholson2022}, it is the first time that it is observed on data.

In contrast to \vobs, no coherent variations in the statistical properties of \vper\ are observed. The GP amplitude remains roughly constant over time, and the correlation between $\lambda_{\mathrm{p}}$ and the spot filling factor (or whole spot area) is now much weaker (Pearson correlation coefficients of 0.55 and 0.39; see Fig.~\ref{fig:correl_lp}) is observed. This confirms that, unlike \vobs, the quasi-periodic activity signal in \vper\ can be modelled by a single GP on time scales of several years (potentially over the full cycle).


\subsection{Impact of systematics}\label{ssec:systematics}

The analysis presented in this study was conducted on CCFs computed from spectra processed by YV1, as described in Sec.~\ref{ssec:post_processing}. Most of the corrected systematics are expected to impact the shape of the CCFs and, therefore, the activity-filtering framework described in Sec.~\ref{ssec:framework}. To quantify this impact, we repeat the analysis presented in Sec.~\ref{ssec:planet-free-sun} with the non-YARARA-processed spectra. The spectra selected in Sec.~\ref{ssec:observations} are normalised and aligned using the procedure described in Sec.~\ref{ssec:post_processing}, and CCFs are computed using the "Kit-Cat" line list of \citet{cretignier2022}. Following the method described in Sec.~\ref{ssec:framework}, we extract the time series of \vpar\ and \vper, using the first 4 components in the basis projection.

We find that, whereas the RMS of \vobs\ remains pretty similar in the post-processed counterpart, the shift-driven RVs \vper\ (resp. shape-driven RVs \vpar) are now significantly more (resp. less) dispersed, with a dispersion of 1.36\,\ms\ RMS (resp. 1.52\,\ms\ RMS). When working independently on the three seasons defined in Tab.~\ref{tab:rms_rvs}, we only see a noticeable decrease in RV RMS in Season~3. Component $\boldsymbol{U_{1}}$ still correlates well with the FWHM of the CCF, but both time series are dominated by 0.5- and 1-yr periodic modulations, induced by the Earth orbital obliquity and eccentricity \citep{cameron2019}. In contrast, no more correlation between $\boldsymbol{U_{1}}$ and \vobs, V$_{\mathrm{s}}$ and S$_{\mathrm{HK}}$ is observed. Moderate correlations (Pearson correlation coefficients up to $\sim$0.6) are observed between these quantities and higher-order components, which reflects the fact that information has been diffused among the SVD components, due to systematics contributing as much, if not more, to the CCF variations.

When the RV time series are modelled with a quasi-periodic 1D GP, we find that \vpar\ is significantly smoother and slowly-evolving than its YV1-processed counterpart. The fit is now poorer, with a residual RMS twice as large. This is due to the fact that \vpar\ is plagued with systematics \citep[e.g. due the 0.5-yr and 1-yr variations in FWHM][]{cameron2019}, which affects the GP modelling. On the other hand, whereas no significant changes in the statistical properties of \vper\ are observed, the values of $\sigma_{\mathrm{j}}$, the uncorrelated jitter term, have increased to $\sim$1\,\ms, which means that the fit is poorer. We conclude that our framework to correct for shape-driven RV variations works best when systematics have been corrected beforehand, through a dedicated post-processing technique. The latter helps to better isolate stellar activity signals in the spectra, which can therefore be more accurately modelled.


\subsection{Activity filtering and planet recovery}

By jointly modelling the stellar activity RV signal in \vobs\ with different activity proxies, either extracted from the input spectra (e.g. CCF FWHM, S$_{\mathrm{HK}}$) or from our framework (i.e. \vpar), we demonstrated that long-period planets with small RV semi-amplitudes could be reliably detected with a multi-dimensional GP framework (see Tab.~\ref{tab:estimates_kp}). This experiment also confirms that modelling the reduced RVs with a one-dimensional GP will likely yield imprecise parameters for long-period planets, whereas multi-dimensional GPs, by increasing the number of constraints on the fitted parameters, will, in most cases, provide a more robust treatment of long-term variations.

Further investigations are also needed to leverage all the information available in the wavelength space (CCFs\footnote{Note that CCFs may not be the most accurate  way of estimating line profiles \citep[in comparison to e.g. profiles obtained via least-squares deconvolution; see][]{donati1997b,lienhard2022}.}, spectra) to extract planet signatures in an optimal way and increase the constraints on their parameters \citep[following the works of, e.g.,][]{dumusque2018,rajpaul2020,cameron2021,almoulla2022a,deBeurs2022,cretignier2022}. In particular, methods like Doppler Imaging \citep{kochukhov2016,luger2021,asensio2022,klein2022}, which model the full absorption line profile (or spectrum), thereby bypassing the computation of RVs, might become robust complements to usual activity modelling techniques. Finally, long-term variations of the activity properties over the cycle (as evidenced in Fig.~\ref{fig:covariance}) are still hard to reproduce by current state-of-the-art modelling tools like GPs. Defining more physically-driven GP kernels \citep[e.g.][]{hara2023} or allowing some hyper-parameters to vary with time are potential avenues for solving this problem.

Yet, even if the best-case scenarios, the RV residuals of the GP fit exhibit RMSs greater than $\sim$0.5\,\ms, significantly larger than the formal RV uncertainties of $\sim$0.1\ms\ for the daily-binned data. This high dispersion is most likely attributable to the long-term instrument stability and to the solar super-granulation. This phenomenon, whose origin remains unclear \citep[see][for a review]{rincon2018}, is expected to induce RV signals on the same order of magnitude as our residuals \citep{meunier2015,almoulla2023}. Moreover, as we do not expect granulation signals to have obvious effects on the shape of the spectral lines, the activity-filtering method presented in Sec.~\ref{ssec:framework} would not correct for them. Similarly, our GP model is not expected to affect super-granulation signals, as it is designed to account for rotationally-modulated activity signals which evolve on significantly longer time scales than super-granulation. Averaging out the RV effect of super-granulation using dedicated sampling strategies, as generally done for oscillations- and granulation-induced RV signals \citep{dumusque2011}, is not a straightforward option here, as it will require the star to be intensively observed on time scales of days. As indicated in \citet{meunier2019}, better understanding the origin of super-granulation, and developing physically- or data-driven methods to model its RV contributions \citep[see]{osullivan2024}, will be an important step to detect Earth-mass planets as part, for example, of PLATO ground-based monitoring or dedicated RV surveys like the Terra Hunting Experiment \citep{hall2018}. As things currently stand, super-granulation signals have to be treated as white noise, which is emphasized by the fact that, in Sec.~\ref{ssec:results_planet}, the precision of the recovered planet semi-amplitudes improves as the squared root of the number of measurements. If this result is confirmed by dedicated studies, it would mean that RV monitoring missions should aim at maximising the number of measurements on each target (ensuring these are spaced at least 1-2\,d apart so that the super-granulation signal is uncorrelated), even at the cost of monitoring fewer stars overall.


\section*{Acknowledgements}

We warmly thank the reviewer, Dr Drake Deming, for valuable suggestions which helped make this analysis clearer and more robust. We also warmly thank A. A. John for pointing out confusing notations in early versions of the manuscript. The HARPS-N project has been funded by the Prodex Program of the Swiss Space Office (SSO), the Harvard University Origins of Life Initiative (HUOLI), the Scottish Universities Physics Alliance (SUPA), the University of Geneva, the Smithsonian Astrophysical Observatory (SAO), and the Italian National Astrophysical Institute (INAF), the University of St Andrews, Queen’s University Belfast, and the University of Edinburgh. We thank the HARPS-N solar team and TNG staﬀ for processing the solar data and maintaining the solar telescope. B.K., S.A., O.B., H.Y. and N.K.O.S. acknowledge funding from the European Research Council under the European Union’s Horizon 2020 research and innovation programme (grant agreement No 865624, GPRV). M.C. acknowledges the SNSF support under grant P500PT\_211024. F.R. is funded by the University of Exeter's College of Engineering, Maths and Physical Sciences, UK. ACC acknowledges support from STFC consolidated grant number ST/V000861/1, and EPSRC grant number EP/Z000181/1 towards the ERC Synergy project REVEAL.

\section*{Data Availability}

This work makes use of the HARPS-N solar RVs, which will be described and made available in Dumusque et al., submitted. The SDO/HMI images are publicly available at \url{https://sdo.gsfc.nasa.gov/data/} and the USAF/NOAA observations are available at \url{http://solarcyclescience.com/index.html}.



\bibliographystyle{mnras}
\bibliography{biblio}

\begin{thebibliography}{}
\makeatletter
\relax
\def\mn@urlcharsother{\let\do\@makeother \do\$\do\&\do\#\do\^\do\_\do\%\do\~}
\def\mn@doi{\begingroup\mn@urlcharsother \@ifnextchar [ {\mn@doi@} {\mn@doi@[]}}
\def\mn@doi@[#1]#2{\def\@tempa{#1}\ifx\@tempa\@empty \href {http://dx.doi.org/#2} {doi:#2}\else \href {http://dx.doi.org/#2} {#1}\fi \endgroup}
\def\mn@eprint#1#2{\mn@eprint@#1:#2::\@nil}
\def\mn@eprint@arXiv#1{\href {http://arxiv.org/abs/#1} {{\tt arXiv:#1}}}
\def\mn@eprint@dblp#1{\href {http://dblp.uni-trier.de/rec/bibtex/#1.xml} {dblp:#1}}
\def\mn@eprint@#1:#2:#3:#4\@nil{\def\@tempa {#1}\def\@tempb {#2}\def\@tempc {#3}\ifx \@tempc \@empty \let \@tempc \@tempb \let \@tempb \@tempa \fi \ifx \@tempb \@empty \def\@tempb {arXiv}\fi \@ifundefined {mn@eprint@\@tempb}{\@tempb:\@tempc}{\expandafter \expandafter \csname mn@eprint@\@tempb\endcsname \expandafter{\@tempc}}}

\bibitem[\protect\citeauthoryear{{Aigrain} \& {Foreman-Mackey}}{{Aigrain} \& {Foreman-Mackey}}{2022}]{aigrain2022}
{Aigrain} S.,  {Foreman-Mackey} D.,  2022, arXiv e-prints, \href {https://ui.adsabs.harvard.edu/abs/2022arXiv220908940A} {p. arXiv:2209.08940}

\bibitem[\protect\citeauthoryear{{Aigrain}, {Pont}  \& {Zucker}}{{Aigrain} et~al.}{2012}]{aigrain2012}
{Aigrain} S.,  {Pont} F.,   {Zucker} S.,  2012, \mn@doi [\mnras] {10.1111/j.1365-2966.2011.19960.x}, \href {https://ui.adsabs.harvard.edu/abs/2012MNRAS.419.3147A} {419, 3147}

\bibitem[\protect\citeauthoryear{{Al Moulla}, {Dumusque}, {Cretignier}, {Zhao}  \& {Valenti}}{{Al Moulla} et~al.}{2022}]{almoulla2022a}
{Al Moulla} K.,  {Dumusque} X.,  {Cretignier} M.,  {Zhao} Y.,   {Valenti} J.~A.,  2022, \mn@doi [\aap] {10.1051/0004-6361/202243276}, \href {https://ui.adsabs.harvard.edu/abs/2022A&A...664A..34A} {664, A34}

\bibitem[\protect\citeauthoryear{{Al Moulla}, {Dumusque}, {Figueira}, {Lo Curto}, {Santos}  \& {Wildi}}{{Al Moulla} et~al.}{2023}]{almoulla2023}
{Al Moulla} K.,  {Dumusque} X.,  {Figueira} P.,  {Lo Curto} G.,  {Santos} N.~C.,   {Wildi} F.,  2023, \mn@doi [\aap] {10.1051/0004-6361/202244663}, \href {https://ui.adsabs.harvard.edu/abs/2023A&A...669A..39A} {669, A39}

\bibitem[\protect\citeauthoryear{{Ambikasaran}, {Foreman-Mackey}, {Greengard}, {Hogg}  \& {O'Neil}}{{Ambikasaran} et~al.}{2015}]{ambikasaran2015}
{Ambikasaran} S.,  {Foreman-Mackey} D.,  {Greengard} L.,  {Hogg} D.~W.,   {O'Neil} M.,  2015, \mn@doi [IEEE Transactions on Pattern Analysis and Machine Intelligence] {10.1109/TPAMI.2015.2448083}, \href {https://ui.adsabs.harvard.edu/abs/2015ITPAM..38..252A} {38, 252}

\bibitem[\protect\citeauthoryear{{Angus}, {Morton}, {Aigrain}, {Foreman-Mackey}  \& {Rajpaul}}{{Angus} et~al.}{2018}]{angus2018}
{Angus} R.,  {Morton} T.,  {Aigrain} S.,  {Foreman-Mackey} D.,   {Rajpaul} V.,  2018, \mn@doi [\mnras] {10.1093/mnras/stx2109}, \href {https://ui.adsabs.harvard.edu/abs/2018MNRAS.474.2094A} {474, 2094}

\bibitem[\protect\citeauthoryear{{Asensio Ramos}, {D{\'\i}az Baso}  \& {Kochukhov}}{{Asensio Ramos} et~al.}{2022}]{asensio2022}
{Asensio Ramos} A.,  {D{\'\i}az Baso} C.~J.,   {Kochukhov} O.,  2022, \mn@doi [\aap] {10.1051/0004-6361/202142027}, \href {https://ui.adsabs.harvard.edu/abs/2022A&A...658A.162A} {658, A162}

\bibitem[\protect\citeauthoryear{{Astropy Collaboration} et~al.,}{{Astropy Collaboration} et~al.}{2013}]{astropy2013}
{Astropy Collaboration} et~al., 2013, \mn@doi [\aap] {10.1051/0004-6361/201322068}, \href {https://ui.adsabs.harvard.edu/abs/2013A&A...558A..33A} {558, A33}

\bibitem[\protect\citeauthoryear{{Astropy Collaboration} et~al.,}{{Astropy Collaboration} et~al.}{2018}]{astropy2018}
{Astropy Collaboration} et~al., 2018, \mn@doi [\aj] {10.3847/1538-3881/aabc4f}, \href {https://ui.adsabs.harvard.edu/abs/2018AJ....156..123A} {156, 123}

\bibitem[\protect\citeauthoryear{{Astropy Collaboration} et~al.,}{{Astropy Collaboration} et~al.}{2022}]{astropy2022}
{Astropy Collaboration} et~al., 2022, \mn@doi [\apj] {10.3847/1538-4357/ac7c74}, \href {https://ui.adsabs.harvard.edu/abs/2022ApJ...935..167A} {935, 167}

\bibitem[\protect\citeauthoryear{{Barrag{\'a}n}, {Gandolfi}  \& {Antoniciello}}{{Barrag{\'a}n} et~al.}{2019}]{barragan2019}
{Barrag{\'a}n} O.,  {Gandolfi} D.,   {Antoniciello} G.,  2019, \mn@doi [\mnras] {10.1093/mnras/sty2472}, \href {https://ui.adsabs.harvard.edu/abs/2019MNRAS.482.1017B} {482, 1017}

\bibitem[\protect\citeauthoryear{{Barrag{\'a}n}, {Aigrain}, {Rajpaul}  \& {Zicher}}{{Barrag{\'a}n} et~al.}{2022a}]{barragan2022A}
{Barrag{\'a}n} O.,  {Aigrain} S.,  {Rajpaul} V.~M.,   {Zicher} N.,  2022a, \mn@doi [\mnras] {10.1093/mnras/stab2889}, \href {https://ui.adsabs.harvard.edu/abs/2022MNRAS.509..866B} {509, 866}

\bibitem[\protect\citeauthoryear{{Barrag{\'a}n} et~al.,}{{Barrag{\'a}n} et~al.}{2022b}]{barragan2022}
{Barrag{\'a}n} O.,  et~al., 2022b, \mn@doi [\mnras] {10.1093/mnras/stac638}, \href {https://ui.adsabs.harvard.edu/abs/2022MNRAS.514.1606B} {514, 1606}

\bibitem[\protect\citeauthoryear{{Batalha}, {Lewis}, {Fortney}, {Batalha}, {Kempton}, {Lewis}  \& {Line}}{{Batalha} et~al.}{2019}]{batalha2019}
{Batalha} N.~E.,  {Lewis} T.,  {Fortney} J.~J.,  {Batalha} N.~M.,  {Kempton} E.,  {Lewis} N.~K.,   {Line} M.~R.,  2019, \mn@doi [\apjl] {10.3847/2041-8213/ab4909}, \href {https://ui.adsabs.harvard.edu/abs/2019ApJ...885L..25B} {885, L25}

\bibitem[\protect\citeauthoryear{{Beck}}{{Beck}}{2000}]{beck2000}
{Beck} J.~G.,  2000, \mn@doi [\solphys] {10.1023/A:1005226402796}, \href {https://ui.adsabs.harvard.edu/abs/2000SoPh..191...47B} {191, 47}

\bibitem[\protect\citeauthoryear{{Berdyugina}}{{Berdyugina}}{2005}]{berdyugina2005}
{Berdyugina} S.~V.,  2005, \mn@doi [Living Reviews in Solar Physics] {10.12942/lrsp-2005-8}, \href {https://ui.adsabs.harvard.edu/abs/2005LRSP....2....8B} {2, 8}

\bibitem[\protect\citeauthoryear{{Boisse}, {Bonfils}  \& {Santos}}{{Boisse} et~al.}{2012}]{boisse2012}
{Boisse} I.,  {Bonfils} X.,   {Santos} N.~C.,  2012, \mn@doi [\aap] {10.1051/0004-6361/201219115}, \href {https://ui.adsabs.harvard.edu/abs/2012A&A...545A.109B} {545, A109}

\bibitem[\protect\citeauthoryear{{Bouchy}, {Pepe}  \& {Queloz}}{{Bouchy} et~al.}{2001}]{bouchy2001}
{Bouchy} F.,  {Pepe} F.,   {Queloz} D.,  2001, \mn@doi [\aap] {10.1051/0004-6361:20010730}, \href {https://ui.adsabs.harvard.edu/abs/2001A&A...374..733B} {374, 733}

\bibitem[\protect\citeauthoryear{{Bourrier} et~al.,}{{Bourrier} et~al.}{2021}]{Bourrier2021}
{Bourrier} V.,  et~al., 2021, \mn@doi [\aap] {10.1051/0004-6361/202141527}, \href {https://ui.adsabs.harvard.edu/abs/2021A&A...654A.152B} {654, A152}

\bibitem[\protect\citeauthoryear{{Cegla} et~al.,}{{Cegla} et~al.}{2018}]{cegla2018}
{Cegla} H.~M.,  et~al., 2018, \mn@doi [\apj] {10.3847/1538-4357/aaddfc}, \href {https://ui.adsabs.harvard.edu/abs/2018ApJ...866...55C} {866, 55}

\bibitem[\protect\citeauthoryear{{Chaplin}, {Cegla}, {Watson}, {Davies}  \& {Ball}}{{Chaplin} et~al.}{2019}]{chaplin2019}
{Chaplin} W.~J.,  {Cegla} H.~M.,  {Watson} C.~A.,  {Davies} G.~R.,   {Ball} W.~H.,  2019, \mn@doi [\aj] {10.3847/1538-3881/ab0c01}, \href {https://ui.adsabs.harvard.edu/abs/2019AJ....157..163C} {157, 163}

\bibitem[\protect\citeauthoryear{{Collier Cameron} et~al.,}{{Collier Cameron} et~al.}{2019}]{cameron2019}
{Collier Cameron} A.,  et~al., 2019, \mn@doi [\mnras] {10.1093/mnras/stz1215}, \href {https://ui.adsabs.harvard.edu/abs/2019MNRAS.487.1082C} {487, 1082}

\bibitem[\protect\citeauthoryear{{Collier Cameron} et~al.,}{{Collier Cameron} et~al.}{2021}]{cameron2021}
{Collier Cameron} A.,  et~al., 2021, \mn@doi [\mnras] {10.1093/mnras/stab1323}, \href {https://ui.adsabs.harvard.edu/abs/2021MNRAS.505.1699C} {505, 1699}

\bibitem[\protect\citeauthoryear{{Cosentino} et~al.,}{{Cosentino} et~al.}{2012}]{cosentino2012}
{Cosentino} R.,  et~al., 2012, in {McLean} I.~S.,  {Ramsay} S.~K.,   {Takami} H.,  eds,  Society of Photo-Optical Instrumentation Engineers (SPIE) Conference Series Vol. 8446, Ground-based and Airborne Instrumentation for Astronomy IV. p. 84461V, \mn@doi{10.1117/12.925738}

\bibitem[\protect\citeauthoryear{{Crass} et~al.,}{{Crass} et~al.}{2021}]{crass2021}
{Crass} J.,  et~al., 2021, arXiv e-prints, \href {https://ui.adsabs.harvard.edu/abs/2021arXiv210714291C} {p. arXiv:2107.14291}

\bibitem[\protect\citeauthoryear{{Cretignier}}{{Cretignier}}{2022}]{Cretignier2022t}
{Cretignier} M.,  2022, PhD thesis, University of Geneva, Switzerland

\bibitem[\protect\citeauthoryear{{Cretignier}, {Dumusque}, {Allart}, {Pepe}  \& {Lovis}}{{Cretignier} et~al.}{2020a}]{cretignier2020}
{Cretignier} M.,  {Dumusque} X.,  {Allart} R.,  {Pepe} F.,   {Lovis} C.,  2020a, \mn@doi [\aap] {10.1051/0004-6361/201936548}, \href {https://ui.adsabs.harvard.edu/abs/2020A&A...633A..76C} {633, A76}

\bibitem[\protect\citeauthoryear{{Cretignier}, {Francfort}, {Dumusque}, {Allart}  \& {Pepe}}{{Cretignier} et~al.}{2020b}]{Cretignier2020b}
{Cretignier} M.,  {Francfort} J.,  {Dumusque} X.,  {Allart} R.,   {Pepe} F.,  2020b, \mn@doi [\aap] {10.1051/0004-6361/202037722}, \href {https://ui.adsabs.harvard.edu/abs/2020A&A...640A..42C} {640, A42}

\bibitem[\protect\citeauthoryear{{Cretignier}, {Dumusque}, {Hara}  \& {Pepe}}{{Cretignier} et~al.}{2021}]{Cretignier2021}
{Cretignier} M.,  {Dumusque} X.,  {Hara} N.~C.,   {Pepe} F.,  2021, \mn@doi [\aap] {10.1051/0004-6361/202140986}, \href {https://ui.adsabs.harvard.edu/abs/2021A&A...653A..43C} {653, A43}

\bibitem[\protect\citeauthoryear{{Cretignier}, {Dumusque}  \& {Pepe}}{{Cretignier} et~al.}{2022}]{cretignier2022}
{Cretignier} M.,  {Dumusque} X.,   {Pepe} F.,  2022, \mn@doi [\aap] {10.1051/0004-6361/202142435}, \href {https://ui.adsabs.harvard.edu/abs/2022A&A...659A..68C} {659, A68}

\bibitem[\protect\citeauthoryear{{Cretignier}, {Dumusque}, {Aigrain}  \& {Pepe}}{{Cretignier} et~al.}{2023}]{cretignier2023}
{Cretignier} M.,  {Dumusque} X.,  {Aigrain} S.,   {Pepe} F.,  2023, \mn@doi [\aap] {10.1051/0004-6361/202347232}, \href {https://ui.adsabs.harvard.edu/abs/2023A&A...678A...2C} {678, A2}

\bibitem[\protect\citeauthoryear{{Cretignier}, {Pietrow}  \& {Aigrain}}{{Cretignier} et~al.}{2024}]{cretignier2024}
{Cretignier} M.,  {Pietrow} A.~G.~M.,   {Aigrain} S.,  2024, \mn@doi [\mnras] {10.1093/mnras/stad3292}, \href {https://ui.adsabs.harvard.edu/abs/2024MNRAS.527.2940C} {527, 2940}

\bibitem[\protect\citeauthoryear{{Dalal}, {Rescigno}  \& {Cretignier}}{{Dalal} et~al.}{prep}]{Dalal2024}
{Dalal} S.,  {Rescigno} F.,   {Cretignier} M.,  in prep., \mnras

\bibitem[\protect\citeauthoryear{{Delisle}, {Unger}, {Hara}  \& {S{\'e}gransan}}{{Delisle} et~al.}{2022}]{delisle2022}
{Delisle} J.~B.,  {Unger} N.,  {Hara} N.~C.,   {S{\'e}gransan} D.,  2022, \mn@doi [\aap] {10.1051/0004-6361/202141949}, \href {https://ui.adsabs.harvard.edu/abs/2022A&A...659A.182D} {659, A182}

\bibitem[\protect\citeauthoryear{{Desort}, {Lagrange}, {Galland}, {Udry}  \& {Mayor}}{{Desort} et~al.}{2007}]{desort2007}
{Desort} M.,  {Lagrange} A.~M.,  {Galland} F.,  {Udry} S.,   {Mayor} M.,  2007, \mn@doi [\aap] {10.1051/0004-6361:20078144}, \href {https://ui.adsabs.harvard.edu/abs/2007A&A...473..983D} {473, 983}

\bibitem[\protect\citeauthoryear{{Donati} \& {Brown}}{{Donati} \& {Brown}}{1997}]{donati1997b}
{Donati} J.~F.,  {Brown} S.~F.,  1997, \aap, \href {https://ui.adsabs.harvard.edu/abs/1997A&A...326.1135D} {326, 1135}

\bibitem[\protect\citeauthoryear{{Dumusque}}{{Dumusque}}{2018}]{dumusque2018}
{Dumusque} X.,  2018, \mn@doi [\aap] {10.1051/0004-6361/201833795}, \href {https://ui.adsabs.harvard.edu/abs/2018A&A...620A..47D} {620, A47}

\bibitem[\protect\citeauthoryear{{Dumusque}, {Udry}, {Lovis}, {Santos}  \& {Monteiro}}{{Dumusque} et~al.}{2011}]{dumusque2011}
{Dumusque} X.,  {Udry} S.,  {Lovis} C.,  {Santos} N.~C.,   {Monteiro} M.~J.~P.~F.~G.,  2011, \mn@doi [\aap] {10.1051/0004-6361/201014097}, \href {https://ui.adsabs.harvard.edu/abs/2011A&A...525A.140D} {525, A140}

\bibitem[\protect\citeauthoryear{{Dumusque}, {Boisse}  \& {Santos}}{{Dumusque} et~al.}{2014}]{dumusque2014}
{Dumusque} X.,  {Boisse} I.,   {Santos} N.~C.,  2014, \mn@doi [\apj] {10.1088/0004-637X/796/2/132}, \href {https://ui.adsabs.harvard.edu/abs/2014ApJ...796..132D} {796, 132}

\bibitem[\protect\citeauthoryear{{Dumusque} et~al.,}{{Dumusque} et~al.}{2015}]{dumusque2015}
{Dumusque} X.,  et~al., 2015, \mn@doi [\apjl] {10.1088/2041-8205/814/2/L21}, \href {https://ui.adsabs.harvard.edu/abs/2015ApJ...814L..21D} {814, L21}

\bibitem[\protect\citeauthoryear{{Dumusque} et~al.,}{{Dumusque} et~al.}{2021}]{dumusque2021}
{Dumusque} X.,  et~al., 2021, \mn@doi [\aap] {10.1051/0004-6361/202039350}, \href {https://ui.adsabs.harvard.edu/abs/2021A&A...648A.103D} {648, A103}

\bibitem[\protect\citeauthoryear{{Ervin} et~al.,}{{Ervin} et~al.}{2022}]{ervin2022}
{Ervin} T.,  et~al., 2022, \mn@doi [\aj] {10.3847/1538-3881/ac67e6}, \href {https://ui.adsabs.harvard.edu/abs/2022AJ....163..272E} {163, 272}

\bibitem[\protect\citeauthoryear{{Faria} et~al.,}{{Faria} et~al.}{2022}]{faria2022}
{Faria} J.~P.,  et~al., 2022, \mn@doi [\aap] {10.1051/0004-6361/202142337}, \href {https://ui.adsabs.harvard.edu/abs/2022A&A...658A.115F} {658, A115}

\bibitem[\protect\citeauthoryear{{Fischer} et~al.,}{{Fischer} et~al.}{2016}]{fischer2016}
{Fischer} D.~A.,  et~al., 2016, \mn@doi [\pasp] {10.1088/1538-3873/128/964/066001}, \href {https://ui.adsabs.harvard.edu/abs/2016PASP..128f6001F} {128, 066001}

\bibitem[\protect\citeauthoryear{{Foreman-Mackey}, {Hogg}, {Lang}  \& {Goodman}}{{Foreman-Mackey} et~al.}{2013}]{foreman2013}
{Foreman-Mackey} D.,  {Hogg} D.~W.,  {Lang} D.,   {Goodman} J.,  2013, \mn@doi [\pasp] {10.1086/670067}, \href {https://ui.adsabs.harvard.edu/abs/2013PASP..125..306F} {125, 306}

\bibitem[\protect\citeauthoryear{{Foukal}}{{Foukal}}{1998}]{foukal1998}
{Foukal} P.,  1998, \mn@doi [\apj] {10.1086/305764}, \href {https://ui.adsabs.harvard.edu/abs/1998ApJ...500..958F} {500, 958}

\bibitem[\protect\citeauthoryear{{Gelman} \& {Rubin}}{{Gelman} \& {Rubin}}{1992}]{gelman1992}
{Gelman} A.,  {Rubin} D.~B.,  1992, \mn@doi [Statistical Science] {10.1214/ss/1177011136}, \href {https://ui.adsabs.harvard.edu/abs/1992StaSc...7..457G} {7, 457}

\bibitem[\protect\citeauthoryear{Gelman, Carlin, Stern  \& Rubin}{Gelman et~al.}{2004}]{gelman2004}
Gelman A.,  Carlin J.~B.,  Stern H.~S.,   Rubin D.~B.,  2004, Bayesian Data Analysis, 2nd ed. edn.
Chapman and Hall/CRC

\bibitem[\protect\citeauthoryear{{Gibson}, {Howard}, {Marcy}, {Edelstein}, {Wishnow}  \& {Poppett}}{{Gibson} et~al.}{2016}]{gibson2016}
{Gibson} S.~R.,  {Howard} A.~W.,  {Marcy} G.~W.,  {Edelstein} J.,  {Wishnow} E.~H.,   {Poppett} C.~L.,  2016, in {Evans} C.~J.,  {Simard} L.,   {Takami} H.,  eds,  Society of Photo-Optical Instrumentation Engineers (SPIE) Conference Series Vol. 9908, Ground-based and Airborne Instrumentation for Astronomy VI. p. 990870, \mn@doi{10.1117/12.2233334}

\bibitem[\protect\citeauthoryear{{Gomes da Silva}, {Santos}, {Bonfils}, {Delfosse}, {Forveille}, {Udry}, {Dumusque}  \& {Lovis}}{{Gomes da Silva} et~al.}{2012}]{gomes2012}
{Gomes da Silva} J.,  {Santos} N.~C.,  {Bonfils} X.,  {Delfosse} X.,  {Forveille} T.,  {Udry} S.,  {Dumusque} X.,   {Lovis} C.,  2012, \mn@doi [\aap] {10.1051/0004-6361/201118598}, \href {https://ui.adsabs.harvard.edu/abs/2012A&A...541A...9G} {541, A9}

\bibitem[\protect\citeauthoryear{{Hall}, {Thompson}, {Handley}  \& {Queloz}}{{Hall} et~al.}{2018}]{hall2018}
{Hall} R.~D.,  {Thompson} S.~J.,  {Handley} W.,   {Queloz} D.,  2018, \mn@doi [\mnras] {10.1093/mnras/sty1464}, \href {https://ui.adsabs.harvard.edu/abs/2018MNRAS.479.2968H} {479, 2968}

\bibitem[\protect\citeauthoryear{{Hara} \& {Delisle}}{{Hara} \& {Delisle}}{2023}]{hara2023}
{Hara} N.~C.,  {Delisle} J.-B.,  2023, \mn@doi [arXiv e-prints] {10.48550/arXiv.2304.08489}, \href {https://ui.adsabs.harvard.edu/abs/2023arXiv230408489H} {p. arXiv:2304.08489}

\bibitem[\protect\citeauthoryear{{Hathaway}}{{Hathaway}}{2010}]{hathaway2010}
{Hathaway} D.~H.,  2010, \mn@doi [Living Reviews in Solar Physics] {10.12942/lrsp-2010-1}, \href {https://ui.adsabs.harvard.edu/abs/2010LRSP....7....1H} {7, 1}

\bibitem[\protect\citeauthoryear{{Hatzes}}{{Hatzes}}{1996}]{hatzes1996}
{Hatzes} A.~P.,  1996, \mn@doi [\pasp] {10.1086/133805}, \href {https://ui.adsabs.harvard.edu/abs/1996PASP..108..839H} {108, 839}

\bibitem[\protect\citeauthoryear{{Haywood} et~al.,}{{Haywood} et~al.}{2014}]{haywood2014}
{Haywood} R.~D.,  et~al., 2014, \mn@doi [\mnras] {10.1093/mnras/stu1320}, \href {https://ui.adsabs.harvard.edu/abs/2014MNRAS.443.2517H} {443, 2517}

\bibitem[\protect\citeauthoryear{{Haywood} et~al.,}{{Haywood} et~al.}{2022}]{haywood2022}
{Haywood} R.~D.,  et~al., 2022, \mn@doi [\apj] {10.3847/1538-4357/ac7c12}, \href {https://ui.adsabs.harvard.edu/abs/2022ApJ...935....6H} {935, 6}

\bibitem[\protect\citeauthoryear{{Hogg}, {Bovy}  \& {Lang}}{{Hogg} et~al.}{2010}]{hogg2010}
{Hogg} D.~W.,  {Bovy} J.,   {Lang} D.,  2010, arXiv e-prints, \href {https://ui.adsabs.harvard.edu/abs/2010arXiv1008.4686H} {p. arXiv:1008.4686}

\bibitem[\protect\citeauthoryear{{Howard} \& {Harvey}}{{Howard} \& {Harvey}}{1970}]{howard1970}
{Howard} R.,  {Harvey} J.,  1970, \mn@doi [\solphys] {10.1007/BF02276562}, \href {https://ui.adsabs.harvard.edu/abs/1970SoPh...12...23H} {12, 23}

\bibitem[\protect\citeauthoryear{{John}, {Collier Cameron}  \& {Wilson}}{{John} et~al.}{2022}]{john2022}
{John} A.~A.,  {Collier Cameron} A.,   {Wilson} T.~G.,  2022, \mn@doi [\mnras] {10.1093/mnras/stac1814}, \href {https://ui.adsabs.harvard.edu/abs/2022MNRAS.515.3975J} {515, 3975}

\bibitem[\protect\citeauthoryear{{John} et~al.,}{{John} et~al.}{2023}]{john2023}
{John} A.~A.,  et~al., 2023, \mn@doi [\mnras] {10.1093/mnras/stad2381}, \href {https://ui.adsabs.harvard.edu/abs/2023MNRAS.525.1687J} {525, 1687}

\bibitem[\protect\citeauthoryear{{Jones}, {Stenning}, {Ford}, {Wolpert}, {Loredo}, {Gilbertson}  \& {Dumusque}}{{Jones} et~al.}{2017}]{jones2017}
{Jones} D.~E.,  {Stenning} D.~C.,  {Ford} E.~B.,  {Wolpert} R.~L.,  {Loredo} T.~J.,  {Gilbertson} C.,   {Dumusque} X.,  2017, arXiv e-prints, \href {https://ui.adsabs.harvard.edu/abs/2017arXiv171101318J} {p. arXiv:1711.01318}

\bibitem[\protect\citeauthoryear{{Jurgenson}, {Fischer}, {McCracken}, {Sawyer}, {Szymkowiak}, {Davis}, {Muller}  \& {Santoro}}{{Jurgenson} et~al.}{2016}]{jurgenson2016}
{Jurgenson} C.,  {Fischer} D.,  {McCracken} T.,  {Sawyer} D.,  {Szymkowiak} A.,  {Davis} A.,  {Muller} G.,   {Santoro} F.,  2016, in {Evans} C.~J.,  {Simard} L.,   {Takami} H.,  eds,  Society of Photo-Optical Instrumentation Engineers (SPIE) Conference Series Vol. 9908, Ground-based and Airborne Instrumentation for Astronomy VI. p. 99086T (\mn@eprint {arXiv} {1606.04413}), \mn@doi{10.1117/12.2233002}

\bibitem[\protect\citeauthoryear{{Klein} \& {Donati}}{{Klein} \& {Donati}}{2019}]{klein2019}
{Klein} B.,  {Donati} J.~F.,  2019, \mn@doi [\mnras] {10.1093/mnras/stz1953}, \href {https://ui.adsabs.harvard.edu/abs/2019MNRAS.488.5114K} {488, 5114}

\bibitem[\protect\citeauthoryear{{Klein} et~al.,}{{Klein} et~al.}{2022}]{klein2022}
{Klein} B.,  et~al., 2022, \mn@doi [\mnras] {10.1093/mnras/stac761}, \href {https://ui.adsabs.harvard.edu/abs/2022MNRAS.512.5067K} {512, 5067}

\bibitem[\protect\citeauthoryear{{Klein} et~al.,}{{Klein} et~al.}{2024}]{klein2024}
{Klein} B.,  et~al., 2024, \mn@doi [\mnras] {10.1093/mnras/stad2607}, \href {https://ui.adsabs.harvard.edu/abs/2024MNRAS.527..544K} {527, 544}

\bibitem[\protect\citeauthoryear{{Kochukhov}}{{Kochukhov}}{2016}]{kochukhov2016}
{Kochukhov} O.,  2016, in {Rozelot} J.-P.,  {Neiner} C.,  eds, , Vol.~914, Lecture Notes in Physics, Berlin Springer Verlag.
p.~177, \mn@doi{10.1007/978-3-319-24151-7\_9}

\bibitem[\protect\citeauthoryear{{Lakeland} et~al.,}{{Lakeland} et~al.}{2024}]{lakeland2024}
{Lakeland} B.~S.,  et~al., 2024, \mn@doi [\mnras] {10.1093/mnras/stad3723}, \href {https://ui.adsabs.harvard.edu/abs/2024MNRAS.527.7681L} {527, 7681}

\bibitem[\protect\citeauthoryear{{Leighton}}{{Leighton}}{1964}]{leighton1964}
{Leighton} R.~B.,  1964, \mn@doi [\apj] {10.1086/148058}, \href {https://ui.adsabs.harvard.edu/abs/1964ApJ...140.1547L} {140, 1547}

\bibitem[\protect\citeauthoryear{{Lienhard}, {Mortier}, {Buchhave}, {Collier Cameron}, {L{\'o}pez-Morales}, {Sozzetti}, {Watson}  \& {Cosentino}}{{Lienhard} et~al.}{2022}]{lienhard2022}
{Lienhard} F.,  {Mortier} A.,  {Buchhave} L.,  {Collier Cameron} A.,  {L{\'o}pez-Morales} M.,  {Sozzetti} A.,  {Watson} C.~A.,   {Cosentino} R.,  2022, \mn@doi [\mnras] {10.1093/mnras/stac1098}, \href {https://ui.adsabs.harvard.edu/abs/2022MNRAS.513.5328L} {513, 5328}

\bibitem[\protect\citeauthoryear{{Lienhard}, {Mortier}, {Cegla}, {Cameron}, {Klein}  \& {Watson}}{{Lienhard} et~al.}{2023}]{lienhard2023}
{Lienhard} F.,  {Mortier} A.,  {Cegla} H.~M.,  {Cameron} A.~C.,  {Klein} B.,   {Watson} C.~A.,  2023, \mn@doi [\mnras] {10.1093/mnras/stad1343}, \href {https://ui.adsabs.harvard.edu/abs/2023MNRAS.522.5862L} {522, 5862}

\bibitem[\protect\citeauthoryear{{Lovis} et~al.,}{{Lovis} et~al.}{2011}]{lovis2011}
{Lovis} C.,  et~al., 2011, arXiv e-prints, \href {https://ui.adsabs.harvard.edu/abs/2011arXiv1107.5325L} {p. arXiv:1107.5325}

\bibitem[\protect\citeauthoryear{{Luger}, {Bedell}, {Foreman-Mackey}, {Crossfield}, {Zhao}  \& {Hogg}}{{Luger} et~al.}{2021}]{luger2021}
{Luger} R.,  {Bedell} M.,  {Foreman-Mackey} D.,  {Crossfield} I. J.~M.,  {Zhao} L.~L.,   {Hogg} D.~W.,  2021, \mn@doi [arXiv e-prints] {10.48550/arXiv.2110.06271}, \href {https://ui.adsabs.harvard.edu/abs/2021arXiv211006271L} {p. arXiv:2110.06271}

\bibitem[\protect\citeauthoryear{{Malavolta}, {Lovis}, {Pepe}, {Sneden}  \& {Udry}}{{Malavolta} et~al.}{2017}]{malavolta2017}
{Malavolta} L.,  {Lovis} C.,  {Pepe} F.,  {Sneden} C.,   {Udry} S.,  2017, \mn@doi [\mnras] {10.1093/mnras/stx1100}, \href {https://ui.adsabs.harvard.edu/abs/2017MNRAS.469.3965M} {469, 3965}

\bibitem[\protect\citeauthoryear{{Mayor} \& {Queloz}}{{Mayor} \& {Queloz}}{1995}]{mayor1995}
{Mayor} M.,  {Queloz} D.,  1995, \mn@doi [\nat] {10.1038/378355a0}, \href {https://ui.adsabs.harvard.edu/abs/1995Natur.378..355M} {378, 355}

\bibitem[\protect\citeauthoryear{{Meunier}}{{Meunier}}{2021}]{meunier2021}
{Meunier} N.,  2021, \mn@doi [arXiv e-prints] {10.48550/arXiv.2104.06072}, \href {https://ui.adsabs.harvard.edu/abs/2021arXiv210406072M} {p. arXiv:2104.06072}

\bibitem[\protect\citeauthoryear{{Meunier} \& {Lagrange}}{{Meunier} \& {Lagrange}}{2019}]{meunier2019}
{Meunier} N.,  {Lagrange} A.~M.,  2019, \mn@doi [\aap] {10.1051/0004-6361/201935099}, \href {https://ui.adsabs.harvard.edu/abs/2019A&A...625L...6M} {625, L6}

\bibitem[\protect\citeauthoryear{{Meunier} \& {Lagrange}}{{Meunier} \& {Lagrange}}{2020}]{meunier2020}
{Meunier} N.,  {Lagrange} A.~M.,  2020, \mn@doi [\aap] {10.1051/0004-6361/201937354}, \href {https://ui.adsabs.harvard.edu/abs/2020A&A...638A..54M} {638, A54}

\bibitem[\protect\citeauthoryear{{Meunier}, {Desort}  \& {Lagrange}}{{Meunier} et~al.}{2010}]{meunier2010}
{Meunier} N.,  {Desort} M.,   {Lagrange} A.~M.,  2010, \mn@doi [\aap] {10.1051/0004-6361/200913551}, \href {https://ui.adsabs.harvard.edu/abs/2010A&A...512A..39M} {512, A39}

\bibitem[\protect\citeauthoryear{{Meunier}, {Lagrange}, {Borgniet}  \& {Rieutord}}{{Meunier} et~al.}{2015}]{meunier2015}
{Meunier} N.,  {Lagrange} A.~M.,  {Borgniet} S.,   {Rieutord} M.,  2015, \mn@doi [\aap] {10.1051/0004-6361/201525721}, \href {https://ui.adsabs.harvard.edu/abs/2015A&A...583A.118M} {583, A118}

\bibitem[\protect\citeauthoryear{{Meunier}, {Pous}, {Sulis}, {Mary}  \& {Lagrange}}{{Meunier} et~al.}{2023}]{meunier2023}
{Meunier} N.,  {Pous} R.,  {Sulis} S.,  {Mary} D.,   {Lagrange} A.~M.,  2023, \mn@doi [\aap] {10.1051/0004-6361/202346218}, \href {https://ui.adsabs.harvard.edu/abs/2023A&A...676A..82M} {676, A82}

\bibitem[\protect\citeauthoryear{{Milbourne} et~al.,}{{Milbourne} et~al.}{2019}]{milbourne2019}
{Milbourne} T.~W.,  et~al., 2019, \mn@doi [\apj] {10.3847/1538-4357/ab064a}, \href {https://ui.adsabs.harvard.edu/abs/2019ApJ...874..107M} {874, 107}

\bibitem[\protect\citeauthoryear{{Mordasini}, {Alibert}, {Georgy}, {Dittkrist}, {Klahr}  \& {Henning}}{{Mordasini} et~al.}{2012}]{mordasini2012}
{Mordasini} C.,  {Alibert} Y.,  {Georgy} C.,  {Dittkrist} K.~M.,  {Klahr} H.,   {Henning} T.,  2012, \mn@doi [\aap] {10.1051/0004-6361/201118464}, \href {https://ui.adsabs.harvard.edu/abs/2012A&A...547A.112M} {547, A112}

\bibitem[\protect\citeauthoryear{{Nicholson} \& {Aigrain}}{{Nicholson} \& {Aigrain}}{2022}]{nicholson2022}
{Nicholson} B.~A.,  {Aigrain} S.,  2022, \mn@doi [\mnras] {10.1093/mnras/stac2097}, \href {https://ui.adsabs.harvard.edu/abs/2022MNRAS.515.5251N} {515, 5251}

\bibitem[\protect\citeauthoryear{{Noyes}, {Hartmann}, {Baliunas}, {Duncan}  \& {Vaughan}}{{Noyes} et~al.}{1984}]{noyes1984}
{Noyes} R.~W.,  {Hartmann} L.~W.,  {Baliunas} S.~L.,  {Duncan} D.~K.,   {Vaughan} A.~H.,  1984, \mn@doi [\apj] {10.1086/161945}, \href {https://ui.adsabs.harvard.edu/abs/1984ApJ...279..763N} {279, 763}

\bibitem[\protect\citeauthoryear{{O'Sullivan} \& {Aigrain}}{{O'Sullivan} \& {Aigrain}}{2024}]{osullivan2024}
{O'Sullivan} N.~K.,  {Aigrain} S.,  2024, \mn@doi [arXiv e-prints] {10.48550/arXiv.2404.11662}, \href {https://ui.adsabs.harvard.edu/abs/2024arXiv240411662O} {p. arXiv:2404.11662}

\bibitem[\protect\citeauthoryear{{Pepe} et~al.,}{{Pepe} et~al.}{2021}]{pepe2021}
{Pepe} F.,  et~al., 2021, \mn@doi [\aap] {10.1051/0004-6361/202038306}, \href {https://ui.adsabs.harvard.edu/abs/2021A&A...645A..96P} {645, A96}

\bibitem[\protect\citeauthoryear{{Pesnell}, {Thompson}  \& {Chamberlin}}{{Pesnell} et~al.}{2012}]{pesnell2012}
{Pesnell} W.~D.,  {Thompson} B.~J.,   {Chamberlin} P.~C.,  2012, \mn@doi [\solphys] {10.1007/s11207-011-9841-3}, \href {https://ui.adsabs.harvard.edu/abs/2012SoPh..275....3P} {275, 3}

\bibitem[\protect\citeauthoryear{{Phillips} et~al.,}{{Phillips} et~al.}{2016}]{phillips2016}
{Phillips} D.~F.,  et~al., 2016, in {Navarro} R.,  {Burge} J.~H.,  eds,  Society of Photo-Optical Instrumentation Engineers (SPIE) Conference Series Vol. 9912, Advances in Optical and Mechanical Technologies for Telescopes and Instrumentation II. p. 99126Z, \mn@doi{10.1117/12.2232452}

\bibitem[\protect\citeauthoryear{{Queloz} et~al.,}{{Queloz} et~al.}{2001}]{queloz2001}
{Queloz} D.,  et~al., 2001, \mn@doi [\aap] {10.1051/0004-6361:20011308}, \href {https://ui.adsabs.harvard.edu/abs/2001A&A...379..279Q} {379, 279}

\bibitem[\protect\citeauthoryear{{Rajpaul}, {Aigrain}, {Osborne}, {Reece}  \& {Roberts}}{{Rajpaul} et~al.}{2015}]{rajpaul2015}
{Rajpaul} V.,  {Aigrain} S.,  {Osborne} M.~A.,  {Reece} S.,   {Roberts} S.,  2015, \mn@doi [\mnras] {10.1093/mnras/stv1428}, \href {https://ui.adsabs.harvard.edu/abs/2015MNRAS.452.2269R} {452, 2269}

\bibitem[\protect\citeauthoryear{{Rajpaul}, {Aigrain}  \& {Buchhave}}{{Rajpaul} et~al.}{2020}]{rajpaul2020}
{Rajpaul} V.~M.,  {Aigrain} S.,   {Buchhave} L.~A.,  2020, \mn@doi [\mnras] {10.1093/mnras/stz3599}, \href {https://ui.adsabs.harvard.edu/abs/2020MNRAS.492.3960R} {492, 3960}

\bibitem[\protect\citeauthoryear{Rasmussen \& Williams}{Rasmussen \& Williams}{2006}]{rasmussen2006}
Rasmussen C.~E.,  Williams C. K.~I.,  2006, Gaussian Processes for Machine Learning.
MIT Press

\bibitem[\protect\citeauthoryear{{Rincon} \& {Rieutord}}{{Rincon} \& {Rieutord}}{2018}]{rincon2018}
{Rincon} F.,  {Rieutord} M.,  2018, \mn@doi [Living Reviews in Solar Physics] {10.1007/s41116-018-0013-5}, \href {https://ui.adsabs.harvard.edu/abs/2018LRSP...15....6R} {15, 6}

\bibitem[\protect\citeauthoryear{{Saar} \& {Donahue}}{{Saar} \& {Donahue}}{1997}]{saar1997}
{Saar} S.~H.,  {Donahue} R.~A.,  1997, \mn@doi [\apj] {10.1086/304392}, \href {https://ui.adsabs.harvard.edu/abs/1997ApJ...485..319S} {485, 319}

\bibitem[\protect\citeauthoryear{{Schwab} et~al.,}{{Schwab} et~al.}{2018}]{schwab2016}
{Schwab} C.,  et~al., 2018, in {Evans} C.~J.,  {Simard} L.,   {Takami} H.,  eds,  Society of Photo-Optical Instrumentation Engineers (SPIE) Conference Series Vol. 10702, Ground-based and Airborne Instrumentation for Astronomy VII. p. 1070271, \mn@doi{10.1117/12.2314420}

\bibitem[\protect\citeauthoryear{{Snodgrass} \& {Ulrich}}{{Snodgrass} \& {Ulrich}}{1990}]{snodgrass1990}
{Snodgrass} H.~B.,  {Ulrich} R.~K.,  1990, \mn@doi [\apj] {10.1086/168467}, \href {https://ui.adsabs.harvard.edu/abs/1990ApJ...351..309S} {351, 309}

\bibitem[\protect\citeauthoryear{{Stalport} et~al.,}{{Stalport} et~al.}{2023}]{Stalport2023}
{Stalport} M.,  et~al., 2023, \mn@doi [\aap] {10.1051/0004-6361/202346887}, \href {https://ui.adsabs.harvard.edu/abs/2023A&A...678A..90S} {678, A90}

\bibitem[\protect\citeauthoryear{{Thompson} et~al.,}{{Thompson} et~al.}{2016}]{thompson2016}
{Thompson} S.~J.,  et~al., 2016, in {Evans} C.~J.,  {Simard} L.,   {Takami} H.,  eds,  Society of Photo-Optical Instrumentation Engineers (SPIE) Conference Series Vol. 9908, Ground-based and Airborne Instrumentation for Astronomy VI. p. 99086F (\mn@eprint {arXiv} {1608.04611}), \mn@doi{10.1117/12.2232111}

\bibitem[\protect\citeauthoryear{{Thompson} et~al.,}{{Thompson} et~al.}{2020}]{thompson2020}
{Thompson} A.~P.~G.,  et~al., 2020, \mn@doi [\mnras] {10.1093/mnras/staa1010}, \href {https://ui.adsabs.harvard.edu/abs/2020MNRAS.494.4279T} {494, 4279}

\bibitem[\protect\citeauthoryear{{Wang}, {Nash}  \& {Sheeley}}{{Wang} et~al.}{1989}]{wang1989}
{Wang} Y.~M.,  {Nash} A.~G.,   {Sheeley} N.~R. J.,  1989, \mn@doi [\apj] {10.1086/168143}, \href {https://ui.adsabs.harvard.edu/abs/1989ApJ...347..529W} {347, 529}

\bibitem[\protect\citeauthoryear{{Wilson} et~al.,}{{Wilson} et~al.}{2022}]{wilson2022}
{Wilson} T.~G.,  et~al., 2022, \mn@doi [\mnras] {10.1093/mnras/stab3799}, \href {https://ui.adsabs.harvard.edu/abs/2022MNRAS.511.1043W} {511, 1043}

\bibitem[\protect\citeauthoryear{{Zechmeister} \& {K{\"u}rster}}{{Zechmeister} \& {K{\"u}rster}}{2009}]{zechmeister2009}
{Zechmeister} M.,  {K{\"u}rster} M.,  2009, \mn@doi [\aap] {10.1051/0004-6361:200811296}, \href {https://ui.adsabs.harvard.edu/abs/2009A&A...496..577Z} {496, 577}

\bibitem[\protect\citeauthoryear{{Zhao} et~al.,}{{Zhao} et~al.}{2022}]{zhao2022}
{Zhao} L.~L.,  et~al., 2022, \mn@doi [\aj] {10.3847/1538-3881/ac5176}, \href {https://ui.adsabs.harvard.edu/abs/2022AJ....163..171Z} {163, 171}

\bibitem[\protect\citeauthoryear{{de Beurs} et~al.,}{{de Beurs} et~al.}{2022}]{deBeurs2022}
{de Beurs} Z.~L.,  et~al., 2022, \mn@doi [\aj] {10.3847/1538-3881/ac738e}, \href {https://ui.adsabs.harvard.edu/abs/2022AJ....164...49D} {164, 49}

\bibitem[\protect\citeauthoryear{{de Beurs} et~al.,}{{de Beurs} et~al.}{2024}]{debeurs2024}
{de Beurs} Z.~L.,  et~al., 2024, \mn@doi [\mnras] {10.1093/mnras/stae207}, \href {https://ui.adsabs.harvard.edu/abs/2024MNRAS.529.1047D} {529, 1047}

\makeatother
\end{thebibliography}



\appendix

\section{YARARA upgrade}
\label{appendix:yarara}

The YARARA pipeline was initially described in \citet{Cretignier2021} as developed for the HARPS spectrograph. Its adaptation for HARPS-N was straighforward given that both instruments are very similar, but peculiarity of the HARPS-N instrument pushed us to slightly modify one of the post-processing recipe. Indeed, the YV1 pipeline was subsenquently improved to better disentangle the change of PSF from stellar activity (see Appendix C in \citet{Stalport2023}). We briefly summarise the method used to measure the variation of the PSF time-series. We assumed that instrumental changes were mainly dominated by symmetrical variations of the line profile, which is valid since any asymmetric change of the line profile would also introduce an RV offset that is not observed. The method therefore consists in (1)~deriving CCFs from the spectra, (2)~correcting the position of their centroid by the RV value obtained from a Gaussian fit, (3)~subtracting the median CCF from each RV-corrected CCF, (4)~ linearly detrending the median-subtracted CCFs from the $S_{HK}$ index\footnote{Where the linears coefficients are determined on the high-pass filter of the signals.}, and (5)~transforming the residuals CCFs $\Delta \text{CCF}(v_i,t)$ into a merged symmetric versus asymmetric $\Delta$ profiles $\Delta$ = [$\Delta \text{CCF}_{\text{sym}}(v_i,t)$, $\Delta \text{CCF}_{\text{asym}}(v_i,t)$]. A PCA is then performed on the transformed residual CCFs and only the components that are significantly symmetric (with a parameter\footnote{see \citet{Stalport2023} for the precise definition of R} $R>3$) were selected. Such a method can be understood as an alternative version of the \texttt{SCALPELS} algorithm \citep{cameron2021} using ACFs. The reason for the usage of the $S_{HK}$ was the almost perfect relation between this index and the filling factor of active regions \citep{cretignier2024} and the relative unsensitivity of these lines to instrumental systematics that mainly affect sharps photospheric lines, but less strongly deep and broad chromospheric lines. 

The time-domain scores of those components are then used as linear predictors to decorrelate the residual spectra time-series from the instrumental PSF change as explained in \citet{Cretignier2021} where we already demonstrated that the method was preserving planetary signals. When we upgraded the recipe in \citet{Stalport2023}, the change of the cryostat was at this time really recent and we did not had enough perspective to understand its effect, but the solar data revealed new properties about the instrumental intervention.  

\begin{figure*}
    \centering
    \includegraphics[width=1.0\linewidth]{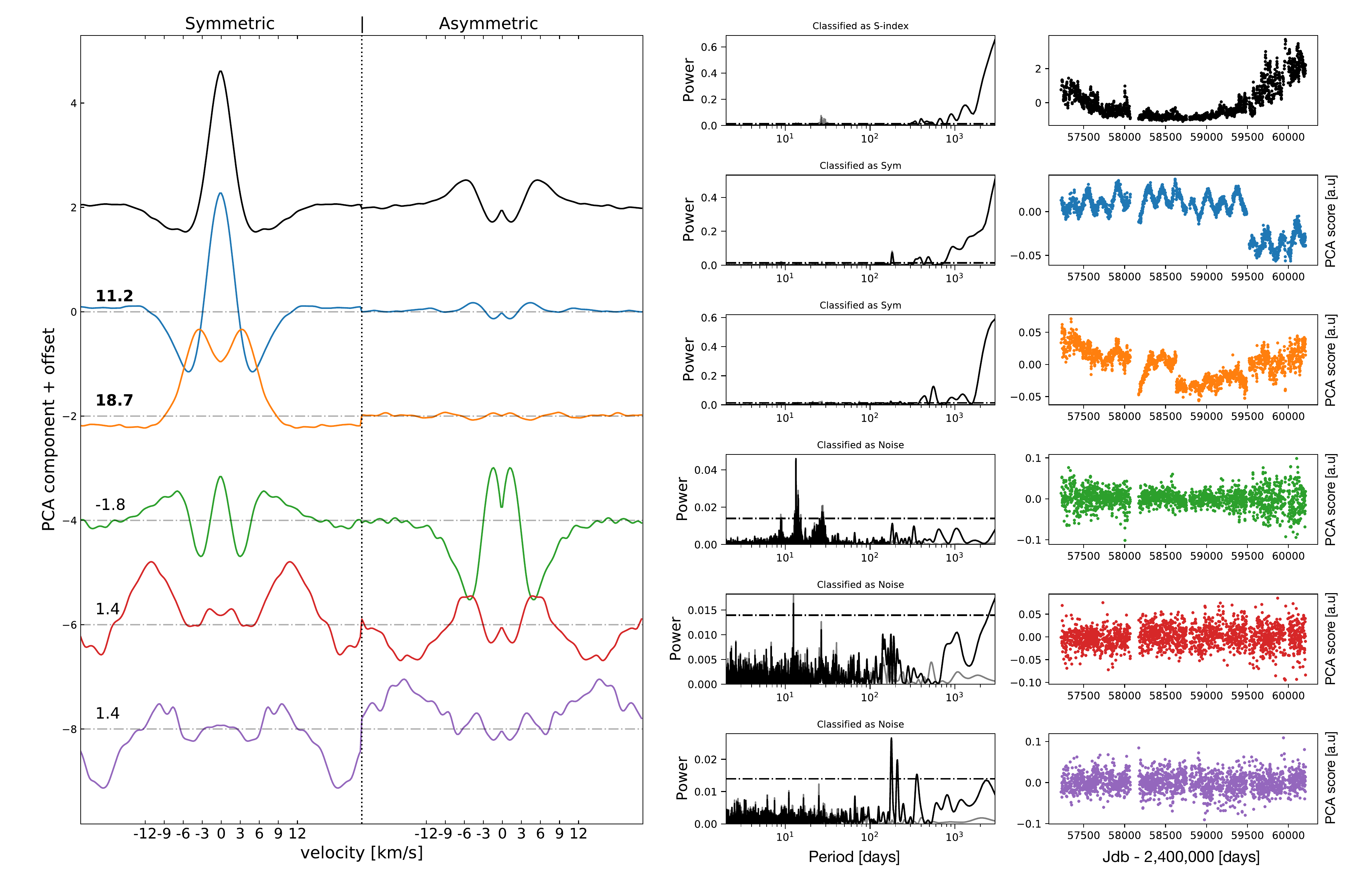}
    \caption{PCA decomposition of the transformed CCFs residuals time-series. The figure is identical to the Fig.C.2 obtained for HD4628 in \citet{Stalport2023} excepted that instead of a CCF obtained with all the stellar lines, we represented here the "blue CCF" ($\lambda<4877$ \AA) analysis. Left: PCA components obtained for the CCF residuals (see main text). The component decorrelate by the $S_{\text{HK}}$ is also shown in black. The vertical dotted line split the symmetric (left) from antisymmetric (right) profiles deformations. The parameter of symmetry $R$ is shown on the left of the profiles and is in bold for significant symmetric profile ($R>3$). Right : Score of the PCA components and corresponding periodograms in days. The time-series coefficient used to correct for the change of the PSF are given by the symmetric components deformation (blue and orange components). }
    \label{fig:psf1}
\end{figure*}

\begin{figure*}
    \centering
    \includegraphics[width=1.0\linewidth]{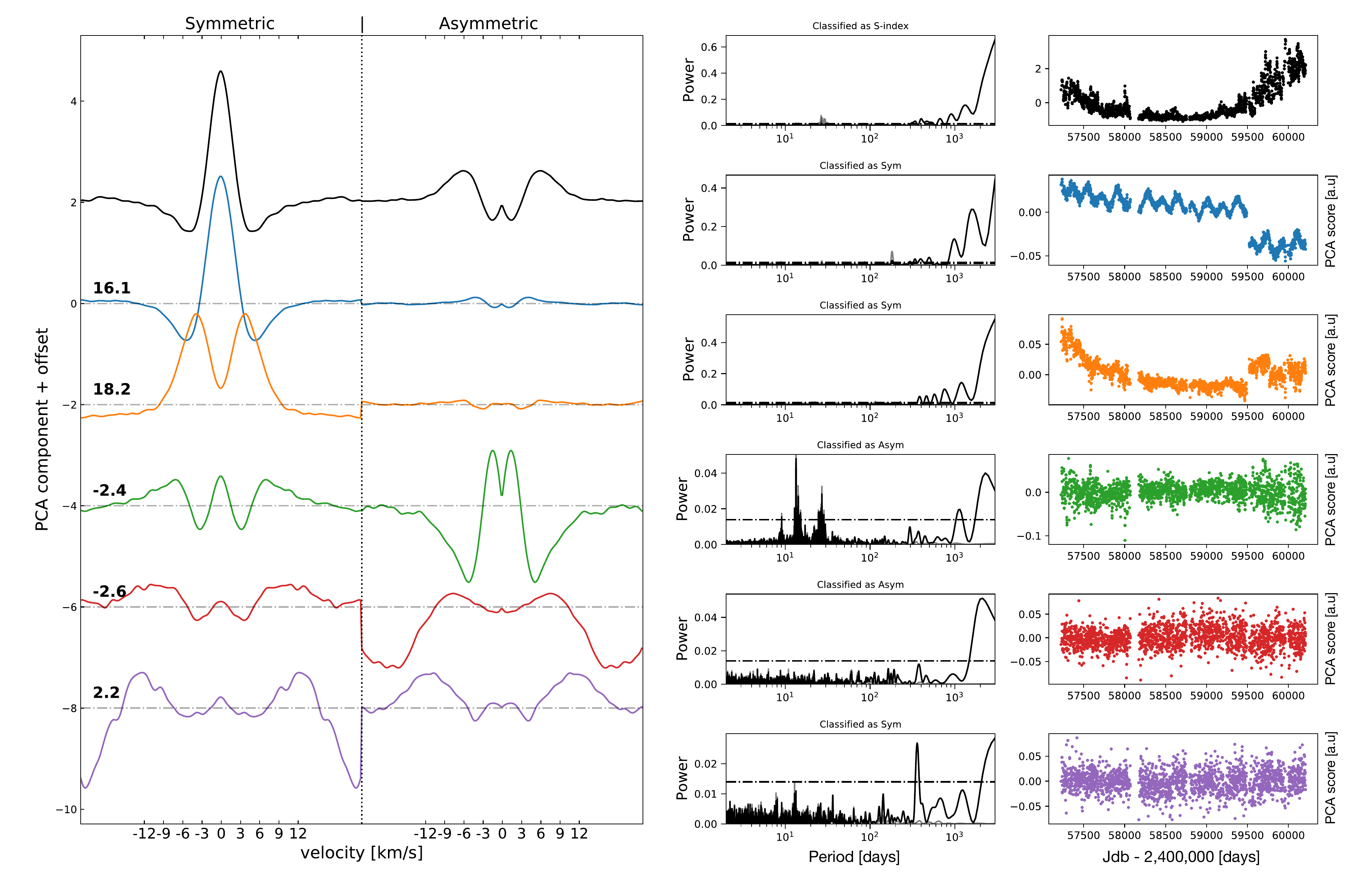}
    \caption{Same as Fig.\ref{fig:psf1} for the green CCF ($4877$ \AA $ < \lambda<5856$ \AA). The jump corresponding for the warm-up of the detector is not visible around BJD=58,629 on contrary to the blue CCF. This indicates that warm-up signatures are mainly visible in the blue part, coherent with the higher contrast of the ghosts.}
    \label{fig:psf2}
\end{figure*}

\begin{figure*}
    \centering
    \includegraphics[width=1.0\linewidth]{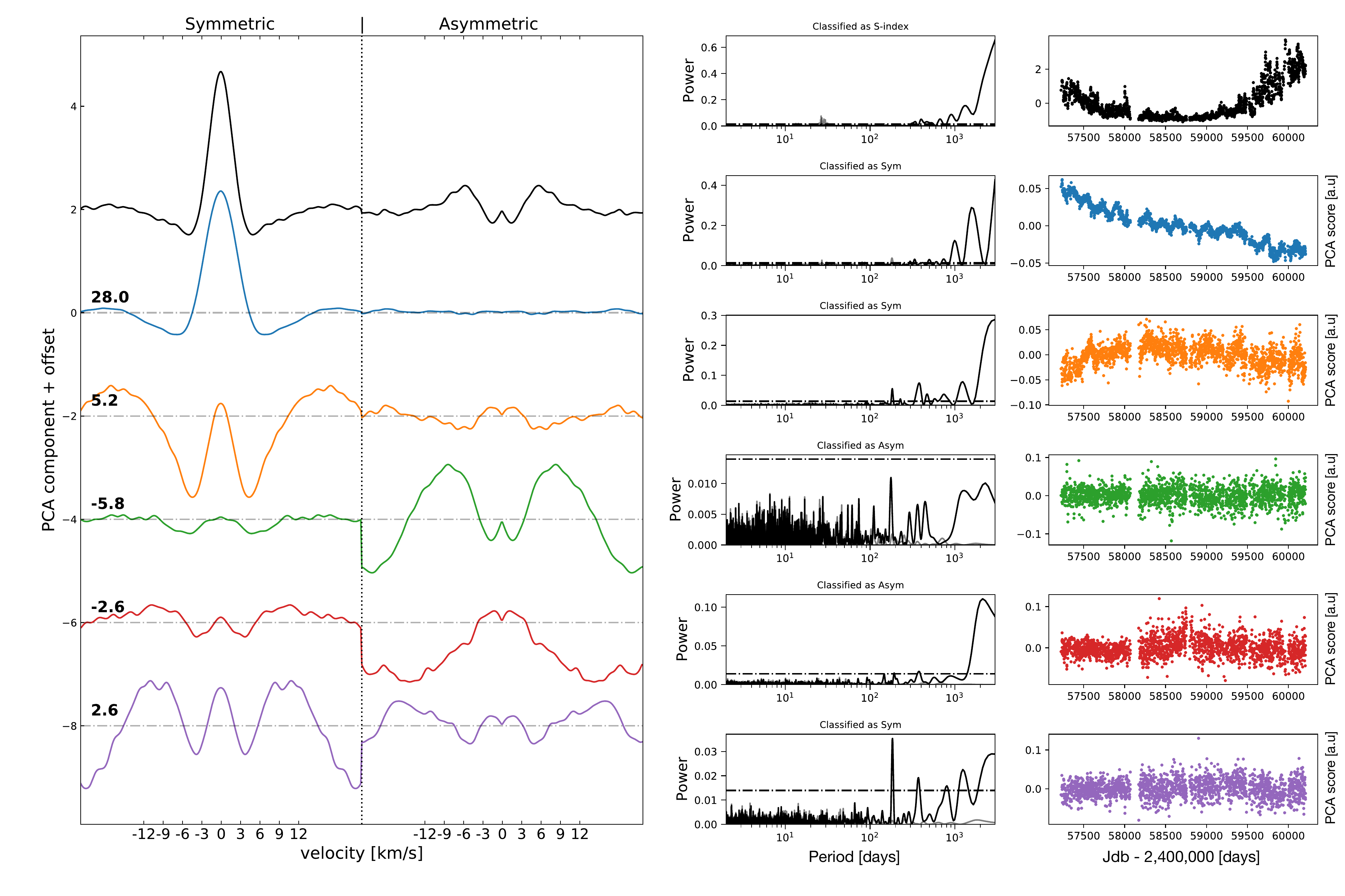}
    \caption{Same as Fig.\ref{fig:psf1} for the red CCF ($\lambda>5856$ \AA). The jump corresponding for the warm-up of the detector is not visible, neither the cryostat change.}
    \label{fig:psf3}
\end{figure*}

The previous recipe was mainly motivated to correct the large instrumental PSF defocus visible in 2012 at the beginning of the instrument's lifetime (before the installation of the solar telescope). Ater the change of the cryostat, we observed that the previous developped recipe was unperfect. Indeed, the change of the PSF was previously obtained with a white CCF using the full wavelength bandpass between 3900 and 6835 \AA, since at first order, all the stellar lines behave in a similar way on the spectrum. However, investigations on the Sun show that some instrumental effects are also very chromatic. As an example, the warm-up of the detector mainly affect the blue part of the spectrum since the contrast of the ghosts is larger in the blue spectral range. Also, the change of the cryostat introduced a larger signal in the green and blue part of the spectrum compared to the red, but the reason for this signature is yet unknown. We therefore modified the recipe developed in \citet{Stalport2023} by computing three colours CCFs (blue, green and red) splited evenly in wavelength and used the obtained vector to decorrelate the spectra residuals time-series as explained in YARARA \citet{Cretignier2021}. The reason for three colour is mainly justified by the usual trade-off sensitivity versus SNR. We displayed the same figure as the one obtained for HD4628 in \citet{Stalport2023} for the solar observations in Fig.\ref{fig:psf1} and Fig.\ref{fig:psf2} with blue and green CCFs respectively. 

The first PCA component of blue and green CCFs reveal the clear signature of the change of v sin i and of the cyostat change around BJD=2,459,491. The second PCA component in the blue contains extra component such as the warm-up at BJD=2,458,629, while the green CCFs contains a long trend with the cryostat offset. This long trend may be correlated with the long decrease in SNR due to ageing of the solar dome. The third PCA component is mainly an antisymmetric variation (therefore not used) and clearly show the stellar activity power at Prot/2.

\section{Planet detection maps}

\begin{figure*}
    \centering
    \includegraphics[width=0.45\linewidth]{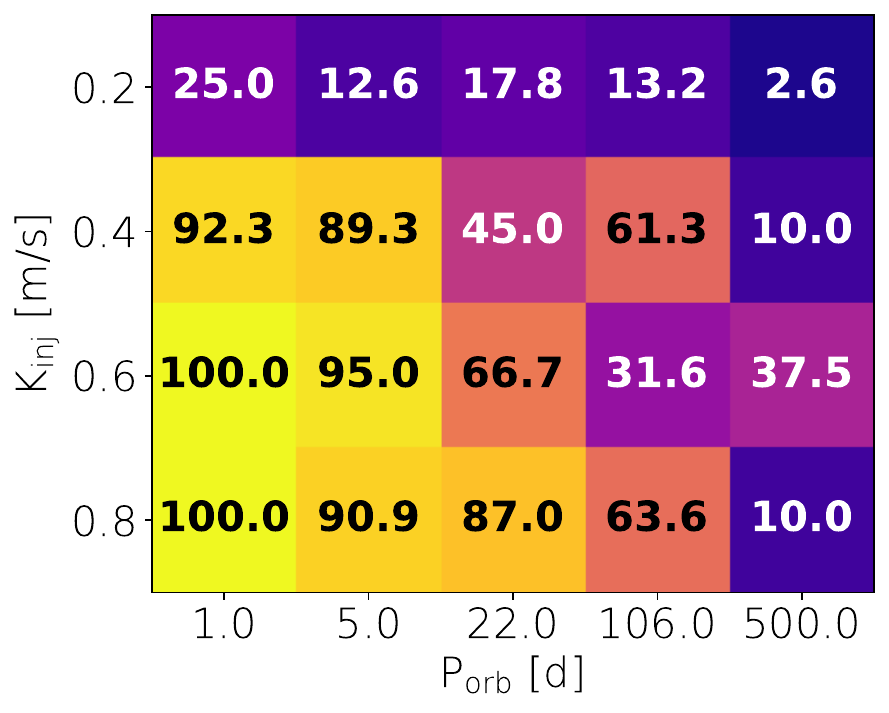} 
    \hspace{0.05\linewidth}
    \includegraphics[width=0.45\linewidth]{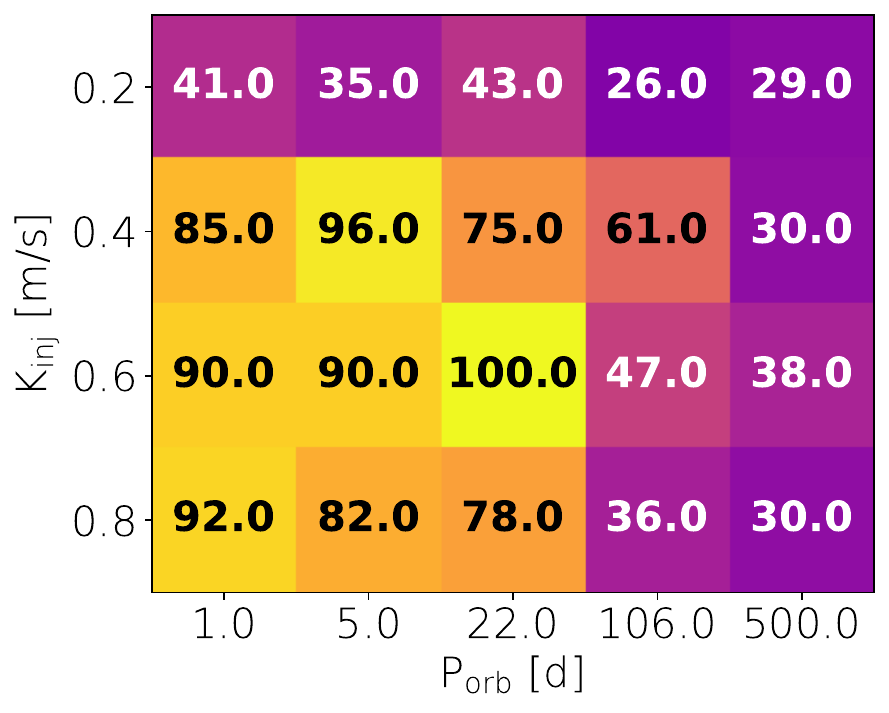}    
    \caption{Planet detectability maps, as a function of the RV semi-amplitude and orbital period, recovered from the HARPS-N solar RVs without activity filtering (left-hand panel) and using the activity-filtering framework of Sec.~\ref{ssec:framework}. In each panel, the color code depicts the planet detection rate (darkblue / yellow indicating detection rates close to 0 / 100\%). The numbers within each cell indicate the detection rate in percent.}
    \label{fig:sensitivity_maps}
\end{figure*}

Fig.~\ref{fig:sensitivity_maps} shows the detection rates of the planets injected in the HARPS-N solar CCFs without and with activity filtering (see Sec.~\ref{ssec:results_planet}), as a function of the planet RV semi-amplitude \kinj\ and orbital period \porb. Planets considered to be detected if the recovered RV semi-amplitude differs by less than 1$\sigma$ from \kinj\ and differs from 0 by at least 3$\sigma$. As outlined in Sec.~\ref{ssec:results_planet}, the activity-filtering framework increases the sensitivity to planets with low semi-amplitude (typically lower than 0.4\,\ms) and larger orbital period (typically greater than $\sim$100\,d). We note a very light decrease of the detection rate of short-period higher-amplitude planet signatures after activity filtering. This can be understood as uncertainties on the RV semi-amplitude are significantly smaller when estimated from the activity-filtered RVs than from the raw RVs (see Fig.~\ref{fig:err_kp}). Note that the best estimates of the RV semi-amplitudes of the planets missed within ths (\kinj,\porb) space still lie within 2$\sigma$ of the injected RV semi-amplitude.

\section{Gaussian Process modelling}

\begin{figure*}
    \centering
    \includegraphics[width=\linewidth]{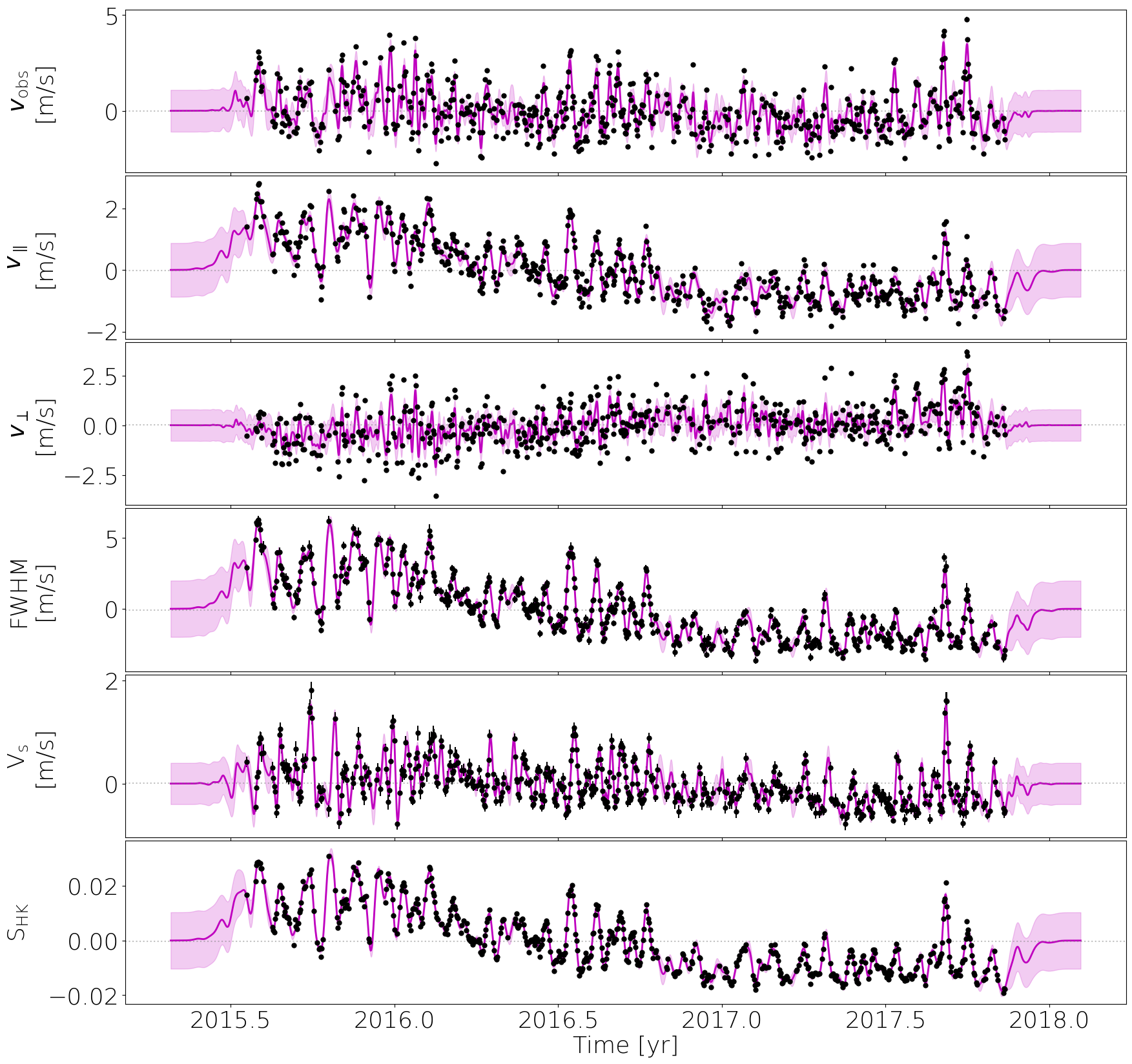}
    \caption{Best-fitting GP prediction to different time series of activity indicators extracted from HARPS-N Solar spectra, during Season 1 (i.e. 2015 to 2018, see Tab.~\ref{tab:rms_rvs}). In each panel, the data points with formal 1$\sigma$ error bars are shown in black dots, and the best-fitting GP prediction (resp. 1$\sigma$ error bands) is indicated by the magenta solid line (resp. shaded bands).}
    \label{fig:fit_season1}
\end{figure*}

\begin{figure*}
    \centering
    \includegraphics[width=\linewidth]{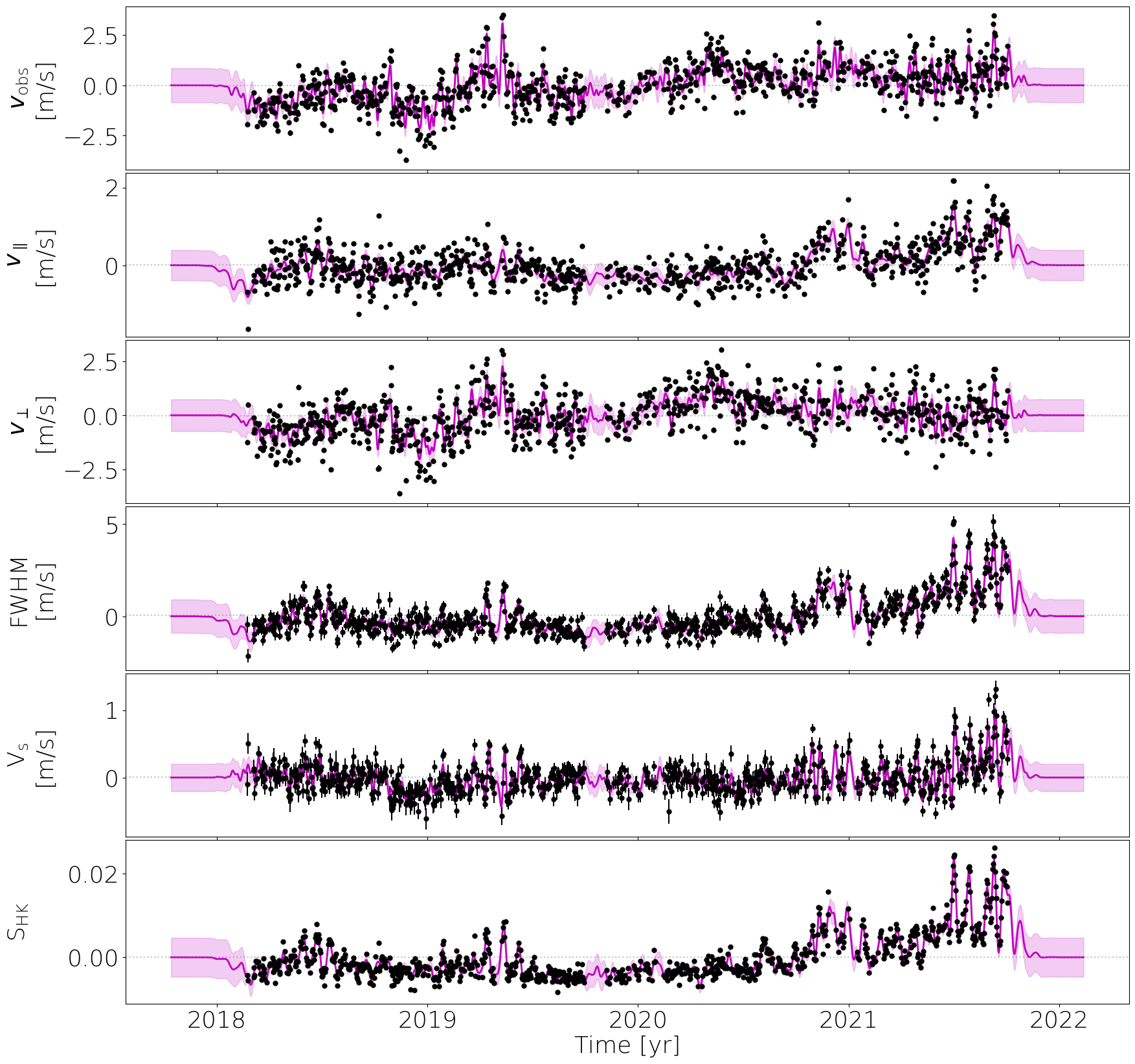}
    \caption{Same as Fig.~\ref{fig:fit_season1}, for the activity indicators observed during Season 2 (i.e. during solar minimum; see Tab.~\ref{tab:rms_rvs}).}
    \label{fig:fit_season2}
\end{figure*}

\begin{figure*}
    \centering
    \includegraphics[width=\linewidth]{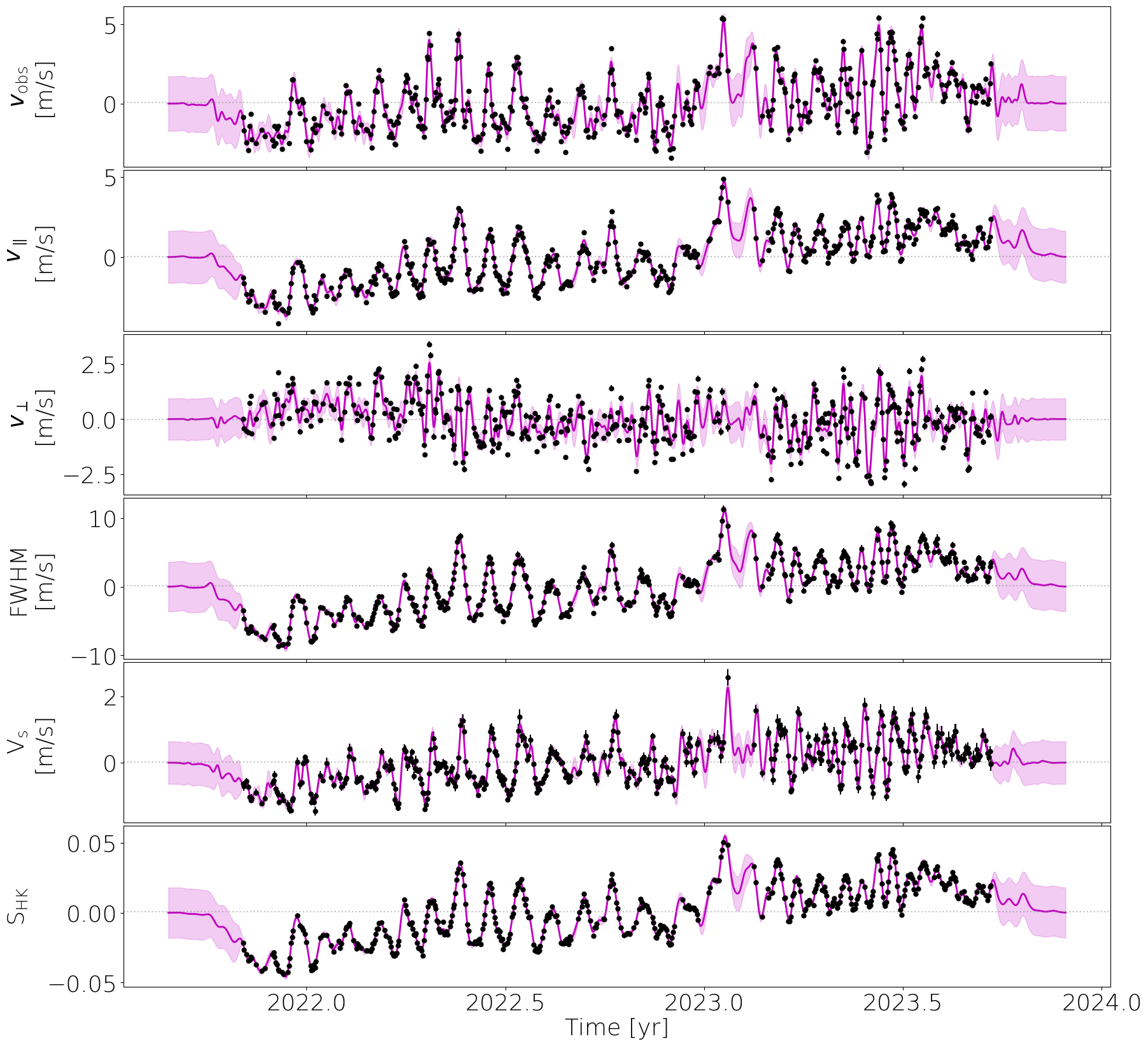}
    \caption{Same as Fig.~\ref{fig:fit_season1}, for the activity indicators observed during Season 3 (i.e. start of cycle 25; see Tab.~\ref{tab:rms_rvs}).}
    \label{fig:fit_season3}
\end{figure*}

\begin{figure*}
    \includegraphics[width=\linewidth]{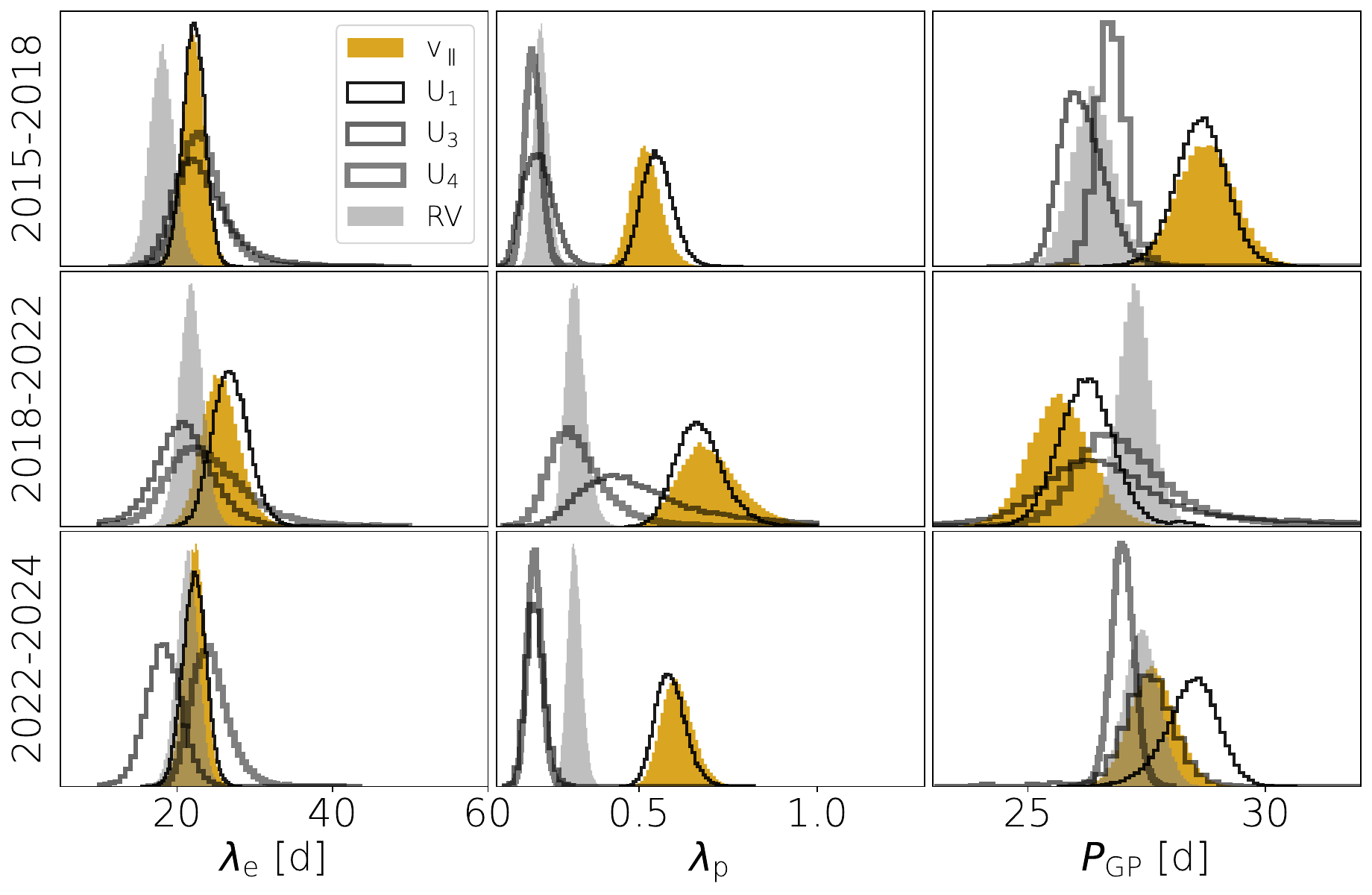}
    \caption{Uni-dimensional posterior distributions of the GP evolution time scale (left-hand column), inverse-harmonic complexity (middle column) and period (right-hand column) obtained by modelling the HARPS-N solar RVs (filled grey histograms), \vper\ (filled yellow histograms), and $\boldsymbol{U_{1}}$, $\boldsymbol{U_{3}}$ and $\boldsymbol{U_{4}}$ (thin, medium thick and thick gray lines).}
    \label{app:fig:posterior_sha}
\end{figure*}

In this appendix, we give additional details of the GP modelling of spectroscopic activity indicators, namely the HARPS-N solar RVs (\vobs), the time series of \vper\ and \vpar, computed in Sec.~\ref{ssec:planet-free-sun}, and usual activity proxies (i.e. FWHM, V$_{\mathrm{s}}$ and S$_{\mathrm{HK}}$). In Fig.~\ref{fig:fit_season1}, Fig.~\ref{fig:fit_season2} and ~\ref{fig:fit_season3}, we show the quasi-periodic 1D GP fit to the different time series in Season~1 (2015-2018), 2 (2018-2021.8) and 3 (2021.8-2024), respectively. The posterior densities of the hyper-parameters of the 1D GP fit to the time series of eigenvectors $\boldsymbol{U_{1}}$, $\boldsymbol{U_{3}}$ and $\boldsymbol{U_{4}}$, extracted in Sec.~\ref{ssec:planet-free-sun}, are shown for all three seasons in Fig.~\ref{app:fig:posterior_sha}. We visually see the change in the solar activity regime from one season to the next. Activity-induced fluctuations are most clearly in the start of cycle~25 (Season~3, Fig.~\ref{fig:fit_season3}), when the Sun is more active. In this regime, rotationally-modulated signals are present in all activity proxies except \vper, where they have been filtered out. These activity-induced fluctuations appear significantly faster-evolving at the end of cycle~24 (Season~1, Fig.~\ref{fig:fit_season1}), which is consistent with the lower GP evolution time scale and inverse harmonic complexity of this season (i.e. $\lambda_{\mathrm{e}}$ and $\lambda_{\mathrm{p}}$ from Tab.~\ref{tab:parameters_1d}). The signal is the most complex over the solar minimum (Season~2, Fig.~\ref{fig:fit_season2}). In this case, the GP fit is controlled by the few activity-induced fluctuations, visible notably in the end of the season, which explains why the best-fit GP parameters in Tab.~\ref{tab:parameters_1d} are similar to those obtained in Season~3. Outside these activity-dominated regions, we note that the FWHM and S$_{\mathrm{HK}}$ index are mostly flat, whereas the RVs still exhibit fluctuations of about 1\,\ms\ peak-to-peak, most likely due to an interplay between activity residuals, granulation and instrument stability.

\begin{figure}
    \centering
    \includegraphics[width=\linewidth]{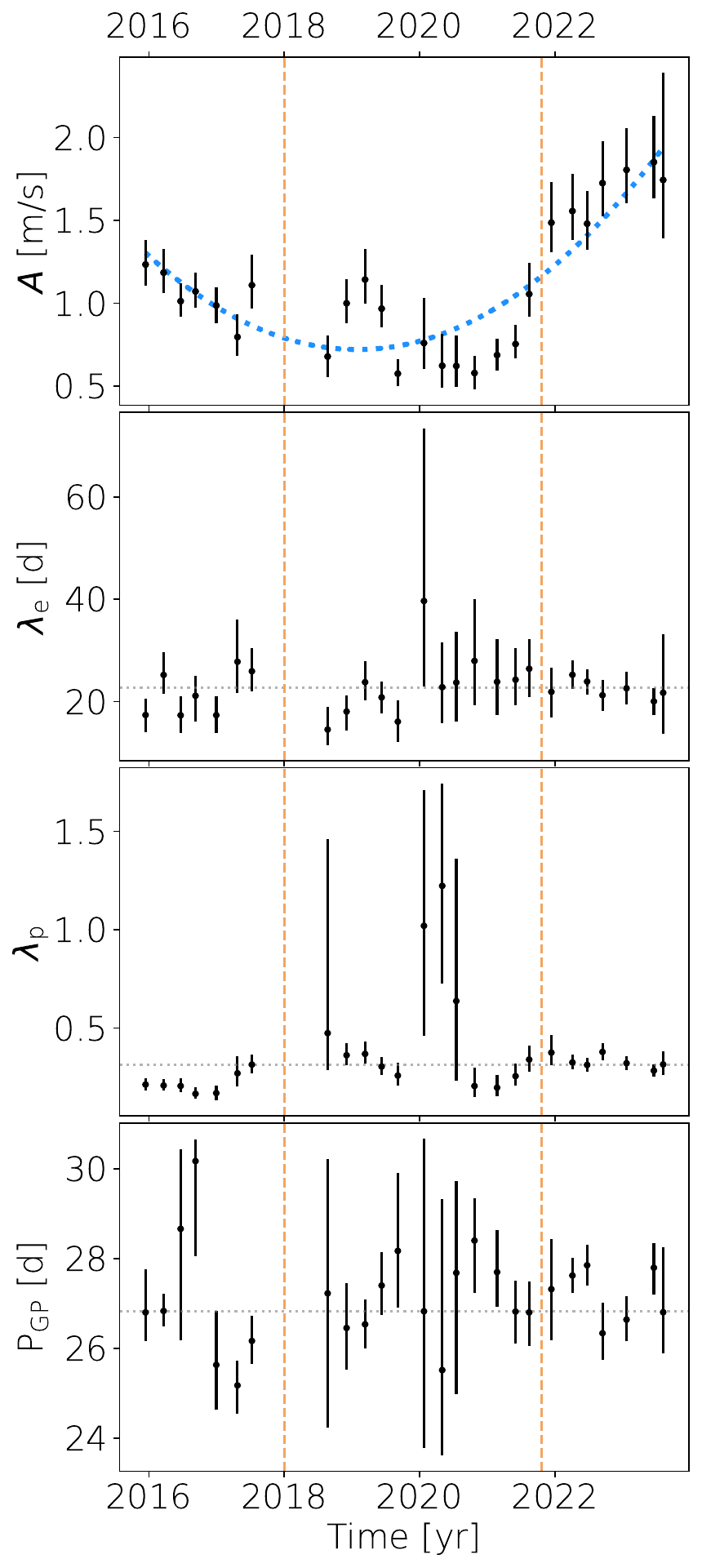}
    \caption{Distribution of the best-fitting GP hyperprameters obtained by modelling 270-d chunks of HARPS-N solar RV. The dotted blue line in the top panel shows the best-fitting parabola to the GP amplitudes. The three seasons defined in Tab.~\ref{tab:rms_rvs} (i.e. end of Cycle~24, solar minimum and start of Cycle~25) are delimited by the vertical dashed lines.}
    \label{fig:distrib_chunks}
\end{figure}


\bsp	
\label{lastpage}
\end{document}